\documentclass[preprint,12pt,3p]{elsarticle}

\usepackage{graphicx}
\usepackage{caption}
\usepackage{subcaption}
\usepackage{psfrag}
\usepackage{array}
\usepackage{multirow}
\usepackage{comment}
\usepackage{amsmath}%
\newcommand{\norm}[1]{\left\lVert#1\right\rVert}
\DeclareMathOperator*{\argmin}{argmin}
\usepackage{xcolor}
\usepackage{amssymb}
\usepackage{amsfonts}%
\usepackage{amssymb}
\usepackage{algorithm}
\usepackage{algpseudocode}
\usepackage{setspace}
\usepackage{setspace}
\usepackage{textcomp}
\usepackage{enumitem}

\usepackage[switch, running]{lineno}

\journal{Computer Methods in Applied Mechanics and Engineering}

\begin{document}
\sloppy
\begin{frontmatter}


\title{Deep-learning-based surrogate flow modeling and geological parameterization for data assimilation in 3D subsurface flow}


\author[label1,label5]{Meng Tang\corref{cor1}}
\address[label1]{367 Panama Street, Stanford, CA, 94305}
\address[label5]{Department of Energy Resources Engineering, Stanford University}

\cortext[cor1]{Corresponding author}

\ead{mengtang@stanford.edu}

\author[label1,label5]{Yimin Liu}

\ead{yiminliu@stanford.edu}

\author[label1,label5]{Louis J. Durlofsky}
\ead{lou@stanford.edu}

\begin{abstract}

Data assimilation in subsurface flow systems is challenging due to the large number of flow simulations often required, and by the need to preserve geological realism in the calibrated (posterior) models. In this work we present a deep-learning-based surrogate model for two-phase flow in 3D subsurface formations. This surrogate model, a 3D recurrent residual U-Net (referred to as recurrent R-U-Net), consists of 3D convolutional and recurrent (convLSTM) neural networks, designed to capture the spatial-temporal information associated with dynamic subsurface flow systems. A CNN-PCA procedure (convolutional neural network post-processing of principal component analysis) for parameterizing complex 3D geomodels is also described. This approach represents a simplified version of a recently developed supervised-learning-based CNN-PCA framework. The recurrent R-U-Net is trained on the simulated dynamic 3D saturation and pressure fields for a set of random `channelized' geomodels (generated using 3D CNN-PCA). Detailed flow predictions demonstrate that the recurrent R-U-Net surrogate model provides accurate results for dynamic states and well responses for new geological realizations, along with accurate flow statistics for an ensemble of new geomodels. The 3D recurrent R-U-Net and CNN-PCA procedures are then used in combination for a challenging data assimilation problem involving a channelized system. Two different algorithms, namely rejection sampling and an ensemble-based method, are successfully applied. The overall methodology described in this paper may enable the assessment and refinement of data assimilation procedures for a range of realistic and challenging subsurface flow problems.

\end{abstract}


\end{frontmatter}

\section{Introduction}

History matching entails the calibration of geological model parameters such that flow simulation predictions are in essential agreement with historical (observed) data. The prediction uncertainty associated with these calibrated (posterior) models is typically less than that from the initial (prior) models, which renders them more useful for reservoir/aquifer management. History matching involves many challenges, including the computational demands associated with the required flow simulations, and the need to preserve geological realism in the posterior models. Thus this application will benefit from the development and use of accurate surrogate models for flow predictions and effective geological parameterizations.

In recent work, a surrogate flow model that applies a recurrent residual U-Net (R-U-Net) procedure was used in conjunction with a deep-learning-based geological parameterization technique, referred to as CNN-PCA \citep{liu2019deep, liu2019multilevel}, to successfully history match 2D channelized systems \citep{tang2020deep}. In this paper, we extend the surrogate dynamic flow model to handle 3D systems. The standalone CNN-PCA parameterization, which enables complex geomodels to be described in terms of a small set of uncorrelated variables, has recently been extended to treat 3D systems \citep{liu20203d}. Here we introduce a simplified yet effective variant of the general 3D CNN-PCA procedure. This parameterization is then combined with the 3D surrogate flow model for uncertainty quantification and history matching.



There have been a number of deep-neural-network architectures developed for surrogate flow modeling in subsurface systems. Zhu and Zabaras~\citep{zhu2018bayesian} devised a fully convolutional encoder-decoder architecture to capture pressure and velocity maps for single-phase flow problems characterized by random 2D Gaussian permeability fields. In later work, an autoregressive strategy was integrated with a fully convolutional encoder-decoder architecture for the prediction of time-dependent transport in 2D problems \citep{mo2018deep}. Tang et al.~\citep{tang2020deep} proposed a combination of a residual U-Net with convLSTM to capture the saturation and pressure evolution in 2D oil-water problems, with wells operating under wellbore pressure control. Mo et al.~\cite{mo2020integration} applied a 2D deep residual dense convolutional neural network to predict solution concentration distributions in 3D groundwater flow problems. In this treatment, the 3D output is a stack of output from 2D convolutional channels. Within the context of optimization (rather than history matching), Jin et al.~\citep{jin2020deep} presented an autoregressive embed-to-control framework to predict well responses and dynamic state evolution with varying well controls, for a fixed geological description.

The approaches cited above all require numerical simulation data to train the neural network. In physics informed surrogate models \citep{han2018solving, berg2018unified, raissi2019physics, sun2020surrogate, mao2020physics}, by contrast, simulation data are not needed for training. With these procedures the governing partial differential equations (PDEs) are approximated by formulating PDE residuals, along with the initial and boundary conditions, as the objective function to be minimized by the neural network \citep{lu2019deepxde}. Issues with existing physics informed procedures may arise, however, if these methods are applied for strongly heterogeneous models involving large numbers of wells (which act as `internal' boundary conditions), or with large sets of geomodels characterized by different detailed property distributions. Both of these capabilities are required for data assimilation in realistic problems. Zhu et al.~\cite{zhu2019physics} proposed a physics-constrained surrogate model for uncertainty quantification with high-dimension discretized fields, where an approximate governing PDE residual is applied as the training objective. To date, this promising approach has only been implemented for steady-state problems. 

Other recent applications of deep learning in subsurface flow settings include the use of a CNN-based surrogate model to predict permeability in tight rocks with strong heterogeneity \citep{tian2020surrogate}, and the use of neural networks in the context of multicomponent thermodynamic flash calculations \citep{zhang2020self}. The latter development could be quite useful in compositional reservoir simulation, as a substantial fraction of the computational effort is often associated with the phase equilibria calculations.


As noted earlier, the effective parameterization of geomodels is also an important component of many data assimilation procedures. With such a treatment, the detailed geological model (e.g., porosity and/or permeability in every grid block in the model) can be represented with many fewer parameters than there are grid blocks. This means that a relatively small set of parameters must be determined during data assimilation, and that posterior geomodels will maintain geological realism. Tools based on deep-learning have been shown to be applicable for such geological parameterizations. Specific approaches include those based on variational autoencoders (VAEs) \citep{Laloy2017, Canchumuni2019a} and generative adversarial networks (GANs) \citep{Chan2017, Chan2018, Dupont2018, Mosser2018, Laloy2018, Laloy2019, Chan2020, Azevedo2020}. Algorithms based on a combination of VAE and GAN have also been devised \citep{mo2020integration}. CNN-PCA methods, which use CNNs to post-process models parameterized using principal component analysis (PCA), have been developed for 2D systems \citep{liu2019deep, liu2019multilevel}, and have recently been extended to 3D \citep{liu20203d}. A variant of this 3D CNN-PCA method will be presented and applied in this paper.

In this work we (1) extend the existing recurrent R-U-Net procedure to provide a surrogate model for dynamic 3D two-phase flow problems, (2) present a simplified 3D CNN-PCA parameterization for binary `channelized' geomodels, and (3) combine the deep-learning-based 3D surrogate flow model and geomodel parameterization to perform uncertainty quantification and data assimilation. The 3D recurrent R-U-Net involves a similar layout to that used for 2D problems, though we now apply 3D CNN and convLSTM architectures. The 3D recurrent R-U-Net surrogate flow model is trained using high-fidelity simulation results for realizations generated by 3D CNN-PCA. The 3D CNN-PCA procedure applied here involves the use of a supervised-learning approach to map PCA models to channelized realizations. The combined methodology is then applied for data assimilation. Two different history matching algorithms -- rejection sampling and ensemble smoother with multiple data assimilation \citep{emerick2013ensemble} -- are implemented and assessed.


This paper proceeds as follows. In Section~\ref{sec:gov_eqn} we present the equations governing two-phase subsurface flow and describe the well model used in this work. Next, in Section~\ref{sec:cnnpca}, we describe the simplified 3D CNN-PCA procedure used for geomodel parameterization. The deep-neural-network architecture and training setup for the 3D recurrent R-U-Net surrogate flow model is presented in Section~\ref{sec:flow_surrogate}. Detailed flow comparisons between model predictions and high-fidelity simulation results appear in Section~\ref{sect:saturation-pressure-map-eval}. The use of the 3D recurrent R-U-Net surrogate and CNN-PCA for data assimilation is considered in Section~\ref{sec:HM_results}. A summary and suggestions for future work are provided in Section~\ref{sec:concl}. Architecture details for 3D CNN-PCA and 3D recurrent R-U-Net are provided in the Appendix.

\section{Governing Equations for Two-phase Flow}
\label{sec:gov_eqn}

We consider the flow of two immiscible fluids in subsurface formations. The governing equations and problem setup are applicable for oil reservoir simulation (e.g., production of oil via water injection), or for the modeling of an environmental remediation project involving water and a nonaqueous phase liquid (NAPL). Our terminology and notation are consistent with the oil-water problem. The system is 3D and includes gravitational effects. 

Phases are denoted by $j=o,~w$, where $o$ indicates oil phase and $w$ water phase. Due to immiscibility, the oil and water components exist only in their respective phases. Mass conservation is expressed as
\begin{equation} 
    -\nabla \cdot (\rho_j \mathbf{v}_j) - q_j^\text{w} = \frac{\partial}{\partial{t}}(\phi \rho_j S_j), \ \ \  j=o,~w.
\label{eq:mass-balance-eq}
\end{equation}
Here $\rho_j$ denotes phase density, $\mathbf{v}_j$ is the phase (Darcy) velocity, $q_j^\text{w}$ is the source/sink term (superscript $\text{w}$ indicates well), $t$ is time, $\phi$ is the rock porosity (volume fraction of the pore space), and $S_j$ is the phase saturation (volume fraction occupied by phase $j$, with $S_o+S_w=1$). Darcy velocity $\mathbf{v}_j$ is given by
\begin{equation} 
     \mathbf{v}_j =  -\frac{\mathbf{k} k_{rj}(S_j)}{\mu_j} (\nabla p_j - \rho_j g \nabla z),
\label{eq:darcy-eq}
\end{equation}
where $\mathbf{k}$ is the absolute permeability tensor (permeability can be viewed as a flow conductivity), $k_{rj}$ is the relative permeability to phase $j$, $\mu_j$ indicates phase viscosity, $p_j$ denotes phase pressure, $g$ is gravitational acceleration, and $z$ is depth. Here, $k_{rj}$ is usually a strongly nonlinear function of phase saturation $S_j$, and we may also have $\mu_j=\mu_j(p_j)$. In this work we neglect capillary pressure effects, which generally have minimal impact at field scales. This means we have $p_o=p_w=p$.

Eqs.~\ref{eq:mass-balance-eq} and \ref{eq:darcy-eq} are typically discretized using finite volume treatments, and the resulting set of nonlinear algebraic equations is solved using Newton's method. The solution provides the pressure and saturation in all grid blocks in the model. The problem is complicated by the fact that permeability and porosity can be highly discontinuous in space (as different geologic features may display very different properties), and because the saturation field develops shocks as a result of the near-hyperbolic character of the water transport equation. In this work, all high-fidelity simulations (HFS), i.e., the numerical solution of Eqs.~\ref{eq:mass-balance-eq} and \ref{eq:darcy-eq}, are performed using Stanford's Automatic Differentiation General Purpose Research Simulator, ADGPRS \cite{zhou2012parallel}.

Well responses (phase flow rates into or out of wells or wellbore pressure) are of primary interest in many subsurface flow problems, and these quantities typically comprise the key data to be assimilated. Wells appear in the finite volume formulation through the discretized representation of the source term $q_j^\text{w}$. In this work we consider 3D models with vertical wells that penetrate multiple blocks. For a grid block containing a well (referred to as a well block), well rates are related to the pressure and saturation in the block through the Peaceman model \citep{peaceman1983interpretation}:
\begin{equation}
\label{eq:well_flow}
    \left(q_j^\text{w}\right)_i = \text{WI}_i  \left(\frac {k_{rj}\rho_j} {\mu_j} \right)_i (p - p^\text{w})_i.
\end{equation}
Here $\left(q_j^\text{w}\right)_i$ is the source/sink term for phase $j$ in block $i$, $p_i$ is well-block pressure, $p^\text{w}_i$ is the wellbore pressure in block $i$, and $\text{WI}_i$ is the well index in block $i$. The well index is given by
\begin{equation}
\label{eq:well_index}
    \text{WI}_i = \frac {2 \pi k_i \Delta z} {\ln \frac{r_0} {r_w}},
\end{equation}
where $k_i$ the grid block permeability (assumed to be isotropic), $\Delta z$ is the block thickness, $r_w$ is the wellbore radius, and $r_0=0.2 \Delta x$, where $\Delta x$ is the block dimension in the $x$-direction (here we assume $\Delta x= \Delta y$). More general expressions for $\text{WI}_i$ exist for cases with anisotropic permeability and/or $\Delta x \neq  \Delta y$.

For a vertical well that penetrates multiple blocks, the wellbore pressure $p^\text{w}_i$ varies with depth $z$ as a result of gravitational effects. Wellbore pressure at the first (uppermost) perforation is referred to as the bottom-hole pressure (BHP). The wellbore pressure in the block below, $p^\text{w}_{i+1}$, is computed using 
\begin{equation}
\label{eq:standard-well}
    p^\text{w}_{i+1} = p^\text{w}_{i} + \rho_{i, i+1}g\Delta z_{i, i+1}.
\end{equation}
Here $\rho_{i, i+1}$ denotes the fluid mixture density between neighboring well-blocks $i$ and $i+1$ and $\Delta z_{i, i+1}$ is the depth difference. The fluid mixture density can be approximated based on the individual phase densities and the phase flow rates. See \citep{jiang2008techniques} for further details.

In all flow results presented in this paper, the treatment described above is applied to compute well rates for both the surrogate model and for the ADGPRS high-fidelity simulations. This procedure has been implemented as a standalone module that accepts BHP, grid geometry and well-block properties and simulation results (well-block permeability, pressure and saturation) as input, and provides $\left(q_j^\text{w}\right)_i$, for $j=o,~w$, for all well blocks $i$, as output. For the ADGPRS runs we could also use the well model in the simulator, which gives very similar results. To assure identical well-model treatments, however, we compute ADGPRS (HFS) well rates from the global simulation results using Eqs.~\ref{eq:well_flow}, \ref{eq:well_index} and \ref{eq:standard-well}.


\section{3D CNN-PCA Low-dimensional Model Representation}
\label{sec:cnnpca}

In recent work, Liu and Durlofsky~\citep{liu20203d} introduced 3D CNN-PCA for the parameterization of complex geomodels in 3D. Here we apply a simplified yet effective variant of this general approach, which performs very well for the geomodels considered in this study. 

The 3D CNN-PCA method represents an extension of the earlier 2D procedure \citep{liu2019deep}. The method post-processes low-dimensional PCA (principal component analysis) models such that large-scale geological structures, characterized in terms of multiple-point spatial statistics, are recovered. Parameterizations of this type are very useful in the context of uncertainty quantification and data assimilation, as we will see later.

The method developed in \citep{liu20203d} includes three loss terms -- reconstruction loss, style loss and hard data loss. Style loss can be important for large models with few wells (and thus limited conditioning data), as it penalizes realizations that do not include the requisite large-scale features. In the system considered here, however, there is a sufficient amount of conditioning data, and models developed using only reconstruction loss and hard data loss are fully adequate. Therefore, here we present this simplified version of 3D CNN-PCA, which includes only these two losses.



A geological realization can be represented as a vector $\mathbf{m} \in \mathbb{R}^{n_b}$ of geologic variables, where $n_b$ is the number of grid blocks in the model. Geological parameterizations map $\mathbf{m}$ to a new low-dimensional variable $\boldsymbol{\xi} \in \mathbb{R}^{l}$, where $l \ll n_b$. The first step in our procedure is to construct a PCA parameterization of the geomodel. As described in detail in \citep{liu2019deep}, this entails generating a set of $n_r$ realizations of $\mathbf{m}$ using geomodeling tools such as Petrel \citep{manual2007petrel}, and assembling these realizations in a centered data matrix $Y \in \mathbb{R}^{n_b \times n_r}$,
\begin{equation}
    \label{eq_center_data_matrix}
    Y = \frac{1}{\sqrt{n_r - 1}}[\mathbf{m}_1 - \bar{\mathbf{m}} \quad \mathbf{m}_2 - \bar{\mathbf{m}} \quad \cdots \quad \mathbf{m}_{n_r} - \bar{\mathbf{m}}],
\end{equation}
where $\mathbf{m}_i \in \mathbb{R}^{n_b}$ represents realization $i$ and $\bar{\mathbf{m}} \in \mathbb{R}^{n_b}$ is the mean of the $n_r$ realizations. Note that these realizations are all conditioned to available data at well locations (this so-called `hard' data could correspond to rock type, porosity, and/or permeability). Singular value decomposition of $Y$ allows us to write $Y = U\Sigma V^{\intercal}$, where $U \in \mathbb{R}^{n_b \times n_r}$ and $V \in \mathbb{R}^{n_r \times n_r}$ are left and right singular matrices, and $\Sigma \in \mathbb{R}^{n_r \times n_r}$ is a diagonal matrix containing the singular values.


New (PCA) realizations $\mathbf{m}_\text{pca} \in \mathbb{R}^{n_b}$ can be generated through application of
\begin{equation}
    \label{eq_pca}
    \mathbf{m}_\text{pca} = \Bar{\mathbf{m}} + U_l \Sigma_l \boldsymbol{\xi}_l,
\end{equation}
where $U_l \in \mathbb{R}^{n_b \times l}$ contains the $l$ columns in $U$ associated with the largest singular values and $\Sigma_l \in \mathbb{R}^{l \times l}$ contains the $l$ largest singular values. Standalone PCA provides accurate geomodels $\mathbf{m}_\text{pca}$ when the initial realizations $\mathbf{m}_i$ follow a multi-Gaussian distribution. With more complex systems such as channelized models, which are characterized by multipoint spatial statistics, PCA does not fully preserve the features appearing in $\mathbf{m}_i$.

In CNN-PCA, we post-process PCA models using a deep convolutional neural network (CNN) to better preserve the complex spatial correlations. This post-processing, referred to as a model transform net, is represented as
\begin{equation}
    \label{eq_cnnpca_fw}
    \mathbf{m}_\text{cnnpca} = f_W(\mathbf{m}_\text{pca}),
\end{equation}
where $f_W$ indicates the model transform net and the subscript $W$ denotes the trainable weights. Here we apply a supervised-learning approach to train the 3D model transform net.

Our method proceeds as follows. Given an original geomodel $\mathbf{m}$, we map $\mathbf{m}$ onto the first $\hat{l}$ principal components,
\begin{equation}
\hat{\boldsymbol{\xi}}_{\hat{l}} = \Sigma^{-1}_{\hat{l}}U^T_{\hat{l}}(\mathbf{m} - \Bar{\mathbf{m}}).
\end{equation}
The model $\mathbf{m}$ can then be approximately reconstructed using
\begin{equation} \label{eq:recon_pca}
    \hat{\mathbf{m}}_\text{pca} = \Bar{\mathbf{m}} + U_{\hat{l}} \Sigma_{\hat{l}} \hat{\boldsymbol{\xi}}_{\hat{l}},
\end{equation}
where $\hat{\mathbf{m}}_\text{pca}\in \mathbb{R}^{n_b}$ is the reconstructed PCA model, and $\hat{l}$ is a low-dimensional variable such that $\hat{l} < l$ (for reasons explained below). The difference between $\hat{\mathbf{m}}_\text{pca}$ and the corresponding $\mathbf{m}$ can be viewed as the reconstruction error. The goal of the supervised training is to determine the parameters in the transform net $f_W$ such that this error is minimized. In other words, we seek to minimize the difference between $f_W(\hat{\mathbf{m}}_\text{pca})$ and $\mathbf{m}$. 

We use the same $n_r$ realizations of $\mathbf{m}$ as were used for constructing the PCA representation to generate the corresponding $\hat{\mathbf{m}}_\text{pca}^i$ models. This forms the training set, with each pair of $(\mathbf{m}^i, \hat{\mathbf{m}}_\text{pca}^i)$, $i=1, \ldots, n_r$, representing one training sample. The supervised learning training loss for training sample $i$ is defined as
\begin{equation}
    L_\text{rec}^i = ||f_W(\hat{\mathbf{m}}_\text{pca}^i) - \mathbf{m}^i||_1.
\end{equation}
This quantity is referred to as the reconstruction loss.

During the test stage, 3D CNN-PCA will be used to generate geomodels for history matching. This entails post-processing new PCA models $\mathbf{m}_\text{pca}(\boldsymbol{\xi}_l)$, with $\boldsymbol{\xi}_l$ determined by the history matching algorithm, or sampled from the standard normal distribution. By using $\hat{l}<l$ during training, the resulting reconstruction error is more representative of what will be encountered at the test stage, where we do not have corresponding pairs $(\mathbf{m}^i, \hat{\mathbf{m}}_\text{pca}^i)$. We found this approach to lead to better performance at test time, when $f_W$ is applied for new PCA models. Note that in \citep{liu20203d}, an additional perturbation is applied in Eq.~\ref{eq:recon_pca}. This does not appear to be required for the system considered here, possibly because our model is slightly less challenging than those considered in \citep{liu20203d}, as discussed earlier.

Hard data loss is also required to ensure that hard data at well locations are honored. This loss is of the form $L_\text{h}^i = [\mathbf{h}^T(\mathbf{m}^i-f_W(\hat{\mathbf{m}}_\text{pca}^i)]^2$, where $\mathbf{h}$ is an indicator vector, with $h_i=1$ indicating the presence of hard data at grid block $i$, and $h_i=0$ the absence of hard data. Note that hard data are captured in the PCA models, but without the inclusion of $L_\text{h}^i$ these data would not necessarily be maintained in the $\mathbf{m}_\text{cnnpca}$ models. The final training loss for training sample $i$ is now
\begin{equation}
    L^i = L_\text{rec}^i + \gamma_h L_\text{h}^i,
\end{equation}
where $\gamma_h$ is a weighting factor for the hard data loss. We determine $\gamma_h$ by considering a range of values and selecting a value that is sufficiently large such that essentially all hard data are honored in the output models. Specifically, in the case considered in this paper, $\gamma_h$ is set such that, over all new 3D CNN-PCA geomodels, hard data in well blocks are honored more than 99.9\% of the time. 

Construction of the 3D model transform net entails replacing the 2D convolutional layers, upsampling layers and padding layers in the 2D model transform net \citep{liu2019deep} with their 3D counterparts. This can be readily accomplished using the Pytorch deep-learning framework \citep{Paszke2017}. The architecture details of the 3D CNN-PCA transform net are provided in the Appendix. 

The underlying geomodels used in this work 
are defined on grids containing $40 \times 40 \times 20$ blocks (total of 32,000~cells). These channelized models are constructed in Petrel \citep{manual2007petrel} using object-based modeling. The models include six fully penetrating wells (two injectors and four producers), and conditioning data are available at all blocks penetrated by wells. All realizations are constrained to honor the hard data at wells.

\begin{figure}[htbp]
     \centering
     \begin{subfigure}[b]{0.32\textwidth}
         \centering
         \includegraphics[trim={7cm 5cm 5cm 4cm},clip, scale=0.3]{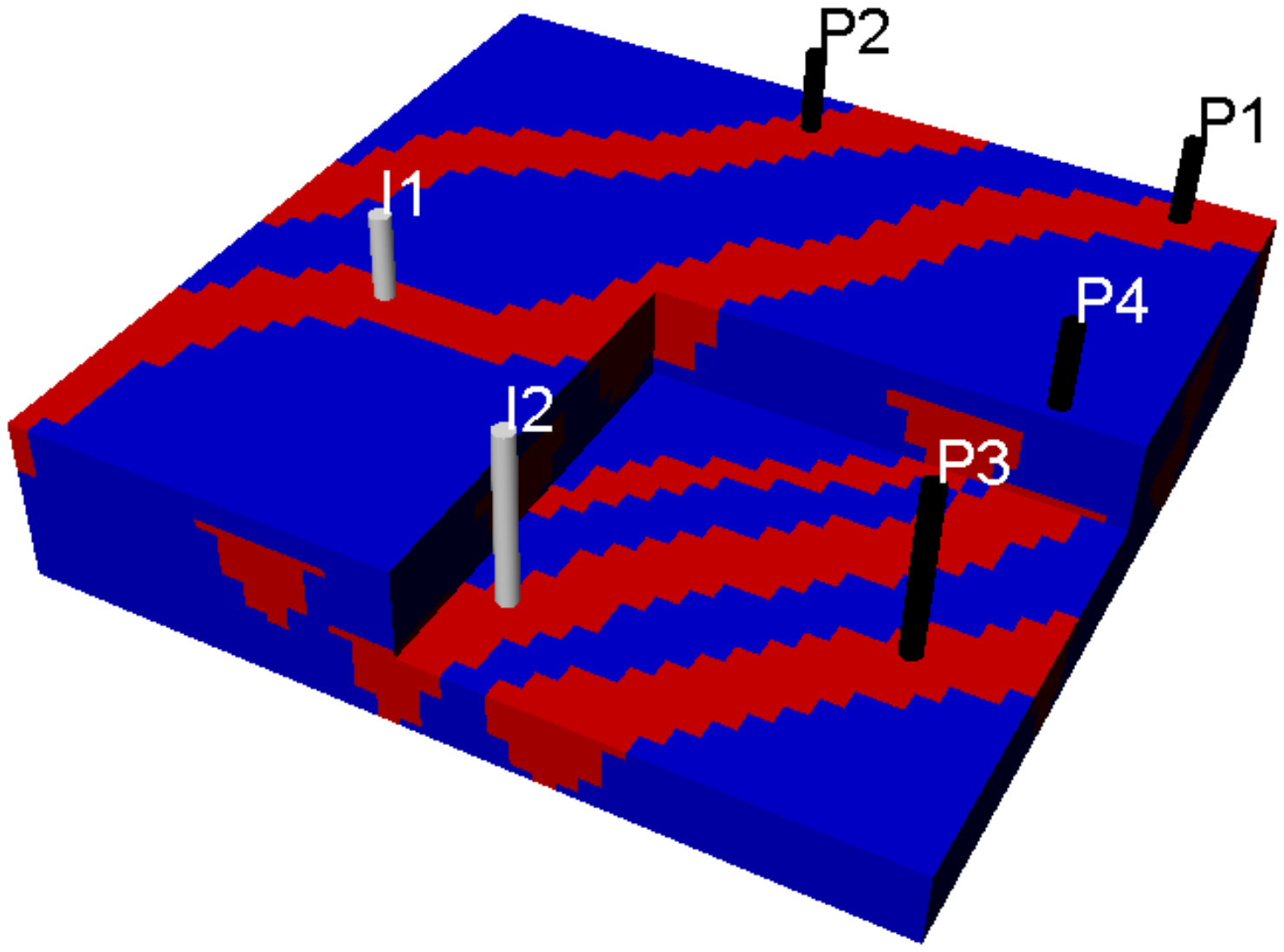}
         \caption{}
         \label{real-31}
     \end{subfigure}
     \begin{subfigure}[b]{0.32\textwidth}
         \centering
         \includegraphics[trim={7cm 5cm 5cm 4cm},clip, scale=0.3]{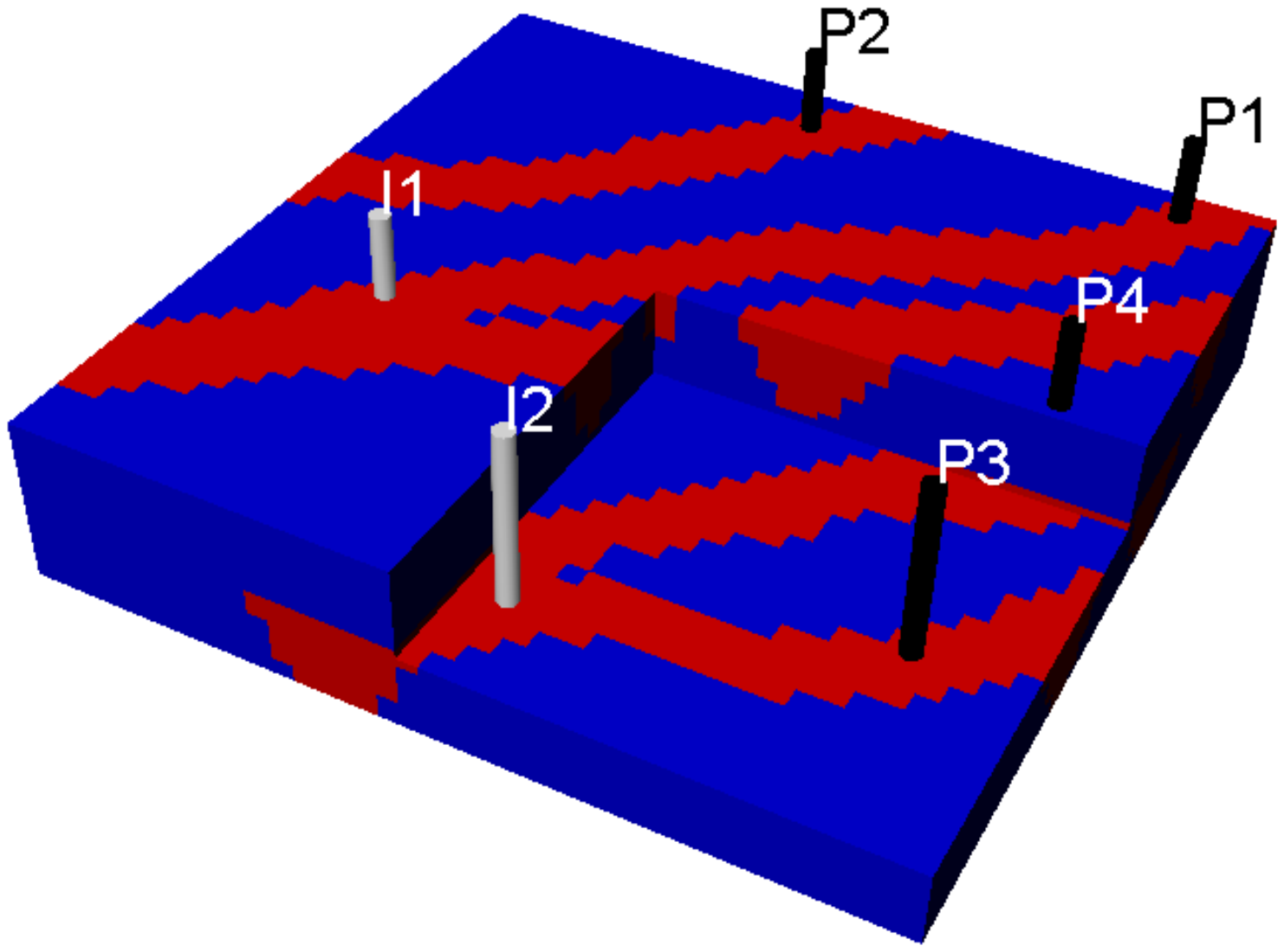}
         \caption{}
         \label{real-41}
     \end{subfigure}
     \begin{subfigure}[b]{0.32\textwidth}
         \centering
         \includegraphics[trim={7cm 5cm 5cm 4cm},clip, scale=0.3]{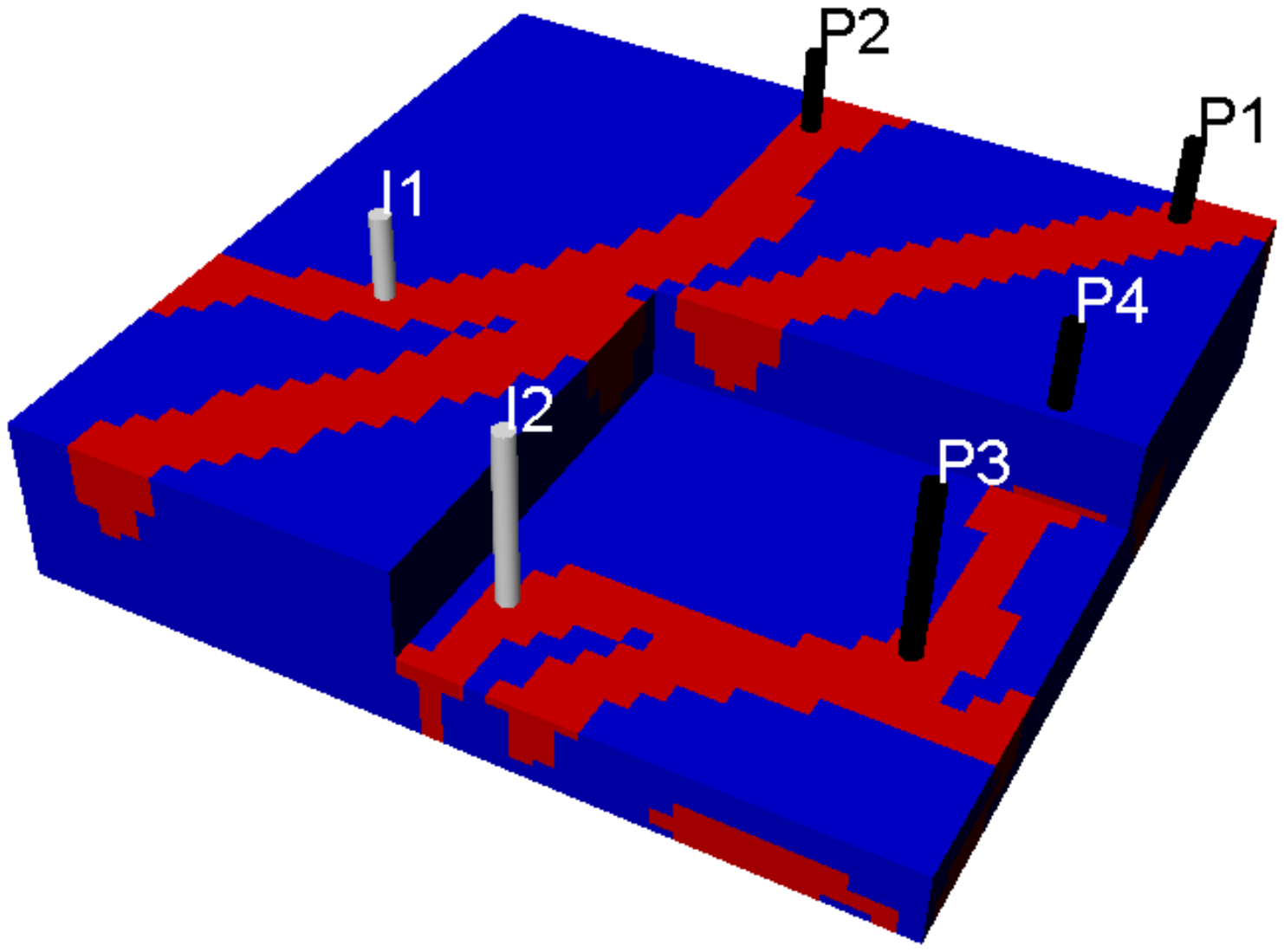}
         \caption{}
         \label{real-51}
     \end{subfigure}
    \caption{Petrel realizations honoring rock-type data at the six wells. Black cylinders denote production wells, and white cylinders denote injection wells.}
    \label{fig:petrel-reals}
\end{figure}

A total of $n_r=3000$ realizations of $\mathbf{m}$ are generated using Petrel. Fig.~\ref{fig:petrel-reals} displays maps of rock type (referred to as facies) for three of these realizations. Although the three models resemble one another in terms of their channelized geology, they clearly differ in the detailed channel locations and the resulting `linkages' between injection and production wells. For example, in the model in Fig.~\ref{fig:petrel-reals}b, wells~I2 and P3 are connected by a high-permeability channel, while in Fig.~\ref{fig:petrel-reals}a these wells are not connected. These differences in connectivity result in very different flow behavior.  

The $n_r=3000$ realizations of $\mathbf{m}$ are used to construct the PCA representation and the corresponding $\hat{\mathbf{m}}_\text{pca}$ models. Based on limited numerical experimentation, we use a reduced dimension of $l=200$ for the new PCA models and a dimension of $\hat{l}=50$ for the reconstructed PCA models. The weighting factor for hard data loss is $\gamma_h = 0.02$. We use a batch size of 4 and train the model transform net for 10~epochs. This requires about 30~minutes using one Tesla V100 GPU. 

\begin{figure}[htbp]
     \centering
     \begin{subfigure}[b]{0.32\textwidth}
         \centering
         \includegraphics[trim={7cm 5cm 5cm 4cm},clip, scale=0.3]{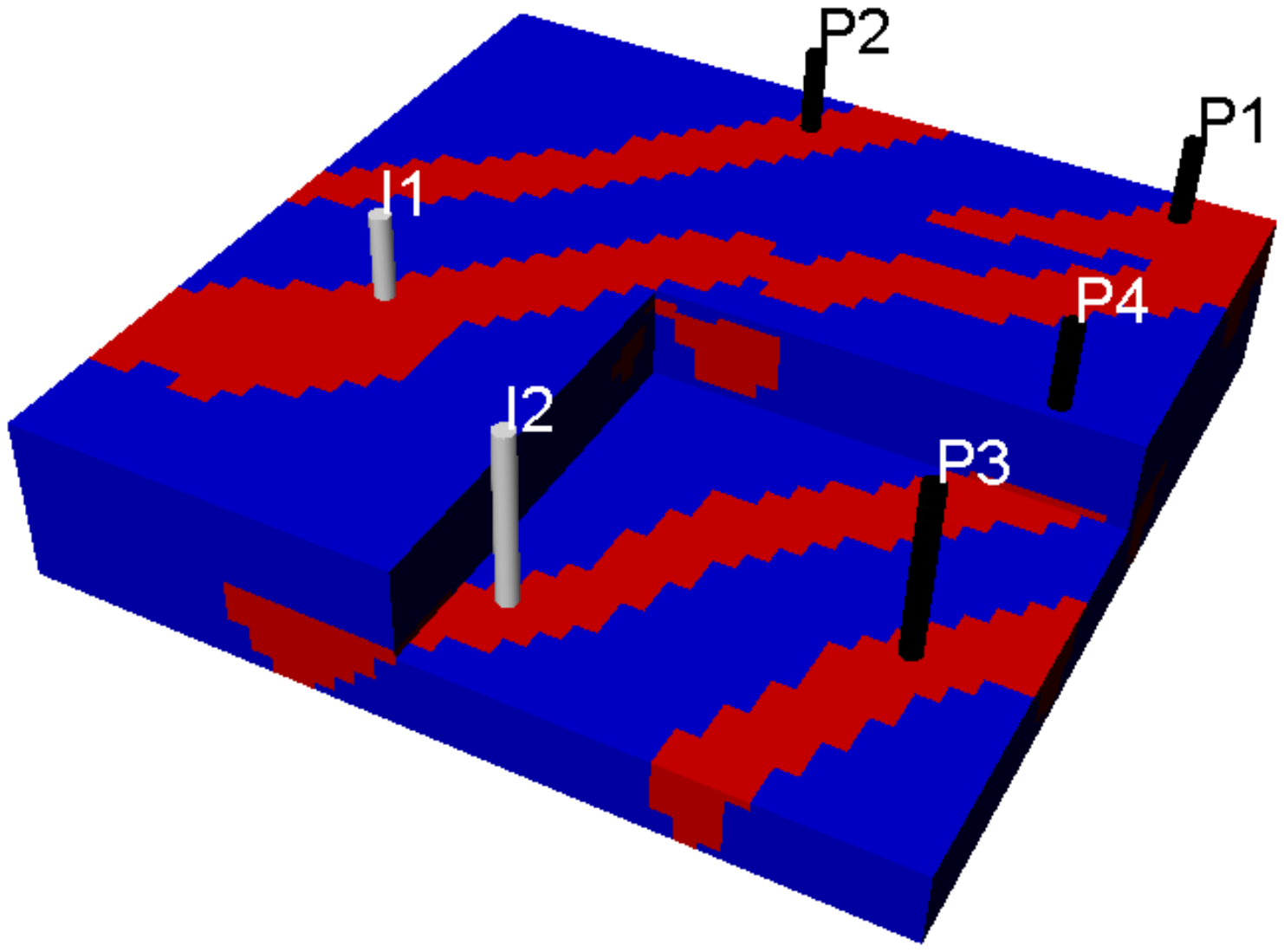}
         \caption{}
         \label{cnnpca-sat-p-eval}
     \end{subfigure}
          \begin{subfigure}[b]{0.32\textwidth}
         \centering
         \includegraphics[trim={7cm 5cm 5cm 4cm},clip, scale=0.3]{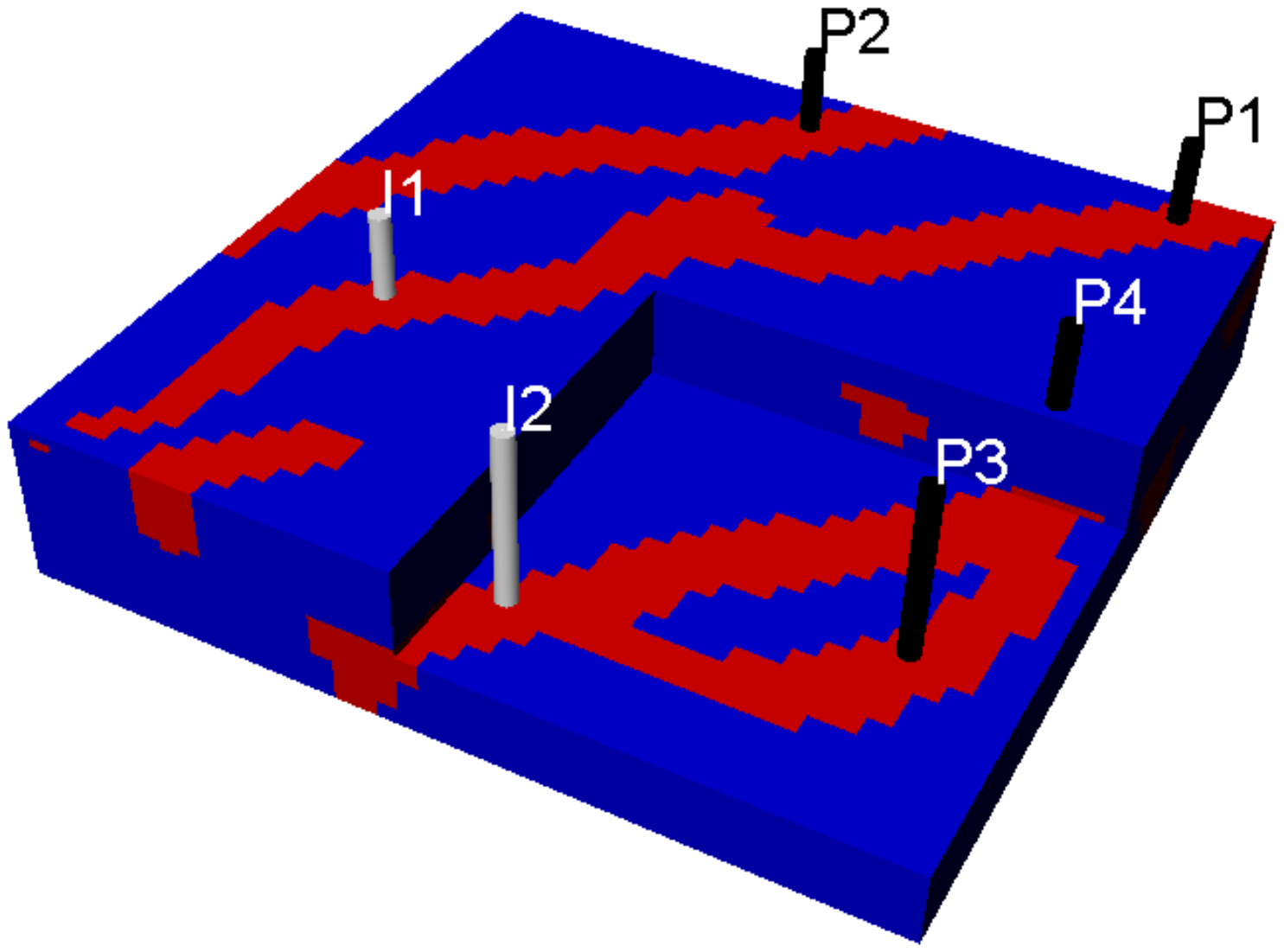}
         \caption{}
         \label{cnnpca-true-model}
     \end{subfigure}
          \begin{subfigure}[b]{0.32\textwidth}
         \centering
         \includegraphics[trim={7cm 5cm 5cm 4cm},clip, scale=0.3]{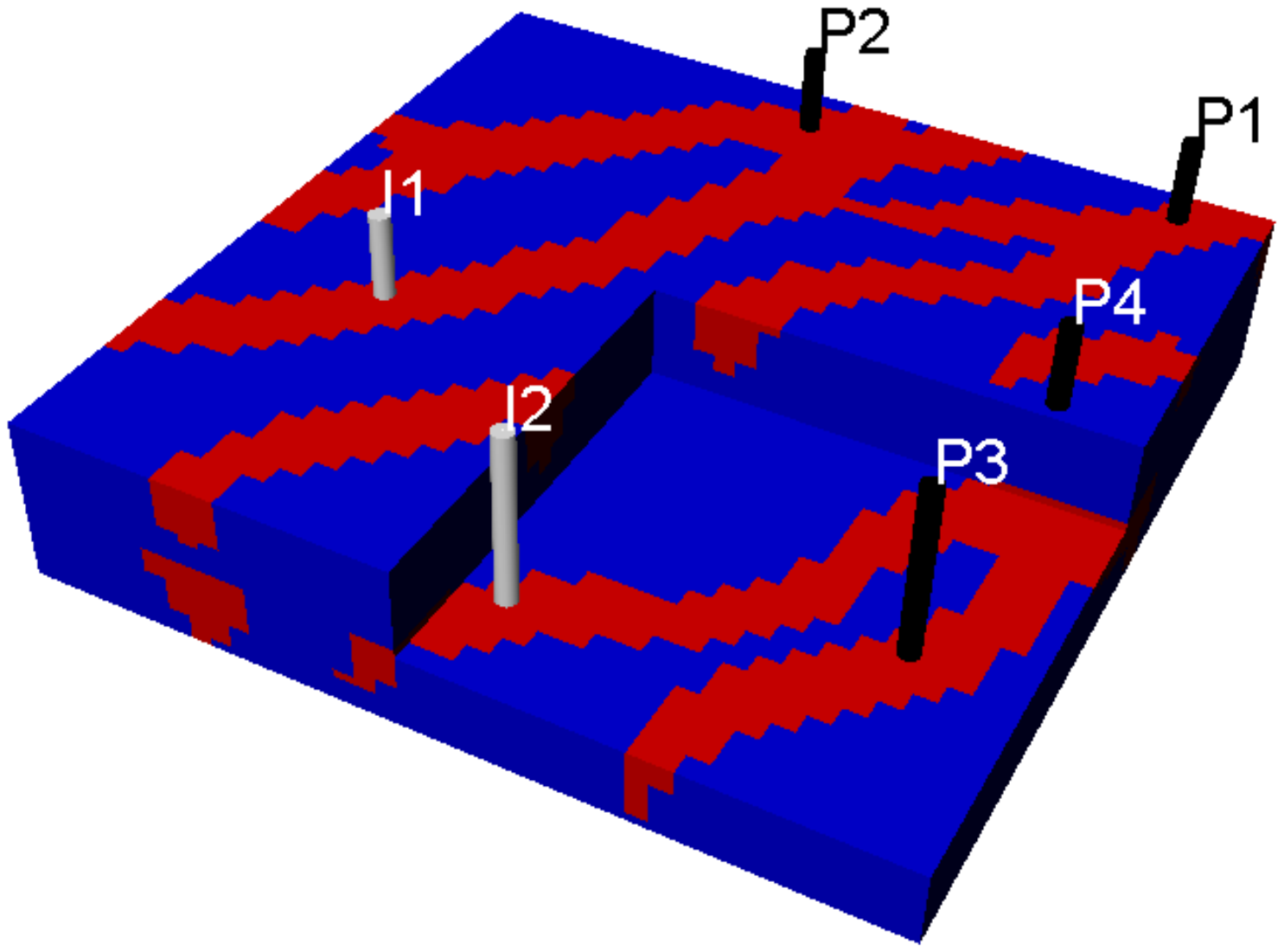}
         \caption{}
         \label{real-8}
     \end{subfigure}
    
    \caption{3D CNN-PCA realizations honoring rock-type data at the six wells.
    Model in (a) is used in the 3D recurrent R-U-Net assessment below. Model in (b) is the true model used in the history matching example. Black cylinders denote production wells, and white cylinders denote injection wells.}
    \label{fig:cnn-pca-reals}
\end{figure}

Three random test-set CNN-PCA models are shown in Fig.~\ref{fig:cnn-pca-reals}. The CNN-PCA facies maps are thresholded such that the average rock-type proportions match those in the original Petrel models. The 3D CNN-PCA realizations display channel features (width, sinuosity) in approximate visual agreement with those in Fig.~\ref{fig:petrel-reals}. We also observe variability in channel connectivity between realizations. In fact, for an oil-water flow problem defined for this model, the resulting flow statistics for 200 random CNN-PCA realizations were found to be in very close agreement with those from 200 reference Petrel geomodels. These results, not presented here for brevity, demonstrate that the level of geological variability, in terms of its impact on flow, is essentially the same between the two sets of models.

\section{Recurrent R-U-Net Surrogate Model and Training}
\label{sec:flow_surrogate}

In this section, we present the detailed 3D recurrent R-U-Net surrogate model and training setup. This surrogate model is intended to provide predictions for the saturation and pressure fields, at particular time steps, for any (random) geological realization generated by the 3D CNN-PCA procedure described in Section~\ref{sec:cnnpca}. 


\subsection{Recurrent R-U-Net Architecture}

In the context of history matching or uncertainty quantification, many simulation runs must be performed to solve the discretized versions of Eq.~\ref{eq:mass-balance-eq} and \ref{eq:darcy-eq} with different geomodels $\mathbf{m}$. We express a simulation run in this setting as
\begin{align}
    \mathbf{x} = f(\mathbf{m}),
    \label{eq-f}
\end{align}
where $f$ denotes the reservoir simulator and $\mathbf{x} \in \mathbb{R}^{n_b \times n_t}$ represents a dynamic state. Here $n_b$ indicates the number of grid blocks and $n_t$ the number of time steps at which solution variables are predicted (here we take $n_t=10$, which is much less than the total number of time steps in the simulation). Because in this work we consider oil-water systems, $\mathbf{x}$ contains either water saturation or pressure variables. We use $\mathbf{x}^t \in \mathbb{R}^{n_b}$ to denote the saturation or pressure field at time step $t$.

\begin{figure}[htbp]
  \centering
  \includegraphics[trim={2cm 2.5cm 1cm 2.5cm},clip, scale=0.6]{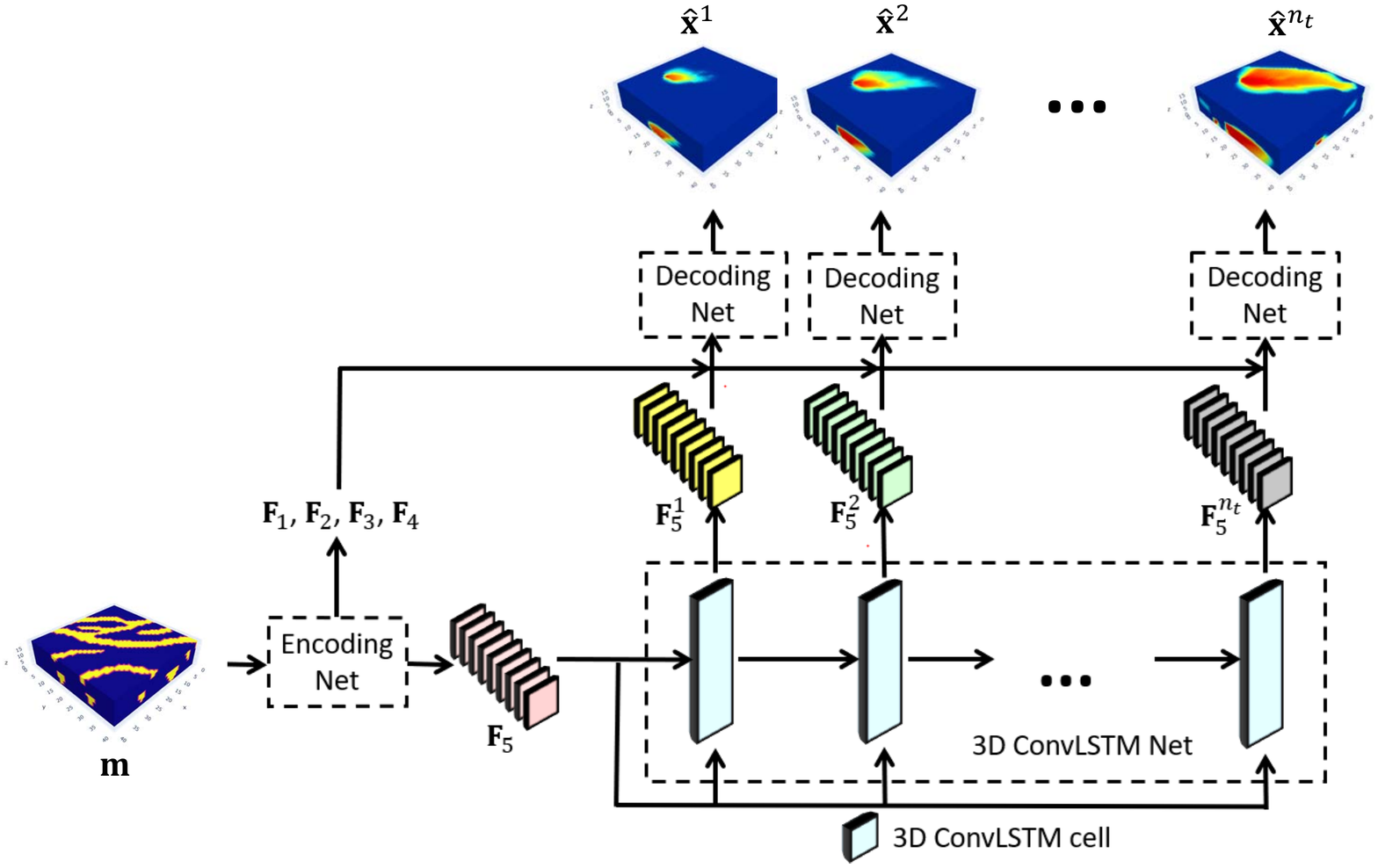}
  \caption{Recurrent R-U-Net architecture incorporating 3D convLSTM into the R-U-Net (see Appendix for detailed specifications). Here the convLSTM net takes the global feature map $\mathbf{F}_5$ from the encoding net and generates a sequence of feature maps $\mathbf{F}_5^t$ $(t=1,\dots, n_t)$. These are decoded, separately, into a sequence of predictions for the states $\hat{\mathbf{x}}^{t}$ $(t=1,\dots, n_t)$ using the same decoding net (figure modified from \citep{tang2020deep}).}
  \label{fig:recurrent-r-u-net}
\end{figure}

The 3D recurrent R-U-Net, illustrated in Fig.~\ref{fig:recurrent-r-u-net}, is applied to replace the expensive numerical simulation $f$. This model is denoted as $\hat{\mathbf{x}} = \hat{f}(\mathbf{m})$, where $\hat{f}$ indicates the surrogate operator and $\hat{\mathbf{x}} \in \mathbb{R}^{n_b \times n_t}$ denotes the predicted states from the surrogate model. The recurrent R-U-Net consists of a 3D residual U-Net \citep{ronneberger2015u}, which itself entails encoding and decoding nets, and convLSTM neural networks \citep{xingjian2015convolutional}. The encoding net extracts feature maps $\mathbf{F}_1$ to $\mathbf{F}_5$ from the input geomodel $\mathbf{m}$. Of these feature maps, $\mathbf{F}_5$ represents $\mathbf{m}$ in the most compressed manner. For this reason, $\mathbf{F}_5$ is propagated in time to provide feature representations of the state maps (i.e., we have $\mathbf{F}_5^1, \cdots, \mathbf{F}_5^{n_t}$) by the 3D convLSTM neural network. Combined with extracted local features $\mathbf{F}_1$ to $\mathbf{F}_4$, the propagated $\mathbf{F}_5^{t}$ $(t=1, \cdots, n_t)$ is upsampled to the corresponding (approximated) state map $\hat{\mathbf{x}}^{t} \in \mathbb{R}^{n_b}$ by the decoding net.

\begin{figure}[htbp]
  \centering
  \includegraphics[trim={3cm 3cm 3cm 3cm},clip, scale=0.6]{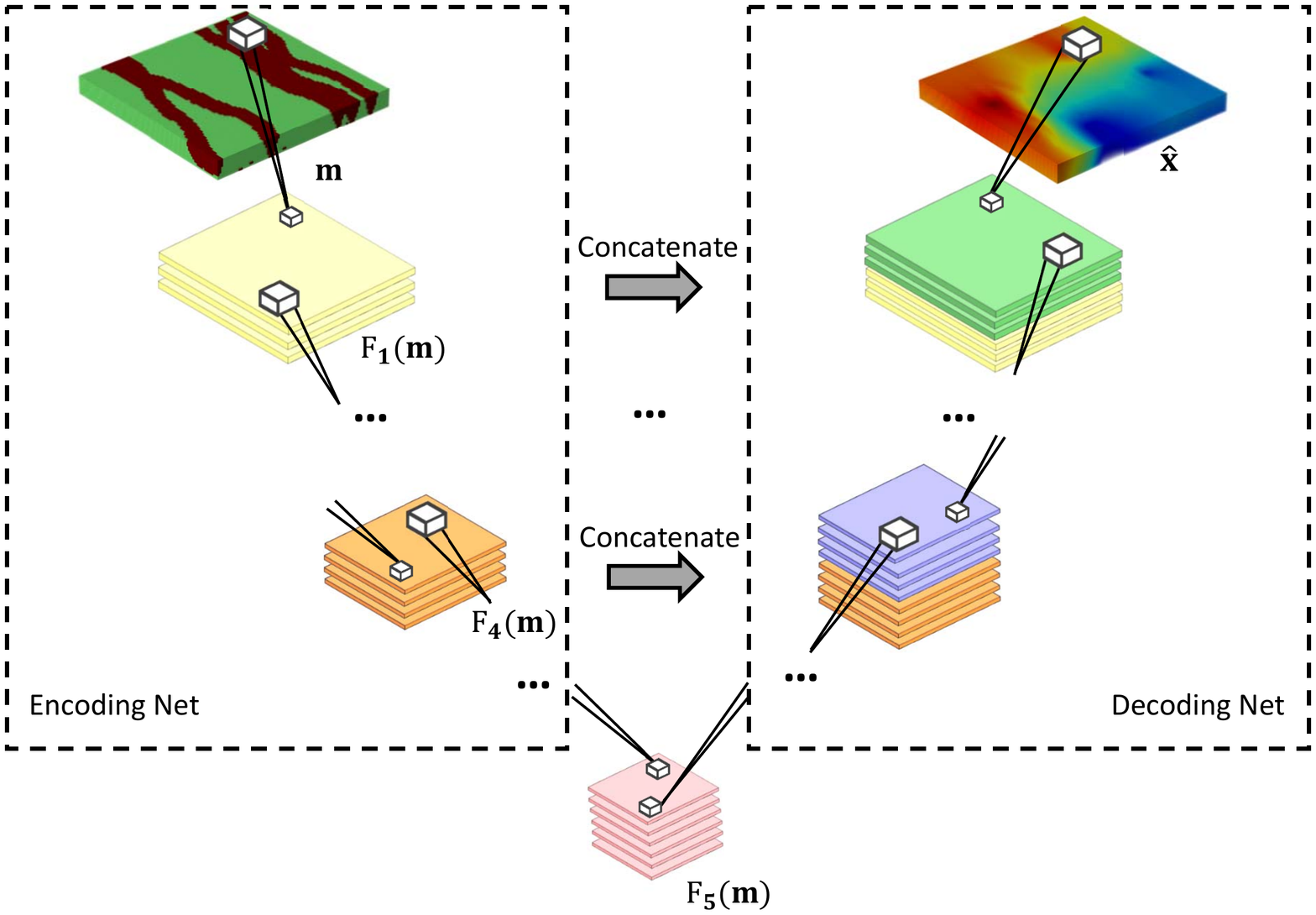}
  \caption{Schematic illustration of 3D residual U-Net (R-U-Net) (detailed architecture is provided in the Appendix). This network entails encoding and decoding nets. Local features $\mathbf{F}_1$ to $\mathbf{F}_4$ extracted in the encoding net are concatenated with the upsampled features in the decoding net to generate predictions for the states.}
  \label{fig:3d-r-u-net}
\end{figure}

The 3D residual U-Net (R-U-Net) is the key module for capturing the spatial correlations between geological properties and the predicted states. A schematic of the 3D R-U-Net is shown in 
Fig.~\ref{fig:3d-r-u-net}. This figure illustrates how local features $\mathbf{F}_1$ to $\mathbf{F}_4$ are combined with the most global feature $\mathbf{F}_5$ during upsampling. The R-U-Net output is the target state (either pressure or saturation) at a particular time step. The concatenation of local features extracted in the encoding net to the corresponding upsampled features in the decoding net facilitates the transfer of multiscale information, which leads to improved state predictions.

The 3D convLSTM net enables the surrogate model to capture temporal dynamics. This network effectively incorporates the advantages of LSTM~\citep{greff2016lstm} in modeling temporal evolution, while maintaining the spatial information extracted by CNNs. The internal cell state $\mathbf{C}^t$ in the 3D convLSTM net carries the spatial-temporal information flow. At time~$t$, this quantity is given by 
\begin{equation}
     \mathbf{C}^t = \mathbf{f}^t \circ \mathbf{C}^{t-1} + \mathbf{i}^t \circ \tilde{\mathbf{C}}^t,
 \end{equation}
where the forget gate $\mathbf{f}^t$ controls which information to discard from the previous cell state $\mathbf{C}^{t-1}$, the input gate $\mathbf{i}^t$ specifies which information to update from the new candidate cell state $\tilde{\mathbf{C}}^t$, and $\circ$ denotes the Hadamard product. The output state at time~$t$, $\mathbf{H}^t$, is computed from $\mathbf{C}^t$ and the output gate $\mathbf{o}^t$ via
   \begin{equation}
     \mathbf{H}^t = \mathbf{o}^t \circ \tanh(\mathbf{C}^t),
 \end{equation}
where $\mathbf{o}^t$ controls which information in the cell state $\mathbf{C}^t$ is transferred to the output state $\mathbf{H}^t$.

The three gates $\mathbf{f}^t$, $\mathbf{i}^t$ and $\mathbf{o}^t$, as well as the new candidate cell state $\tilde{\mathbf{C}}^t$, are determined from the previous output state $\mathbf{H}^{t-1}$ and the current input $\boldsymbol\chi^t$ using the following expressions 
 \begin{equation}
     \mathbf{f}^t = \sigma(\mathbf{W}_{xf} * {\boldsymbol\chi}^t + \mathbf{W}_{hf}*\mathbf{H}^{t-1} + \mathbf{b}_f),
 \end{equation}
  \begin{equation}
     \mathbf{i}^t = \sigma(\mathbf{W}_{xi} * {\boldsymbol\chi}^t + \mathbf{W}_{hi}*\mathbf{H}^{t-1} + \mathbf{b}_i),
 \end{equation}
  \begin{equation}
     \mathbf{o}^t = \sigma(\mathbf{W}_{xo} * {\boldsymbol\chi}^t + \mathbf{W}_{ho}*\mathbf{H}^{t-1} + \mathbf{b}_o),
 \end{equation}
  \begin{equation}
     \tilde{\mathbf{C}}^t = \tanh(\mathbf{W}_{xc} * {\boldsymbol\chi}^t + \mathbf{W}_{hc}*\mathbf{H}^{t-1} + \mathbf{b}_c),
 \end{equation}
where $\mathbf{W}$ and $\mathbf{b}$ are 3D convolution kernel weight and bias terms. The parameters associated with these quantities, which are tuned during the training process, are shared across convLSTM cells. This reduces the number of tunable parameters that must be learned during training.

The detailed architecture of the overall recurrent R-U-Net is provided in the Appendix. The architectures of the encoding and decoding nets are very similar to those described in \citep{tang2020deep}, with 2D convolutional kernels replaced with 3D convolutional kernels. This combination of convolutional and recurrent neural network architectures enables the surrogate model to accurately capture  spatial-temporal information in the subsurface flow problems of interest.

\subsection{Flow Problem Setup and Network Training}
\label{sec:flow_problem}

The recurrent R-U-Net is trained using saturation and pressure solutions from a set of high-fidelity simulation runs. We now describe the simulation setup. 

The models contain $40 \times 40 \times 20$ grid blocks, with each block of dimensions $20$~m $\times$ 20~m $\times$ 2~m in the $x$, $y$ and $z$ directions. The well locations are shown in Figs.~\ref{fig:petrel-reals} and \ref{fig:cnn-pca-reals}. One injector and two of the producers are completed (open to flow) in the top 10 layers, while the other injector and two producers are completed in the bottom 10 layers. The rock type for grid block~$i$ is denoted by $m_i$ ($m_i =1$ for sand and $m_i = 0$ for mud). For grid blocks corresponding to sand, we specify $k_i = 2000$~md and $\phi_i = 0.25$, and for those corresponding to mud, $k_i = 20$~md and $\phi_i = 0.1$.

\begin{figure}[htbp]
  \centering
  \includegraphics[scale=0.5]{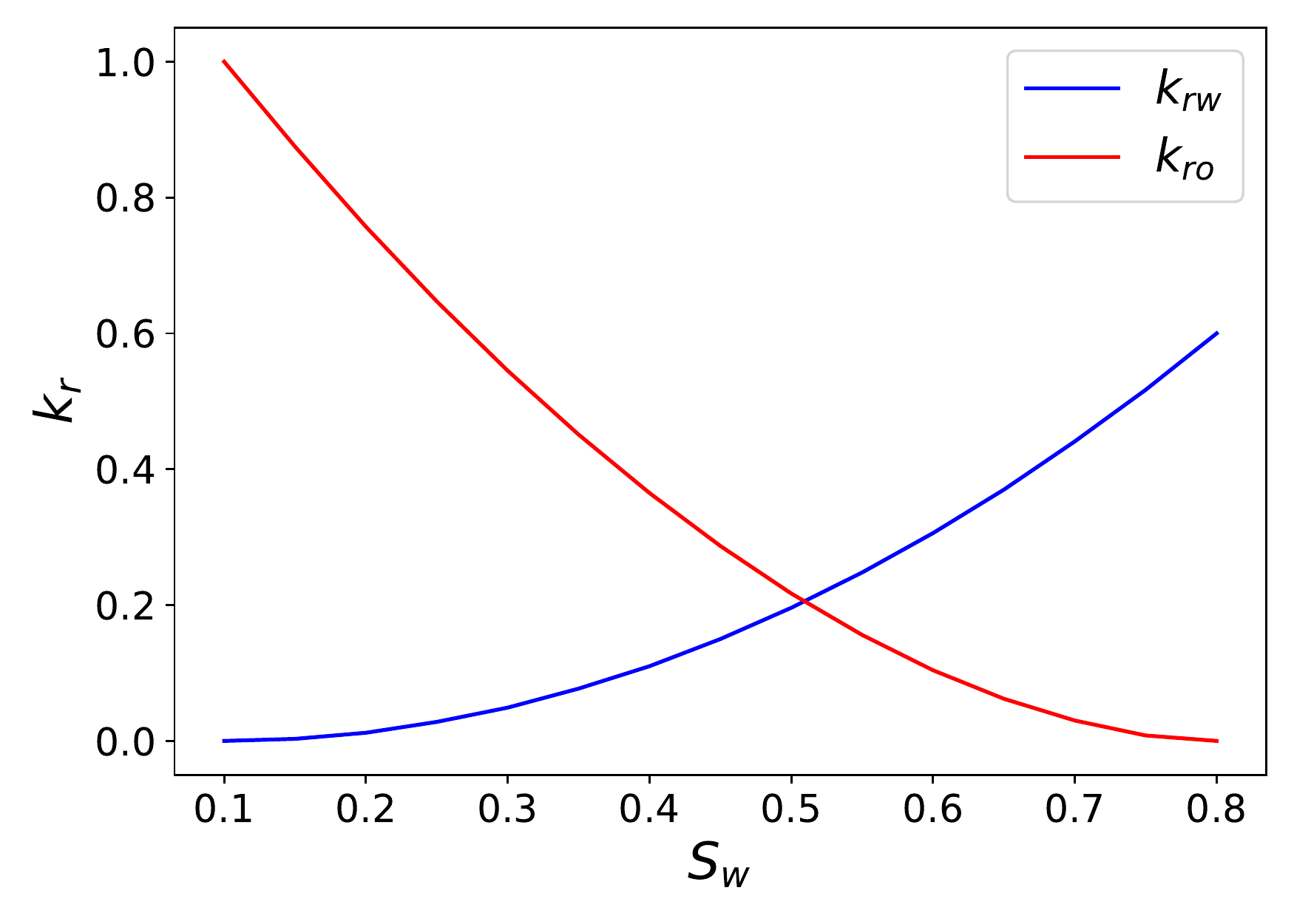}
  \caption{Oil-water relative permeability curves used in all simulations.}
  \label{fig:chap2-rel-perm}
\end{figure}

Oil-water flow is simulated over a time frame of 1000~days. All wells operate with bottom-hole pressure (BHP) specified, with injector BHPs set to 330~bar and producer BHPs to 310~bar. Oil-water relative permeability curves are shown in Fig.~\ref{fig:chap2-rel-perm}. At a pressure of 325~bar, oil viscosity is 1.14~cp. Water viscosity is 0.31~cp, independent of pressure. Oil and water densities are 787~kg/m$^3$ and 1038~kg/m$^3$, respectively, at reservoir conditions.

As discussed in \cite{tang2020deep}, problems with BHPs specified are often more challenging for surrogate models than those with fixed rates. This is because, with BHPs prescribed, different volumes of fluid (as a function of time) are injected into, and flow through, each geomodel. This additional variability, along with the high degree of sensitivity in the well rates to well-block quantities, evident from Eq.~\ref{eq:well_flow}, acts to significantly complicate the surrogate model.

The high-fidelity simulations are all performed using ADGPRS \citep{zhou2012parallel}. Well rates are computed from the states and wellbore pressure in each well block using the procedure described in Section~\ref{sec:gov_eqn} (see Eqs.~\ref{eq:well_flow}, \ref{eq:well_index} and \ref{eq:standard-well}). The problem setup, fluid properties and well settings described here are used in all flow simulations in this paper.

A total of 2500 simulation runs are performed, and the resulting HFS saturation and pressure solutions are used for training. The neural network parameters are tuned, in an automated manner, to minimize the $L^p$ norm of the difference between the recurrent R-U-Net output $\hat{\mathbf{x}}$ and the simulation reference $\mathbf{x}$. Here $\hat{\mathbf{x}}^t_i$ and $\mathbf{x}^t_i$ denote either saturation or pressure in every grid block, at time $t$, for training sample $i$. Training thus entails finding the tunable network parameters $\boldsymbol\theta$ (with the optimum denoted by $\boldsymbol\theta^*$) via
\begin{equation}
\boldsymbol\theta^* = \argmin_{\boldsymbol\theta} \frac{1}{n_s}\frac{1}{n_t} \sum_{i=1}^{n_s} \sum_{t=1}^{n_t} ||{\hat {\mathbf x}}^t_i - \mathbf{{x}}_{i}^{t}||_{p}^p + \lambda\frac{1}{n_s}\frac{1}{n_t}\frac{1}{n_\text{w}} \sum_{i=1}^{n_s} \sum_{t=1}^{n_t} \sum_{\text{w}=1}^{n_\text{w}} ||{\hat {x}}^{t, \text{w}}_{i} - {x}_{i}^{t, \text{w}}||_{p}^p.
\label{eq:loss-function}
\end{equation}
Here, $n_s$ denotes the number of training samples, $n_t$ indicates the number of time steps considered, and $n_w$ is the number of well blocks. The second term on the right-hand side of Eq.~\ref{eq:loss-function} acts to add weight (with weighting factor $\lambda$) to well-block quantities, indicated by superscript w. As discussed earlier, these values are essential for predicting well rates, which are the primary quantities used for data assimilation. 

We found that better predictions of the global saturation and pressure fields were achieved by using the $L^2$ norm for saturation and the $L^1$ norm for pressure. Therefore, two separate recurrent R-U-Nets are used in this work. We use a batch size of 8 and train each of the recurrent R-U-Nets for 260~epochs using the Adam optimizer~\citep{kingma2014adam} with a learning rate decay schedule. These trainings each require about 7~hours using a Tesla V100 GPU. As discussed in \cite{tang2020deep}, training could also be accomplished using a single network, with appropriate (tuned) weights for the pressure and saturation losses. The use of multiple GPUs would accelerate training.


Proper data preparation and scaling contribute to the effective training of the recurrent R-U-Net. In this work, we use the same data pre-processing technique as described in \cite{tang2020deep}. Specifically, the input binary geomodel is naturally represented by rock-type block values (0 denotes mud and 1 denotes sand), and saturation maps are physically constrained between 0 and 1, so these sets of values are used directly. Pressure maps are normalized at each time step based on the maximum and minimum grid-block pressures observed at that time step in the high-fidelity simulation. We found this detrending treatment to lead to better overall performance than normalization based on the time-invariant injector and producer BHPs. See \cite{tang2020deep} for details.




\section{Evaluation of Recurrent R-U-Net Performance}
\label{sect:saturation-pressure-map-eval}

A total of 400 new CNN-PCA geomodels (test cases) are used to evaluate the performance of the 3D recurrent R-U-Net. The surrogate model predicts pressure and saturation fields at $n_t=10$ time steps (50, 100, 150, 300, 400, 500, 600, 700, 850 and 1000~days). We first present predictions for saturation at three particular time steps, 50, 400 and 1000~days, for a single geomodel. This  geomodel, shown in Fig.~\ref{fig:cnn-pca-reals}(a), is characterized by saturation and pressure prediction errors (quantified below) that are slightly greater than the median errors over the 400 test cases. Thus we view this case as representative.

\begin{figure}[htbp]
     \centering
     \begin{subfigure}[b]{0.32\textwidth}
         \centering
         \includegraphics[trim={4.5cm 3.5cm 4cm 3cm},clip, scale=0.25]{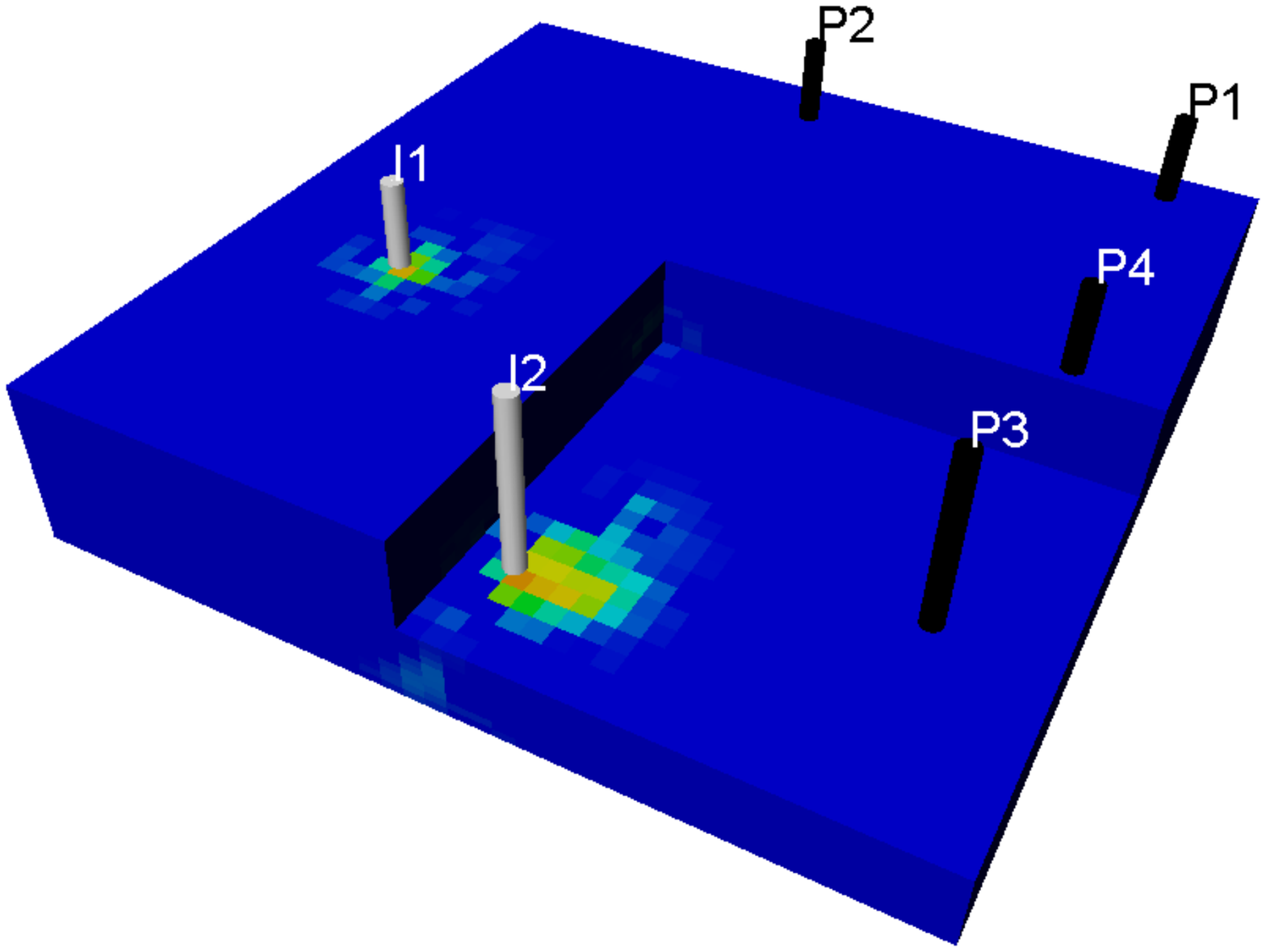}
         \caption{50~days (surr)}
         \label{real-32}
     \end{subfigure}
     \begin{subfigure}[b]{0.32\textwidth}
         \centering
         \includegraphics[trim={4.5cm 3.5cm 4cm 3cm},clip, scale=0.25]{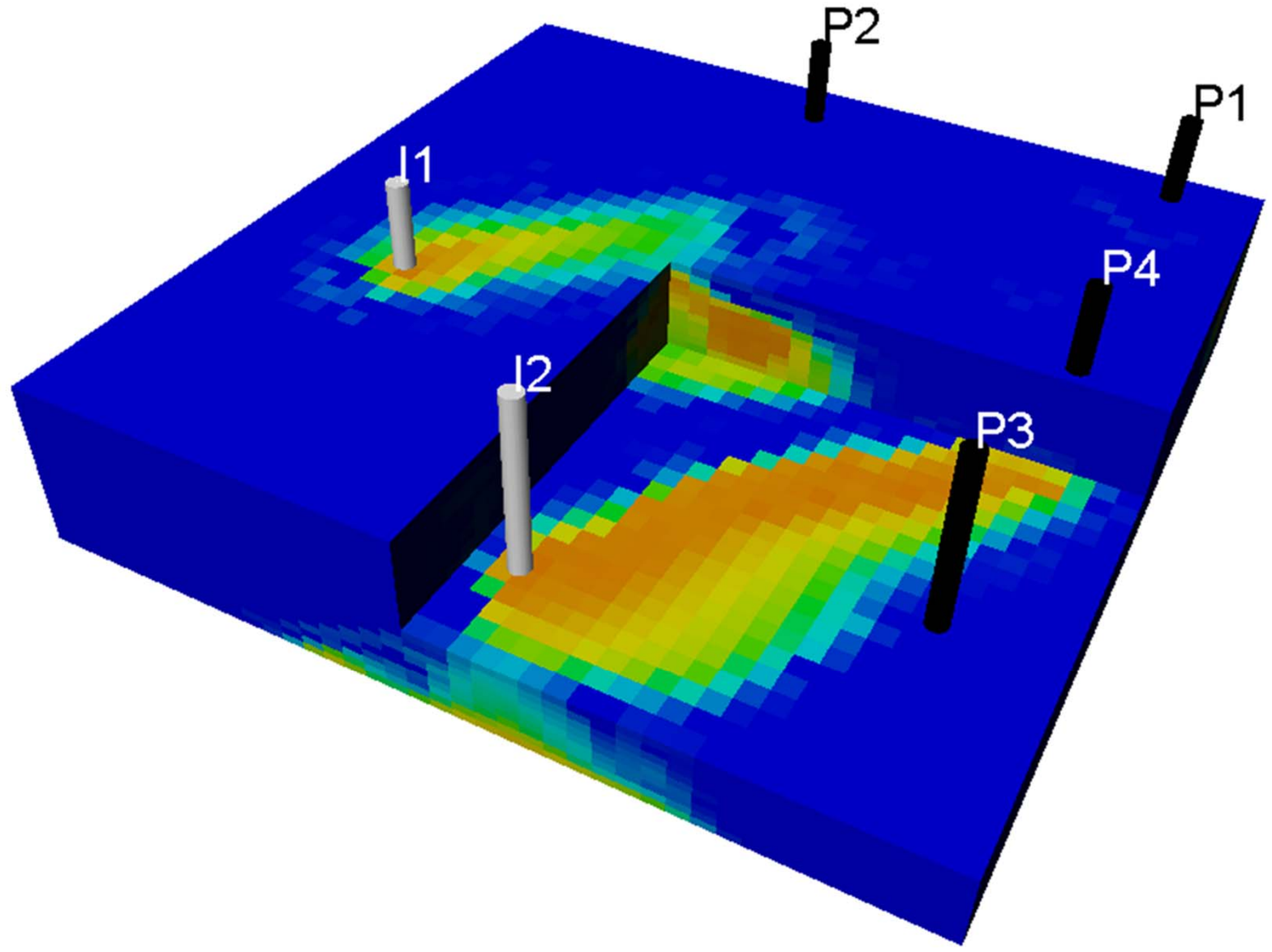}
         \caption{400~days (surr)}
         \label{real-42}
     \end{subfigure}
     \begin{subfigure}[b]{0.32\textwidth}
         \centering
         \includegraphics[trim={4.5cm 3.5cm 4cm 3cm},clip, scale=0.25]{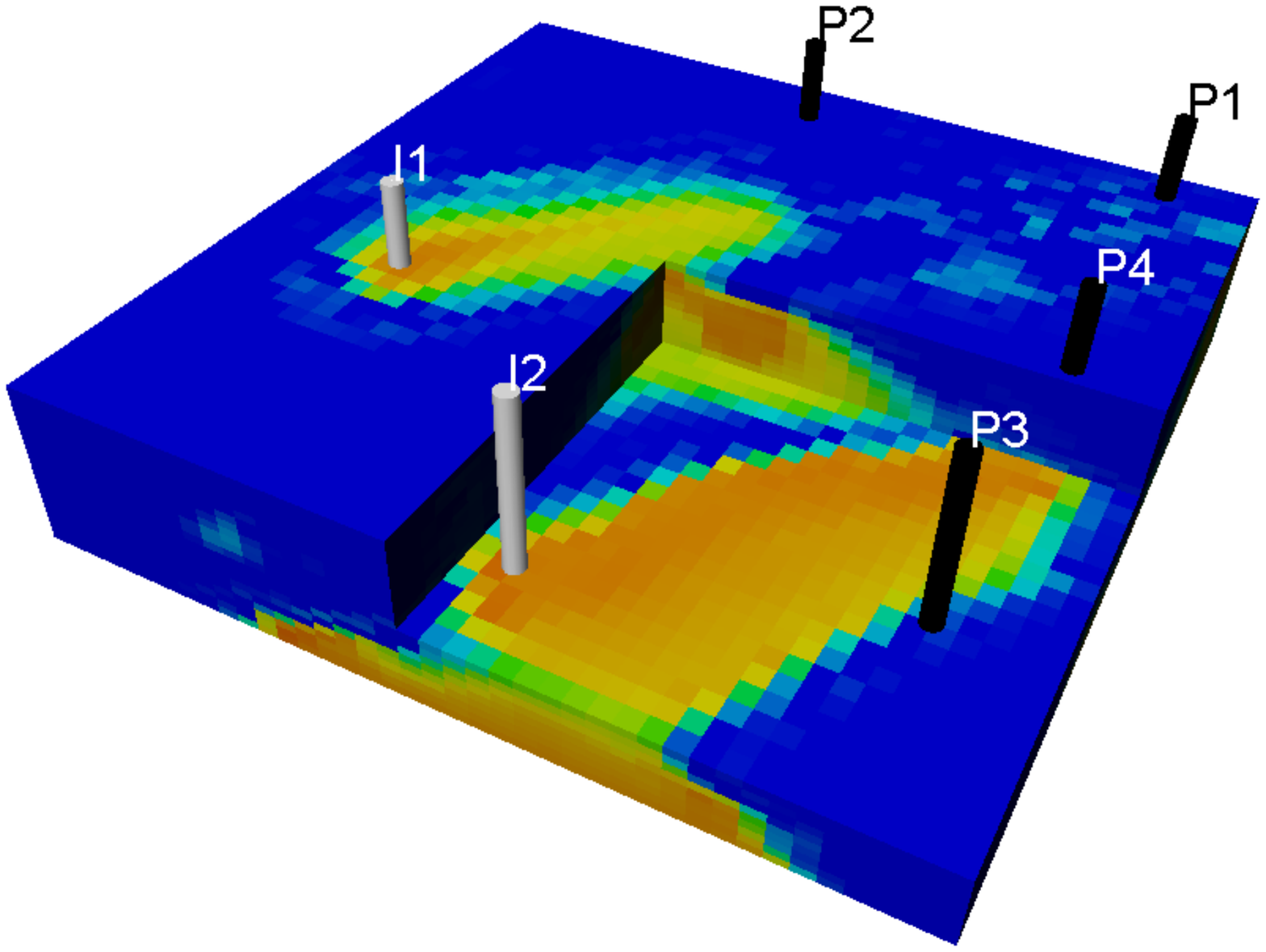}
         \caption{1000~days (surr)}
         \label{real-52}
     \end{subfigure}
     
     \begin{subfigure}[b]{0.32\textwidth}
         \centering
         \includegraphics[trim={4.5cm 3.5cm 4cm 3cm},clip, scale=0.25]{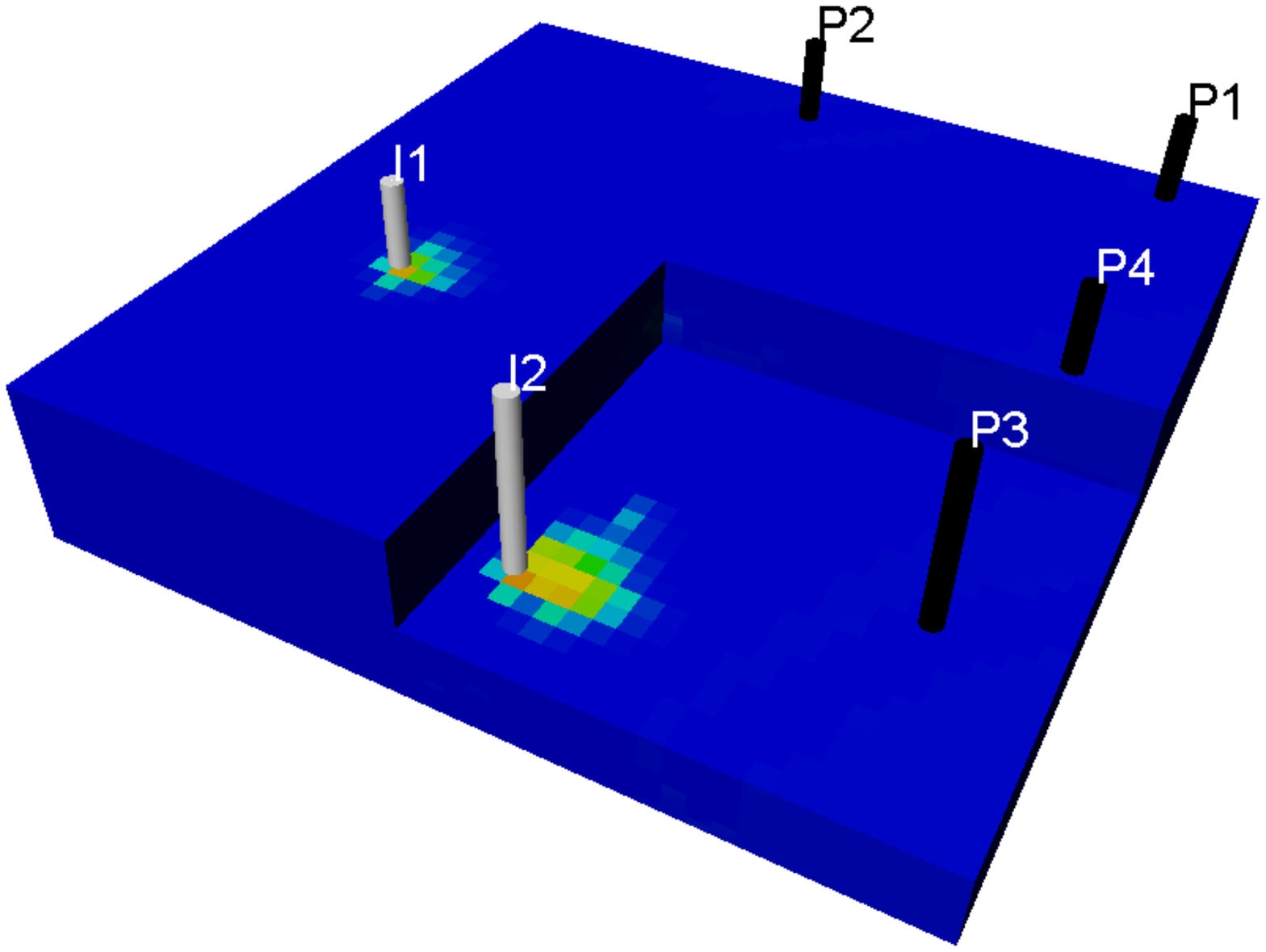}
         \caption{50~days (sim)}
         \label{real-62}
     \end{subfigure}
          \begin{subfigure}[b]{0.32\textwidth}
         \centering
         \includegraphics[trim={4.5cm 3.5cm 4cm 3cm},clip, scale=0.25]{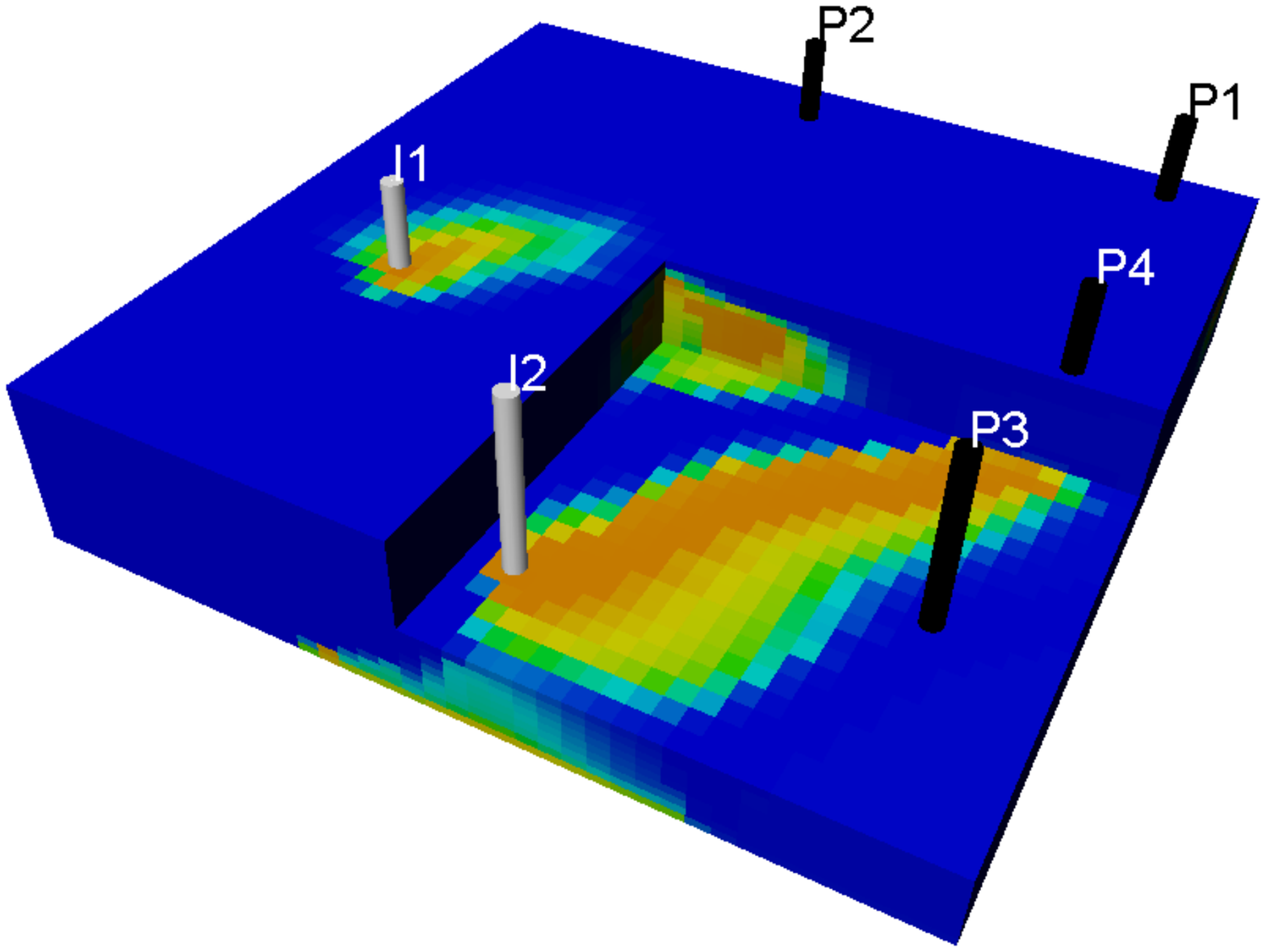}
         \caption{400~days (sim)}
         \label{real-72}
     \end{subfigure}
          \begin{subfigure}[b]{0.32\textwidth}
         \centering
         \includegraphics[trim={4.5cm 3.5cm 4cm 3cm},clip, scale=0.25]{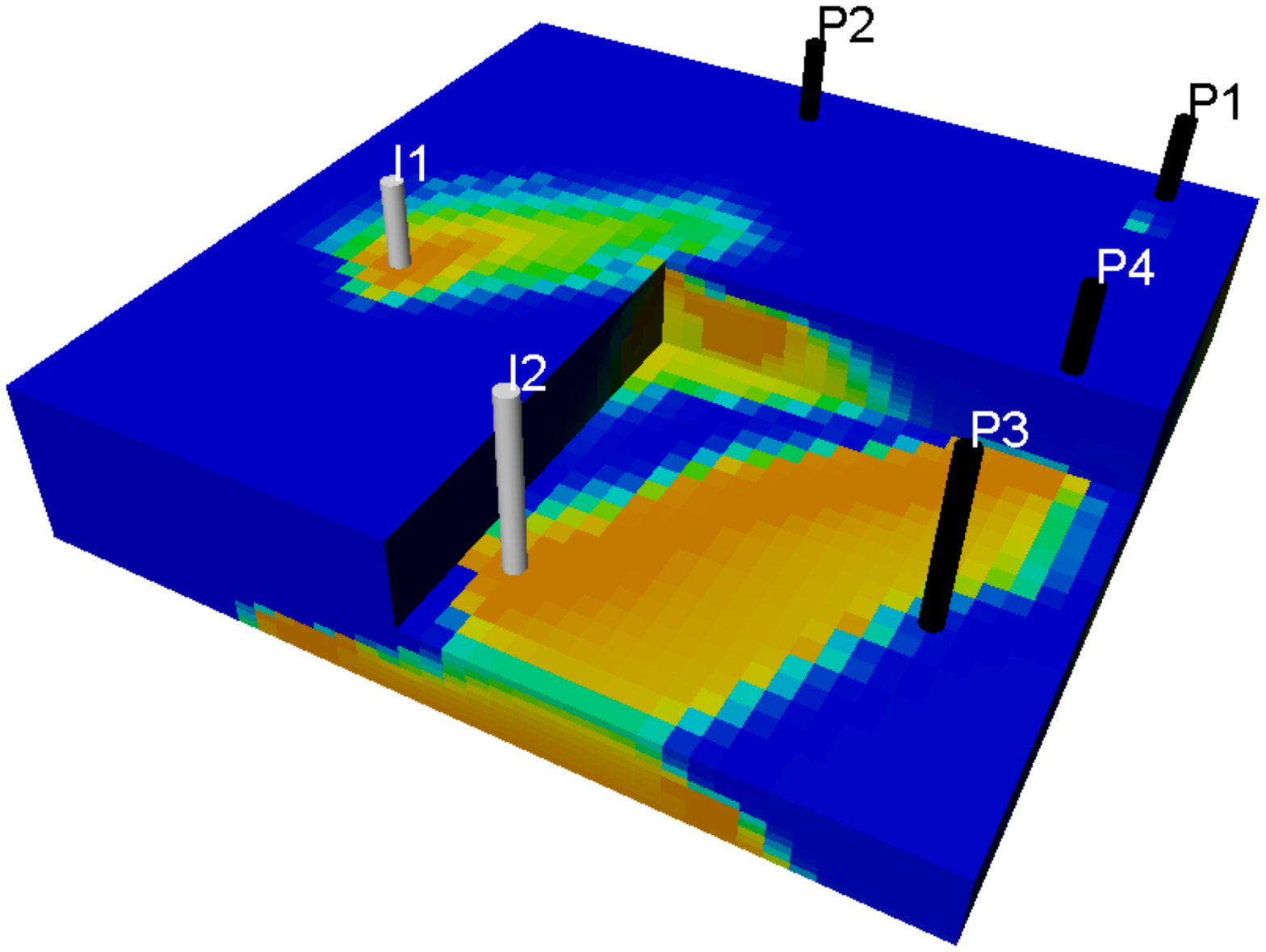}
         \caption{1000~days (sim)}
         \label{real-82}
     \end{subfigure}
    \caption{Saturation fields from 3D recurrent R-U-Net surrogate model (top row) and high-fidelity simulator (bottom row) at three different times.}
    \label{fig:sat-map-single}
\end{figure}

The predicted and reference saturation fields are shown in Fig.~\ref{fig:sat-map-single}.
The top row displays the recurrent R-U-Net 3D state map predictions and the bottom row shows the high-fidelity (ADGPRS) simulation results.
The progress of the saturation front is, as expected, strongly impacted by the high-permeability channels. The 3D recurrent R-U-Net is able to capture the  evolution of the saturation field reasonably accurately, though some error is evident. Differences can be observed, for example, in the top layer, in the progress of the front originating from injector~I1 (compare Fig.~\ref{fig:sat-map-single}(c) to Fig.~\ref{fig:sat-map-single}(f)). We also see some water in the vicinity of producer~P2 in Fig.~\ref{fig:sat-map-single}(c) that does not appear in Fig.~\ref{fig:sat-map-single}(f). We note finally that the 3D aspect of the saturation field appears to be predicted accurately, as is evident from the cross-sectional views in Fig.~\ref{fig:sat-map-single}.

\begin{figure}[htbp]
     \centering
      \begin{subfigure}[b]{0.32\textwidth}
         \centering
         \includegraphics[trim={4.5cm 3.5cm 4cm 3cm},clip, scale=0.25]{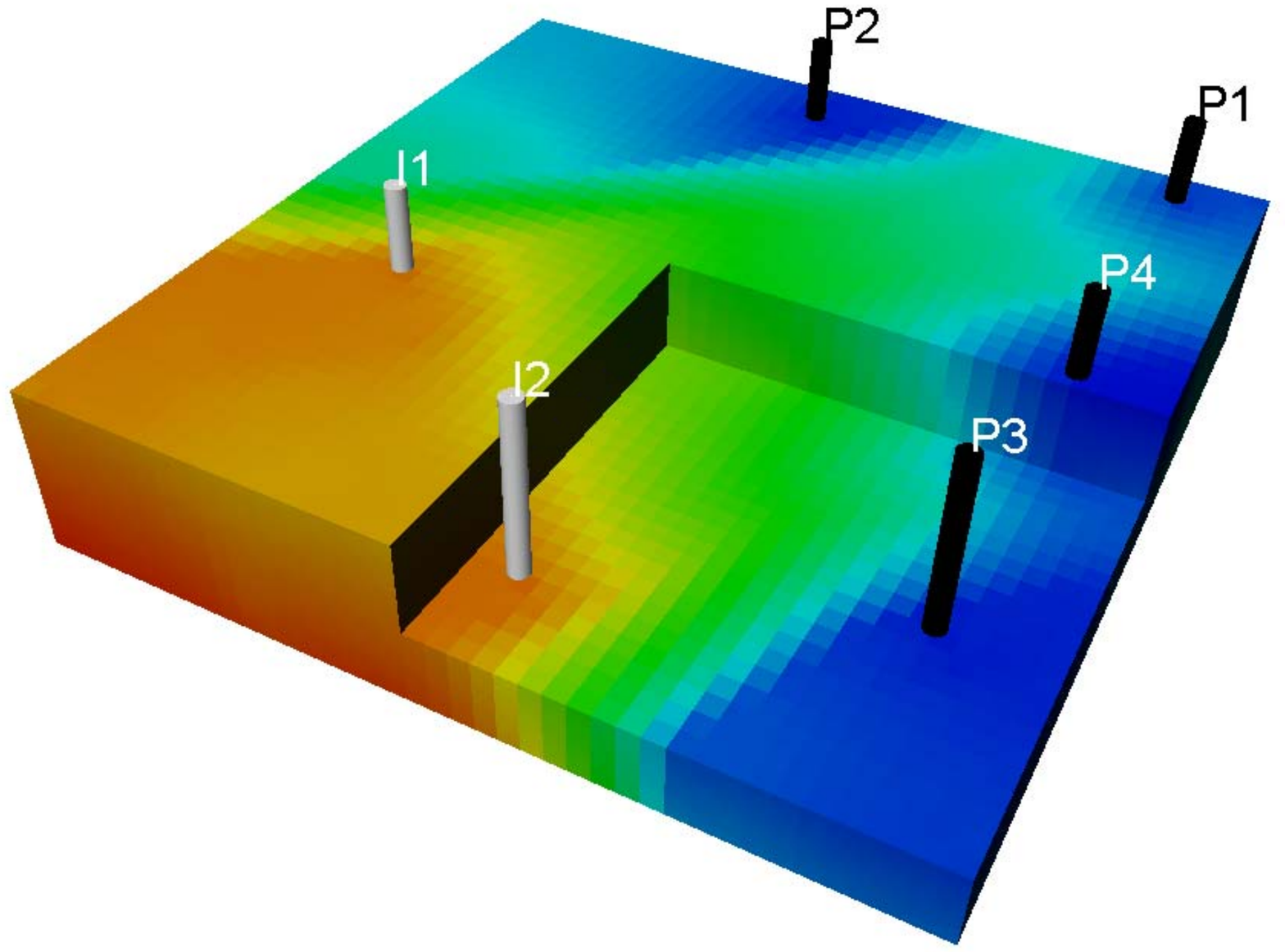}
         \caption{Realization~1 (surr)}
         \label{real-33}
     \end{subfigure}
     \begin{subfigure}[b]{0.32\textwidth}
         \centering
         \includegraphics[trim={4.5cm 3.5cm 4cm 3cm},clip, scale=0.25]{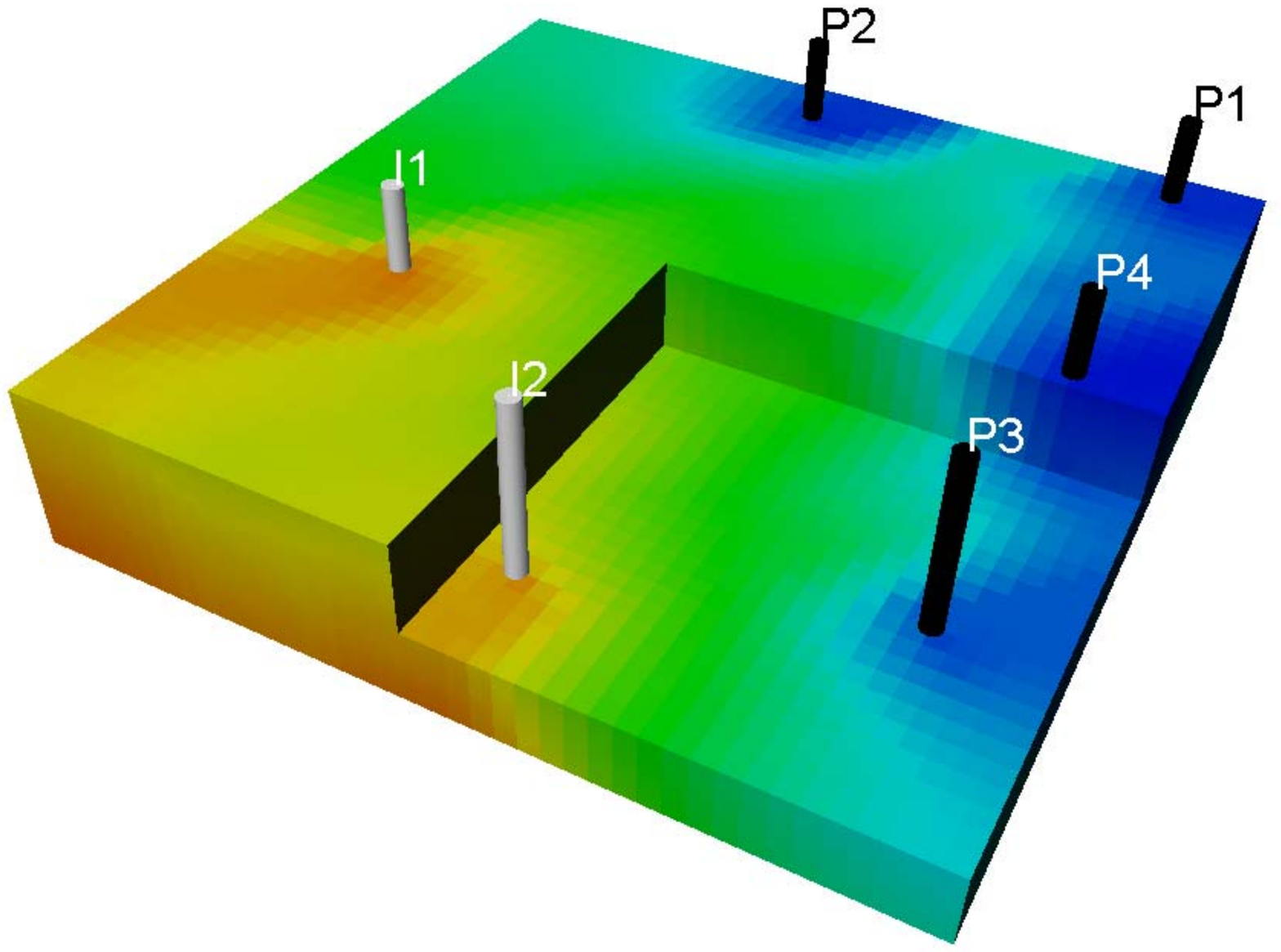}
         \caption{Realization~2 (surr)}
         \label{real-43}
     \end{subfigure}
     \begin{subfigure}[b]{0.32\textwidth}
         \centering
         \includegraphics[trim={4.5cm 3.5cm 4cm 3cm},clip, scale=0.25]{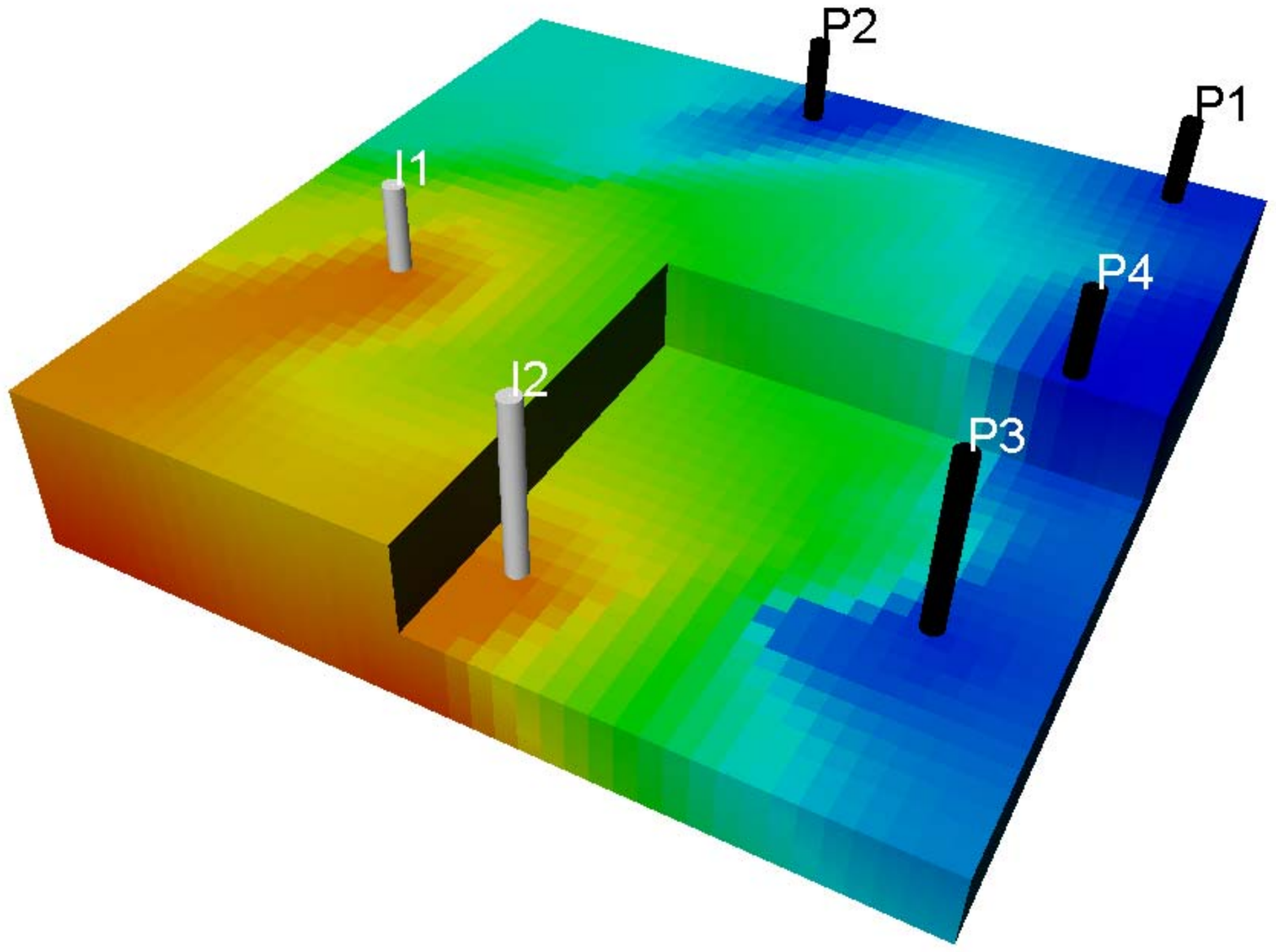}
         \caption{Realization~3 (surr)}
         \label{real-53}
     \end{subfigure}
     
     \begin{subfigure}[b]{0.32\textwidth}
         \centering
         \includegraphics[trim={4.5cm 3.5cm 4cm 3cm},clip, scale=0.25]{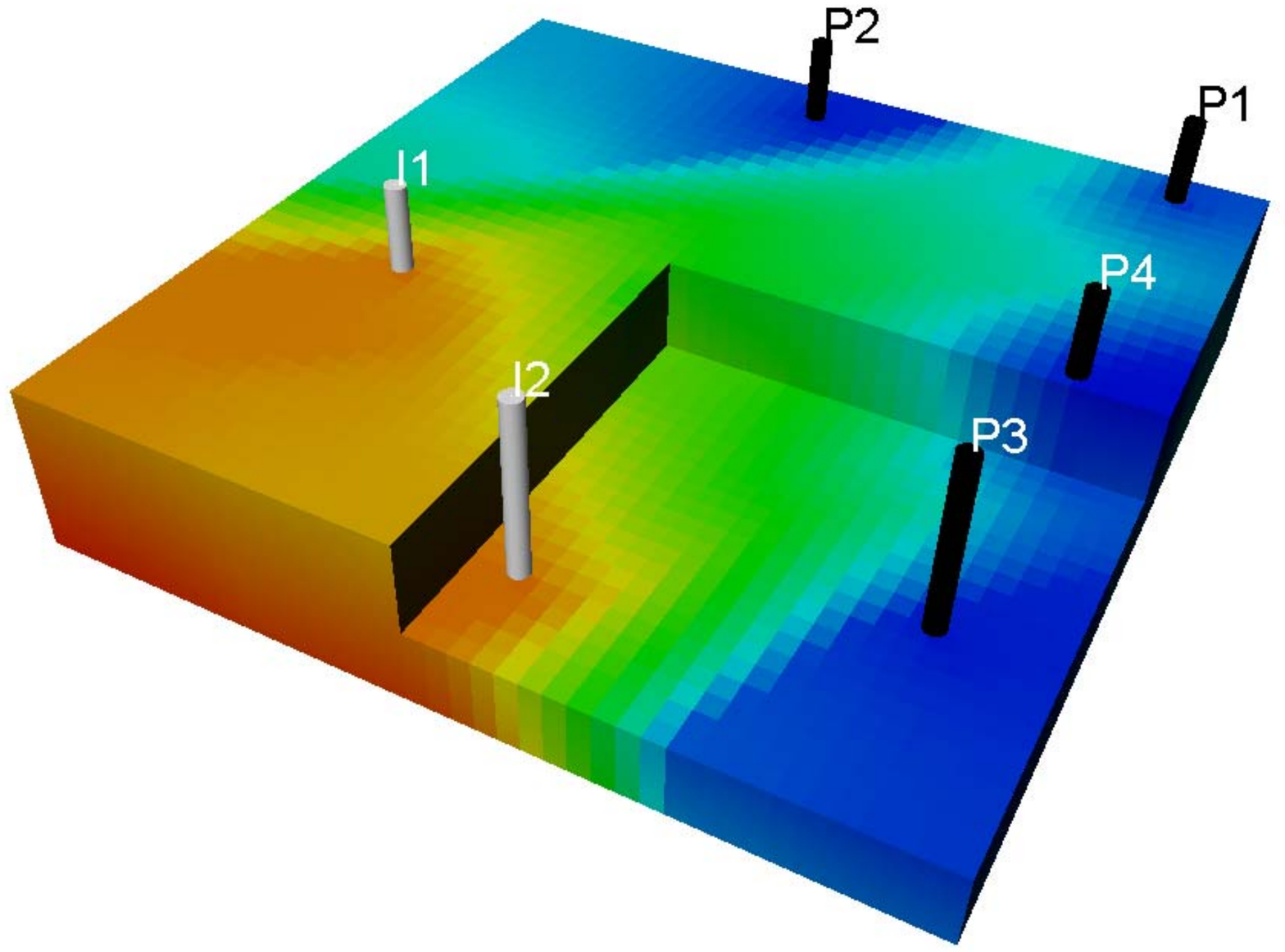}
         \caption{Realization~1 (sim)}
         \label{real-63}
     \end{subfigure}
          \begin{subfigure}[b]{0.32\textwidth}
         \centering
         \includegraphics[trim={4.5cm 3.5cm 4cm 3cm},clip, scale=0.25]{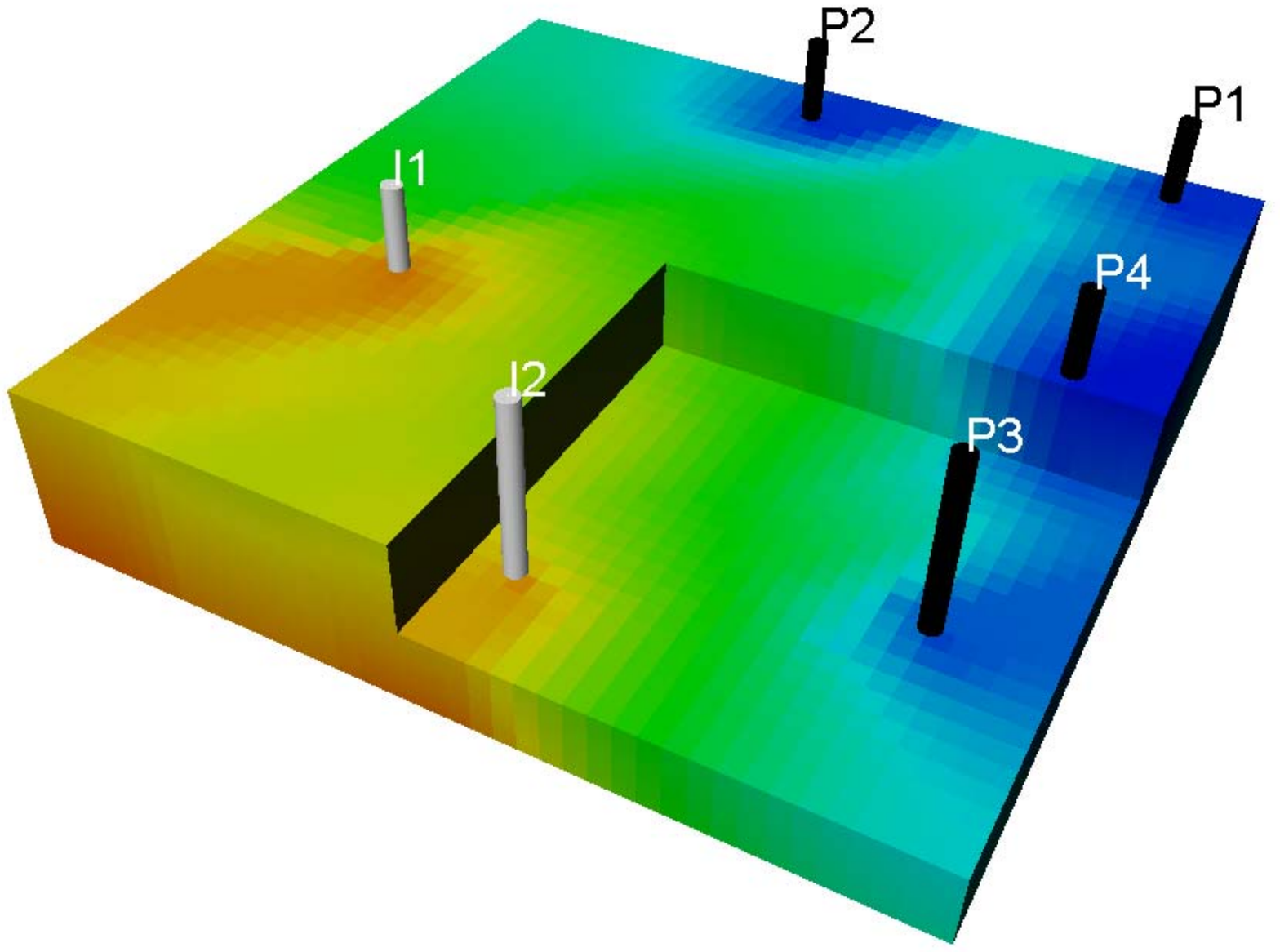}
         \caption{Realization~2 (sim)}
         \label{real-73}
     \end{subfigure}
          \begin{subfigure}[b]{0.32\textwidth}
         \centering
         \includegraphics[trim={4.5cm 3.5cm 4cm 3cm},clip, scale=0.25]{figs/chap2/p-sim-day400-r1.pdf}
         \caption{Realization~3 (sim)}
         \label{real-83}
     \end{subfigure}
    \caption{Pressure maps from 3D recurrent R-U-Net surrogate model (top row) and high-fidelity simulator (bottom row) for three different test-case realizations. All results are at 400~days. Saturation results in Fig.~\ref{fig:sat-map-single} are for Realization~1, shown in Fig.~\ref{fig:cnn-pca-reals}(a).}
    \label{fig:pressure-map-single}
\end{figure}

Because the global pressure field changes only slightly in time, instead of showing the evolution of pressure for a single realization, we instead present the pressure fields for three different test cases. These results are displayed in Fig.~\ref{fig:pressure-map-single}. All results are at 400~days. The global pressure fields for the three different realizations show significant differences, though the surrogate model provides an accurate visual 3D representation in all cases.

In addition to visual comparisons, it is of interest to quantify the errors associated with the saturation and pressure predictions. We thus compute relative saturation and pressure errors, given as follows:
\begin{equation}
{\delta_S = \frac{1}{n_{e}n_b n_t}\sum_{i=1}^{n_{e}}\sum_{j=1}^{n_b}\sum_{t=1}^{n_t} \frac{\norm{{(\hat{S}_w)}_{i, j}^{t} - {(S_w)}_{i,j}^{t}}}{{(S_w)}_{i,j}^{t}}},
\label{eq:sat-relative-error-time-t}
\end{equation}

\begin{equation}
{\delta_p = \frac{1}{n_{e}n_b n_t}\sum_{i=1}^{n_{e}}\sum_{j=1}^{n_b}\sum_{t=1}^{n_t} \frac{\norm{\hat{p}_{i,j}^{t} - p_{i,j}^{t}}}{p_{i,\text{max}}^t - p_{i,\text{min}}^t}},
\label{eq:pressure-relative-error-time-t}
\end{equation}
where $n_e=400$ is the number of test models considered and $p_{i,\text{max}}^t$ and $p_{i,\text{min}}^t$ are the maximum and minimum grid block pressures for test-case $i$ at time step $t$. The denominator in Eq.~\ref{eq:sat-relative-error-time-t} is well behaved since the initial (and minimum) saturation is 0.1. 

Applying Eqs.~\ref{eq:sat-relative-error-time-t} and \ref{eq:pressure-relative-error-time-t}, we obtain $\delta_S = 5.7\%$ and $\delta_p = 0.7\%$. These relatively low error values indicate that the 3D recurrent R-U-Net provides accurate global saturation and pressure predictions over the full ensemble of new geomodels. We next consider prediction accuracy in well-rate quantities.



\begin{figure}[htbp]
     \centering
     \begin{subfigure}[b]{0.36\textwidth}
         \centering
         \includegraphics[width=\textwidth]{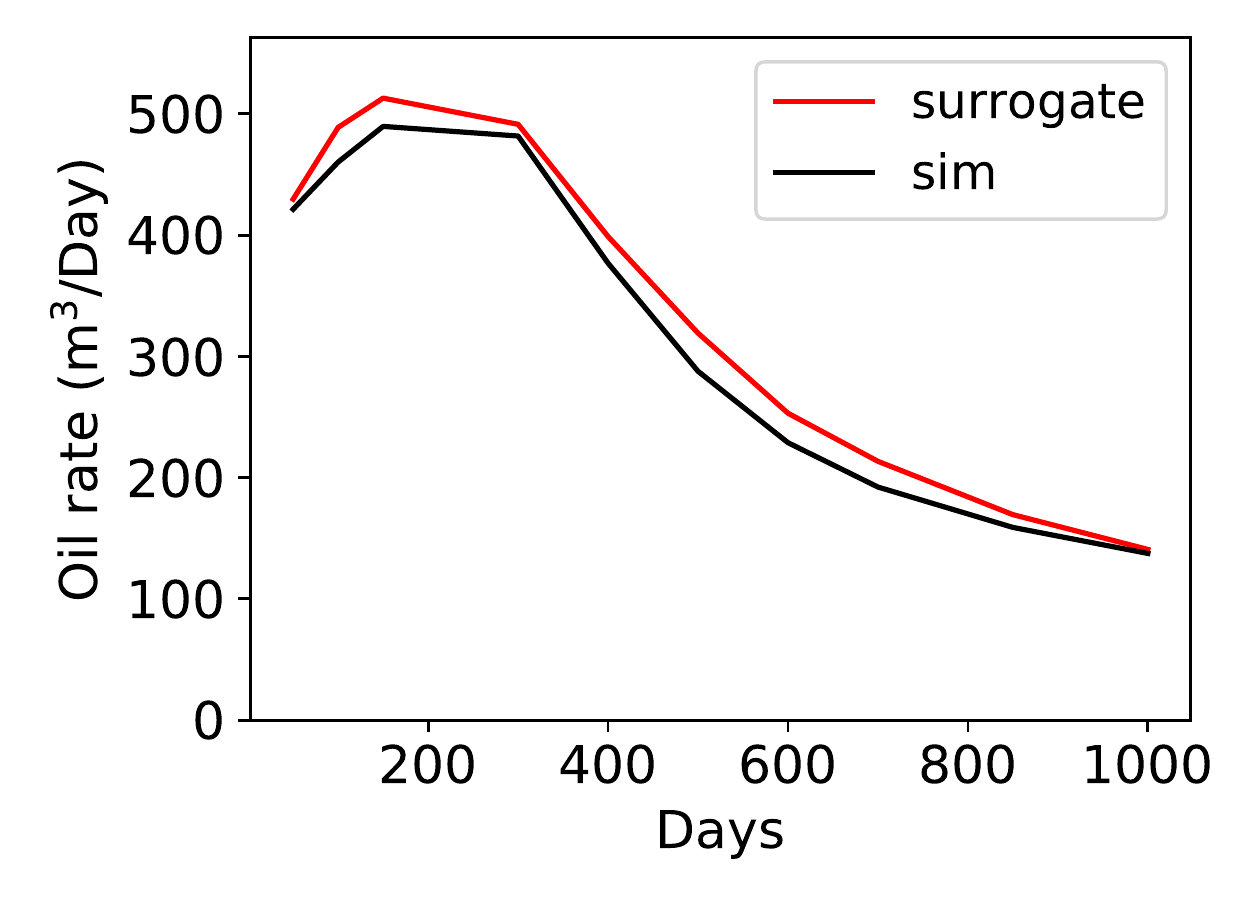}
         \caption{P1 oil rate}
         \label{pfs-case3-orate-w3}
     \end{subfigure}
     \begin{subfigure}[b]{0.36\textwidth}
         \centering
         \includegraphics[width=\textwidth]{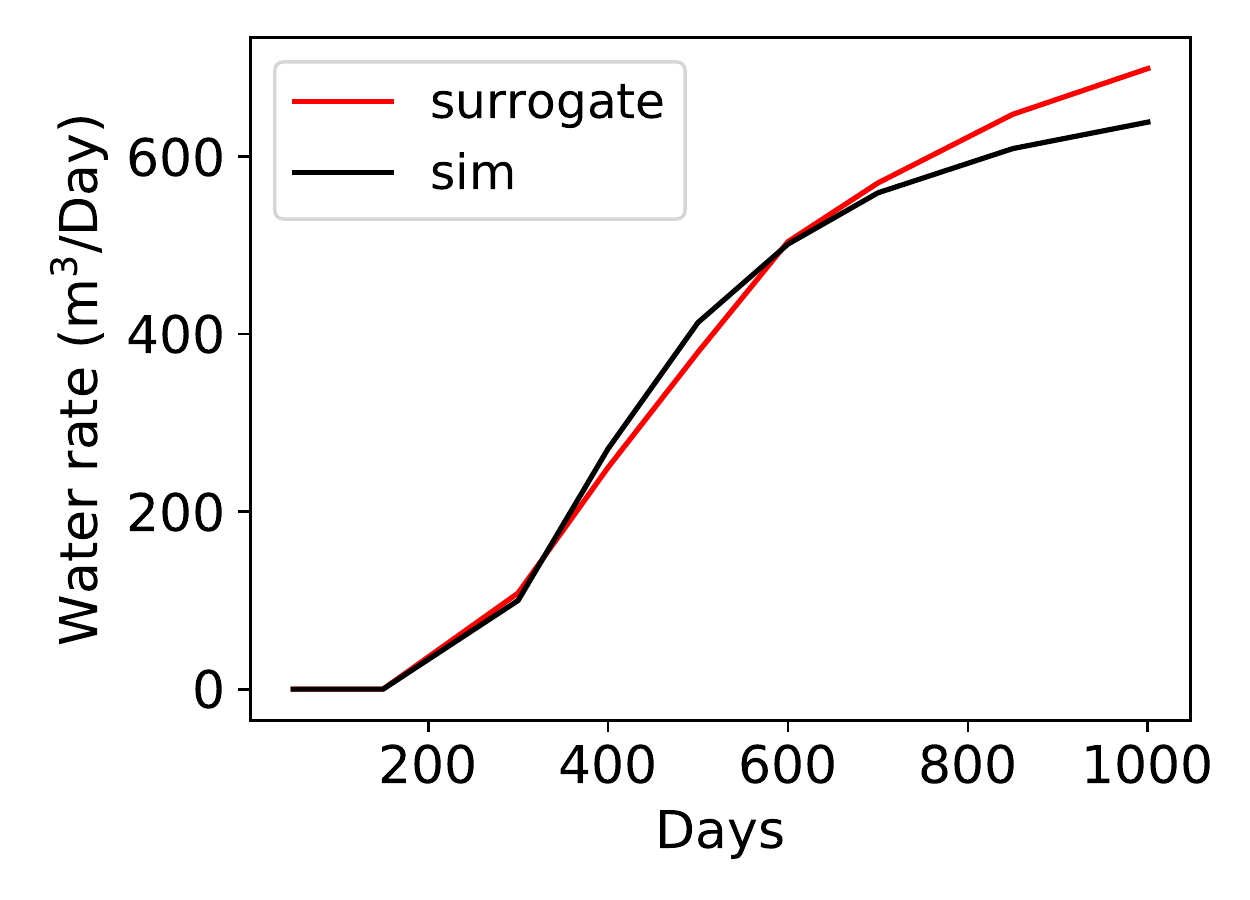}
         \caption{P1 water rate}
         \label{pfs-case3-wrate-w3}
     \end{subfigure}
     
     \begin{subfigure}[b]{0.36\textwidth}
         \centering
         \includegraphics[width=\textwidth]{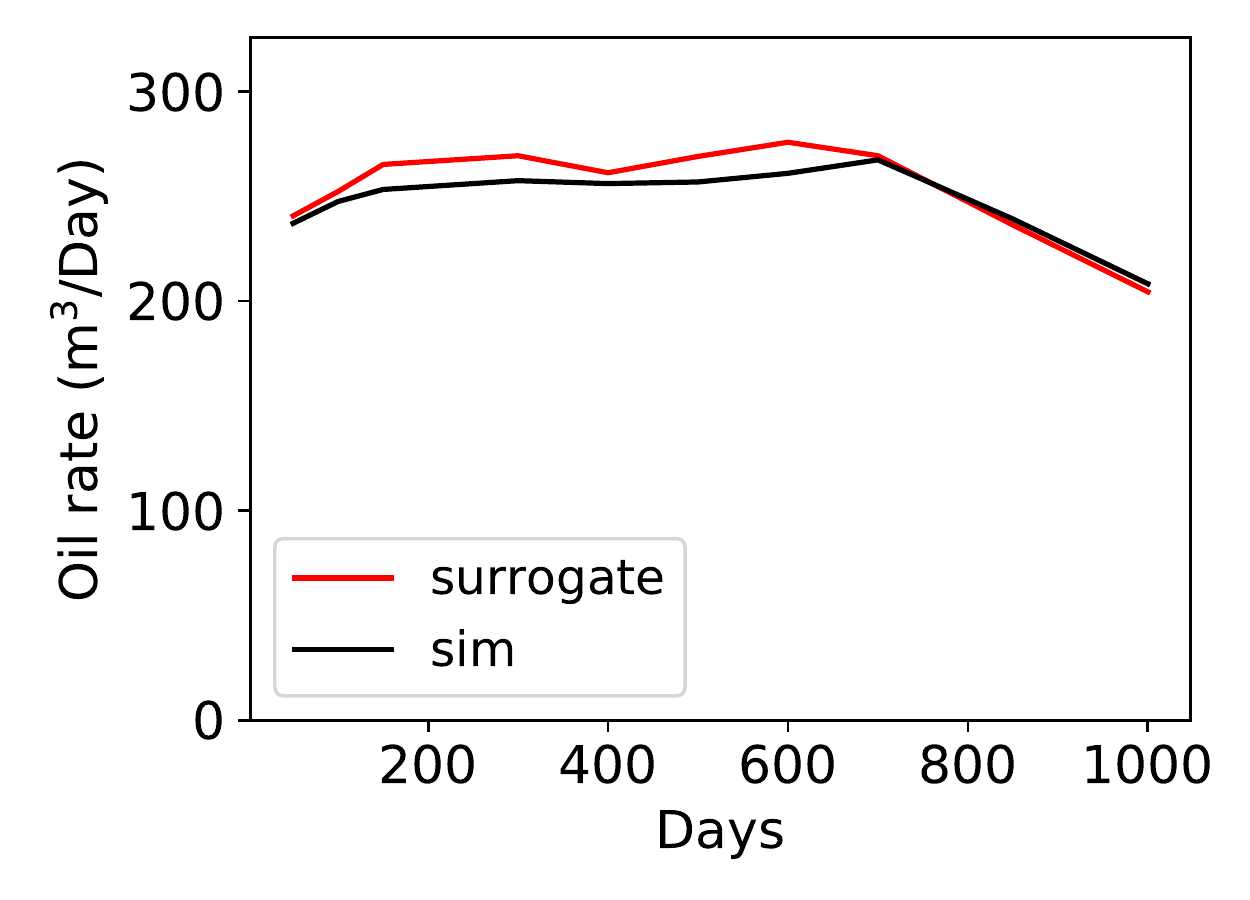}
         \caption{P2 oil rate}
         \label{pfs-case3-orate-w5}
     \end{subfigure}
     \begin{subfigure}[b]{0.36\textwidth}
         \centering
         \includegraphics[width=\textwidth]{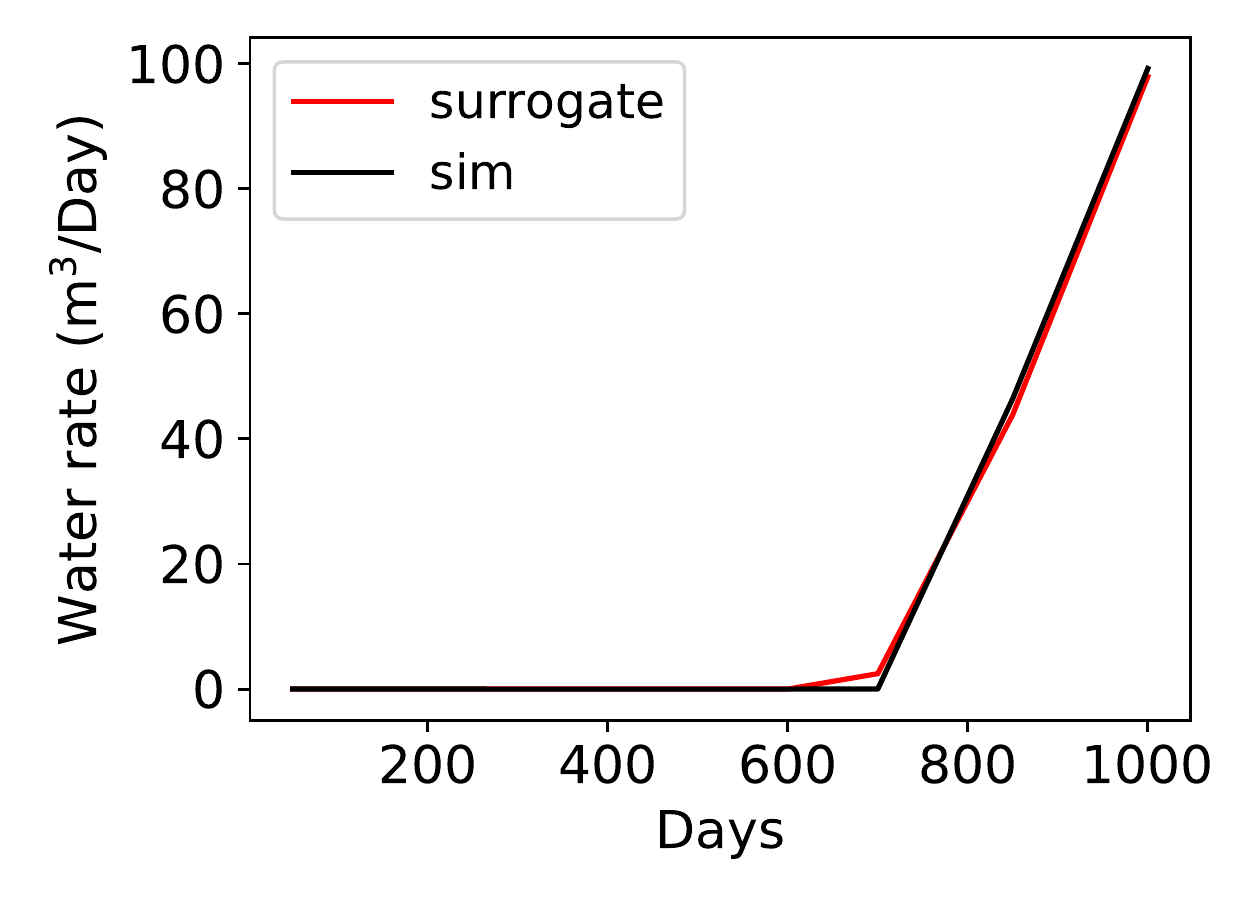}
         \caption{P2 water rate}
         \label{pfs-case3-wrate-w5}
     \end{subfigure}
     
    \begin{subfigure}[b]{0.36\textwidth}
         \centering
         \includegraphics[width=\textwidth]{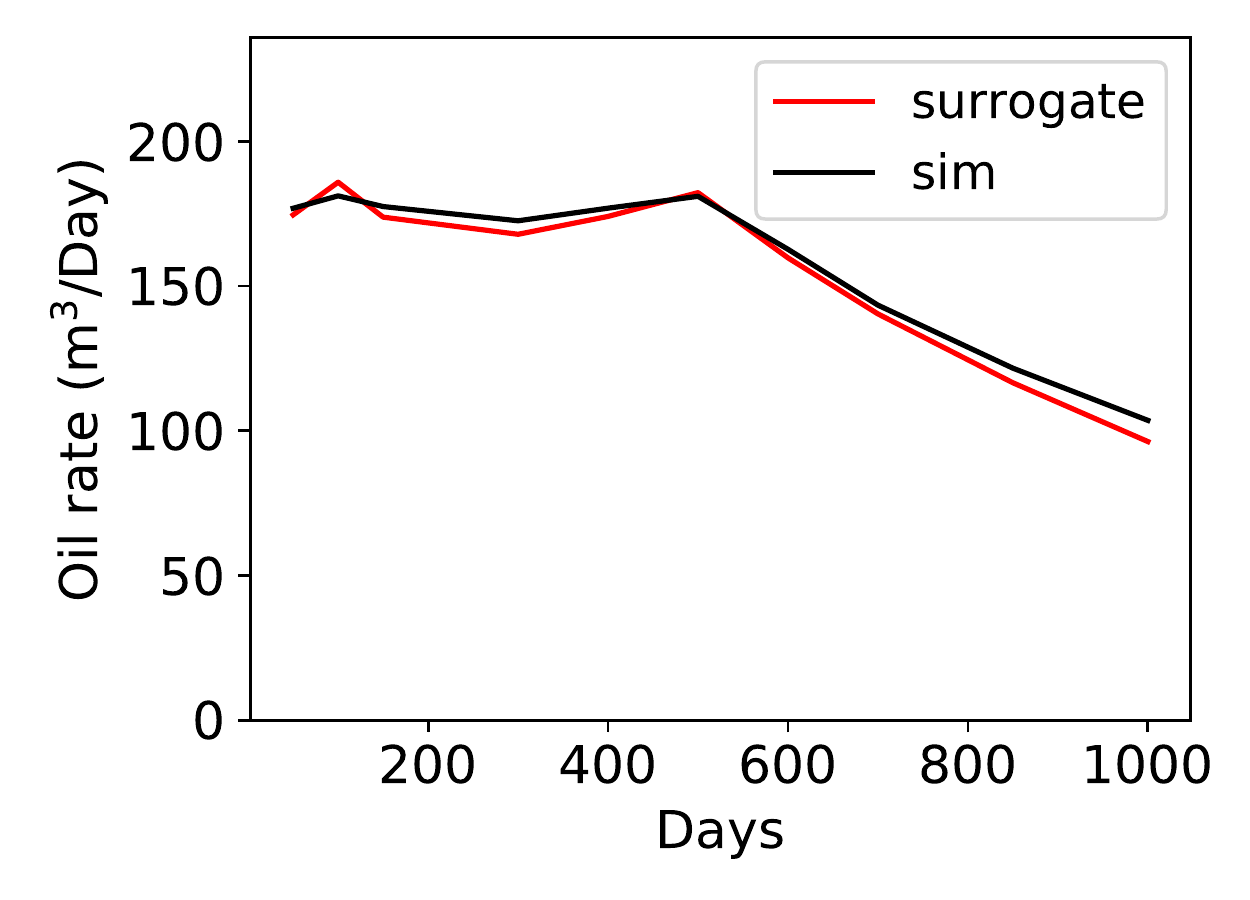}
         \caption{P3 oil rate}
         \label{pfs-case3-orate-w11}
     \end{subfigure}
          \begin{subfigure}[b]{0.36\textwidth}
         \centering
         \includegraphics[width=\textwidth]{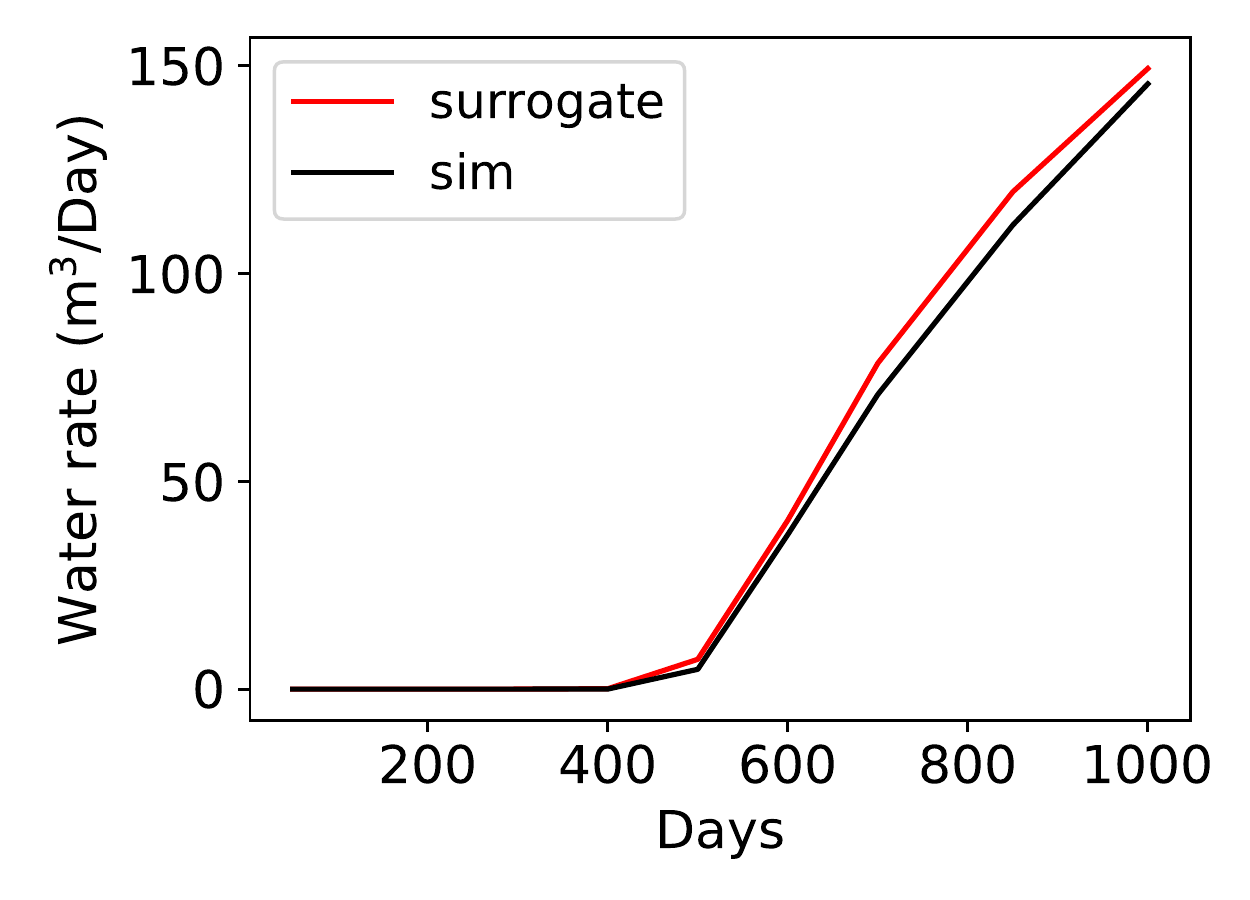}
         \caption{P3 water rate}
         \label{pfs-case3-wrate-w11}
     \end{subfigure}
     
     \begin{subfigure}[b]{0.36\textwidth}
         \centering
         \includegraphics[width=\textwidth]{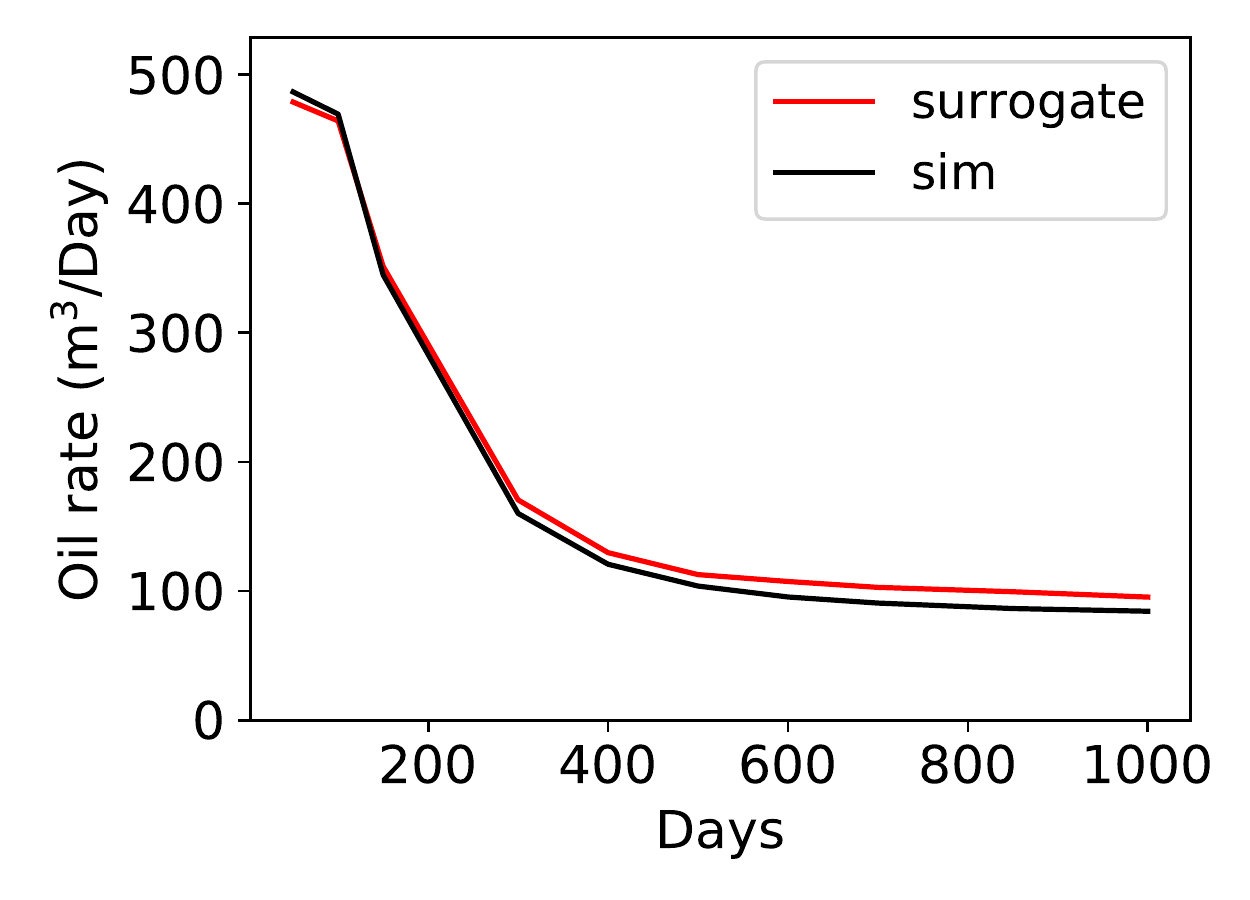}
         \caption{P4 oil rate}
         \label{pfs-case3-orate-w12}
     \end{subfigure}
          \begin{subfigure}[b]{0.36\textwidth}
         \centering
         \includegraphics[width=\textwidth]{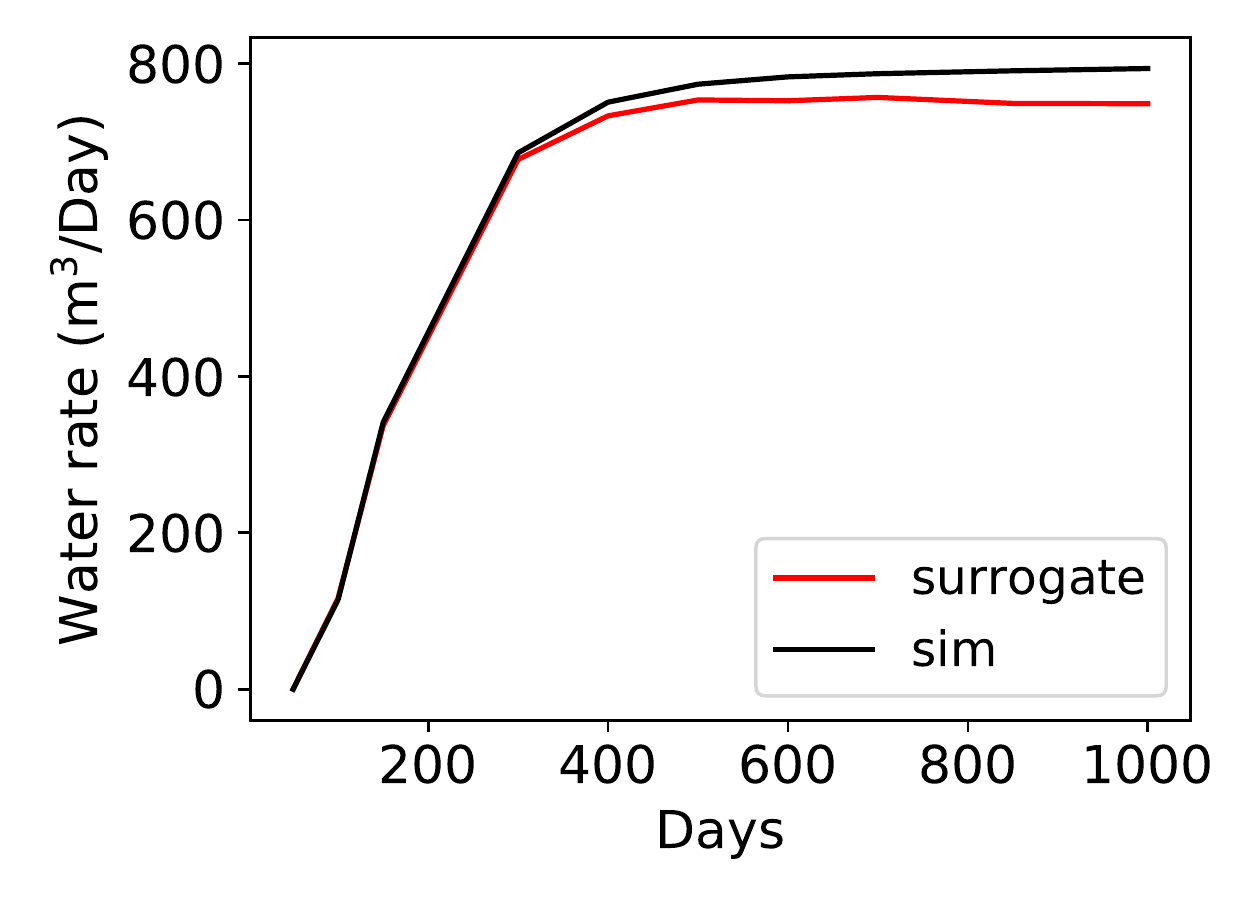}
         \caption{P4 water rate}
         \label{pfs-case3-wrate-w12}
     \end{subfigure}
    
    \caption{Comparison of oil (left) and water (right) production rates, for all four production wells, for the geomodel shown in Fig.~\ref{fig:cnn-pca-reals}(a). Red and black curves represent results from the 3D recurrent R-U-Net surrogate model and high-fidelity simulation, respectively.}
    \label{fig:producer-flow-single-case}
\end{figure}

Well rates are computed for both the surrogate and HFS models using the treatment described in Section~\ref{sec:gov_eqn}. Oil and water production rates for the four producers, as a function of time, are shown in Fig.~\ref{fig:producer-flow-single-case}. These results are for the 3D CNN-PCA geomodel shown in Fig.~\ref{fig:cnn-pca-reals}(a), with corresponding saturation fields shown in Fig.~\ref{fig:sat-map-single}. Results are generated at $n_t=10$ time steps, though in these and subsequent figures the points are connected linearly to provide continuous curves. The general level of agreement in well rates is quite close, and trends and water breakthrough (i.e., appearance of water at production wells) are captured with reasonable accuracy by the surrogate model. Slight discrepancies are apparent in some instances, however, such as in late-time water rate in Fig.~\ref{fig:producer-flow-single-case}(b) and (h).

\begin{figure}[htbp]
     \centering
     \begin{subfigure}[b]{0.36\textwidth}
         \centering
         \includegraphics[width=\textwidth]{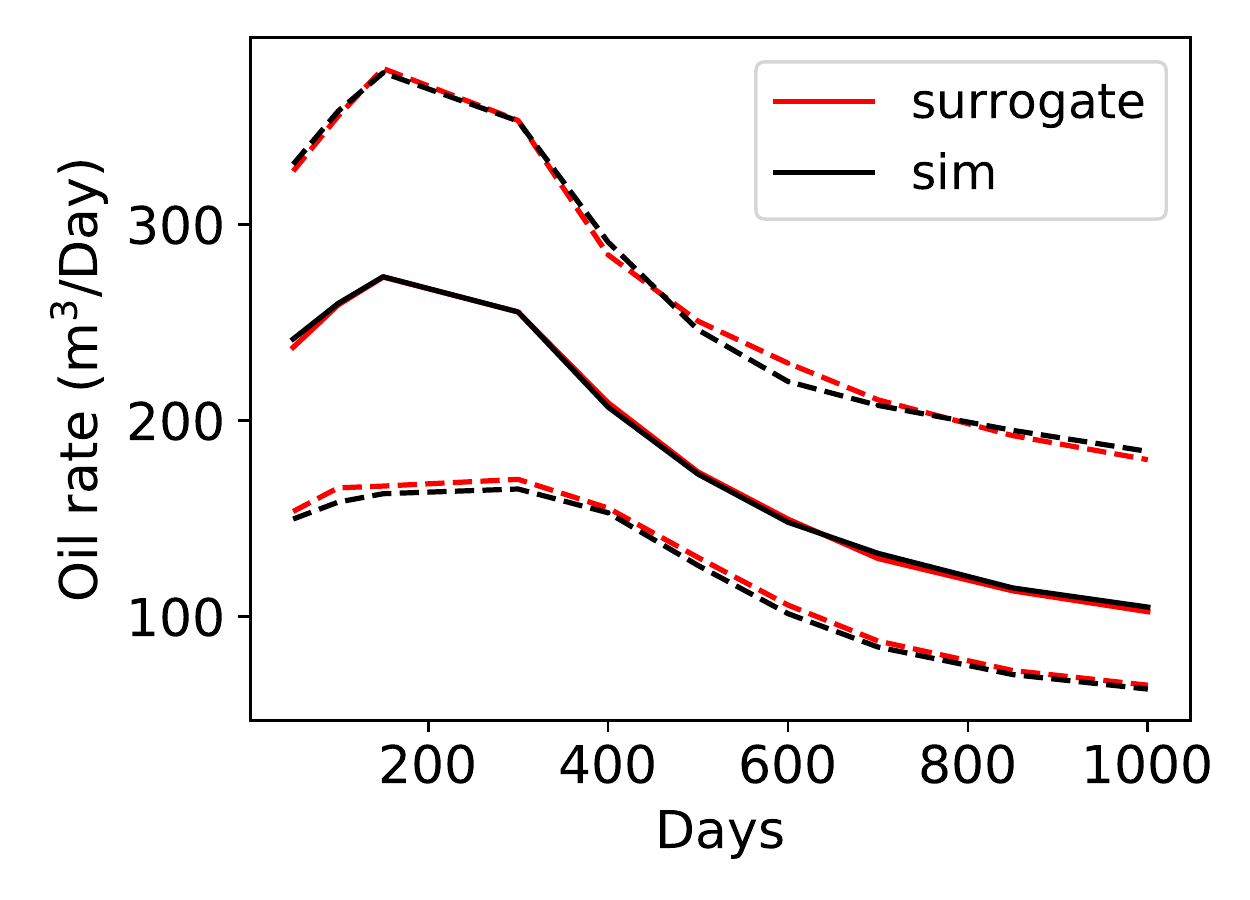}
         \caption{P1 oil rate}
         \label{pfs-orate-w1}
     \end{subfigure}
     \begin{subfigure}[b]{0.36\textwidth}
         \centering
         \includegraphics[width=\textwidth]{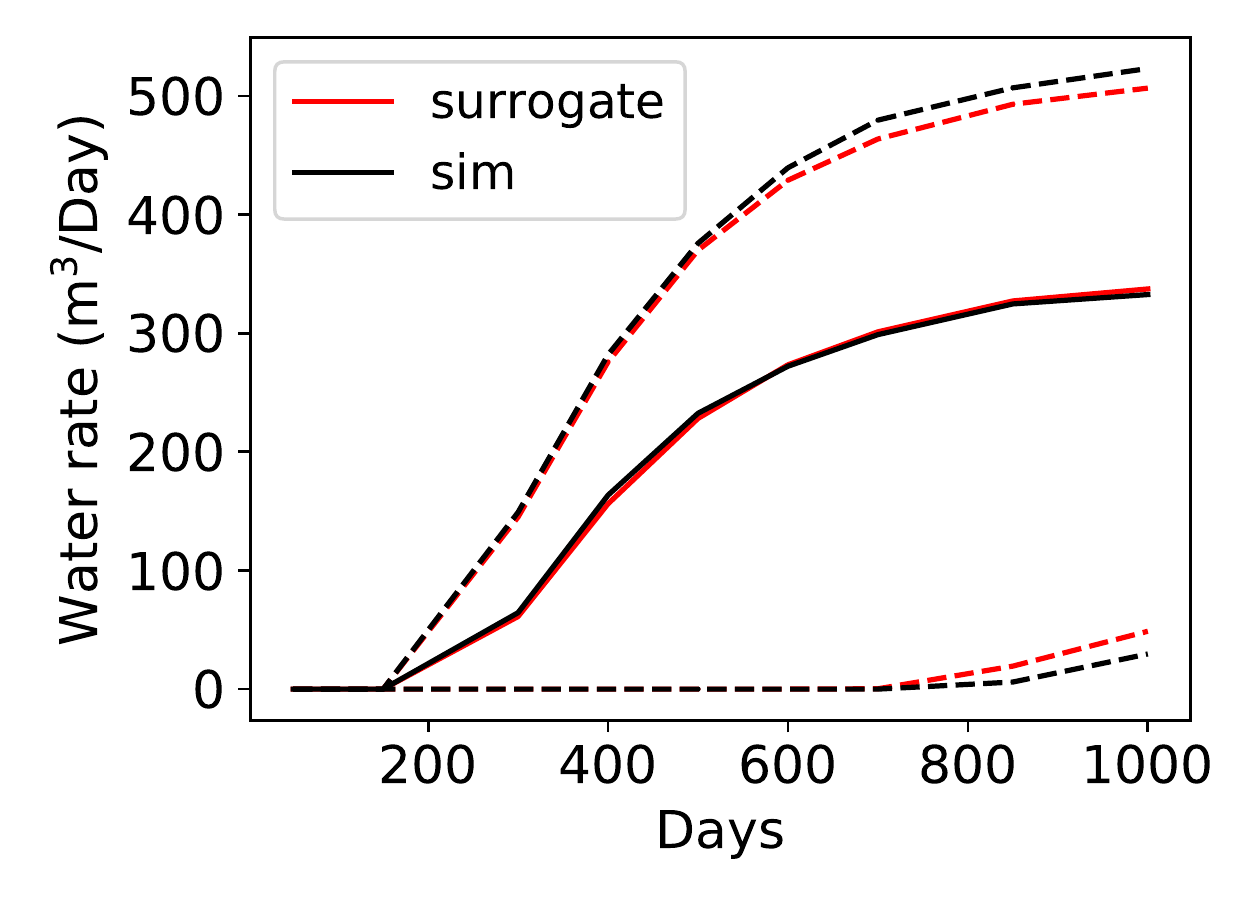}
         \caption{P1 water rate}
         \label{pfs-wrate-w1}
     \end{subfigure}
     
     \begin{subfigure}[b]{0.36\textwidth}
         \centering
         \includegraphics[width=\textwidth]{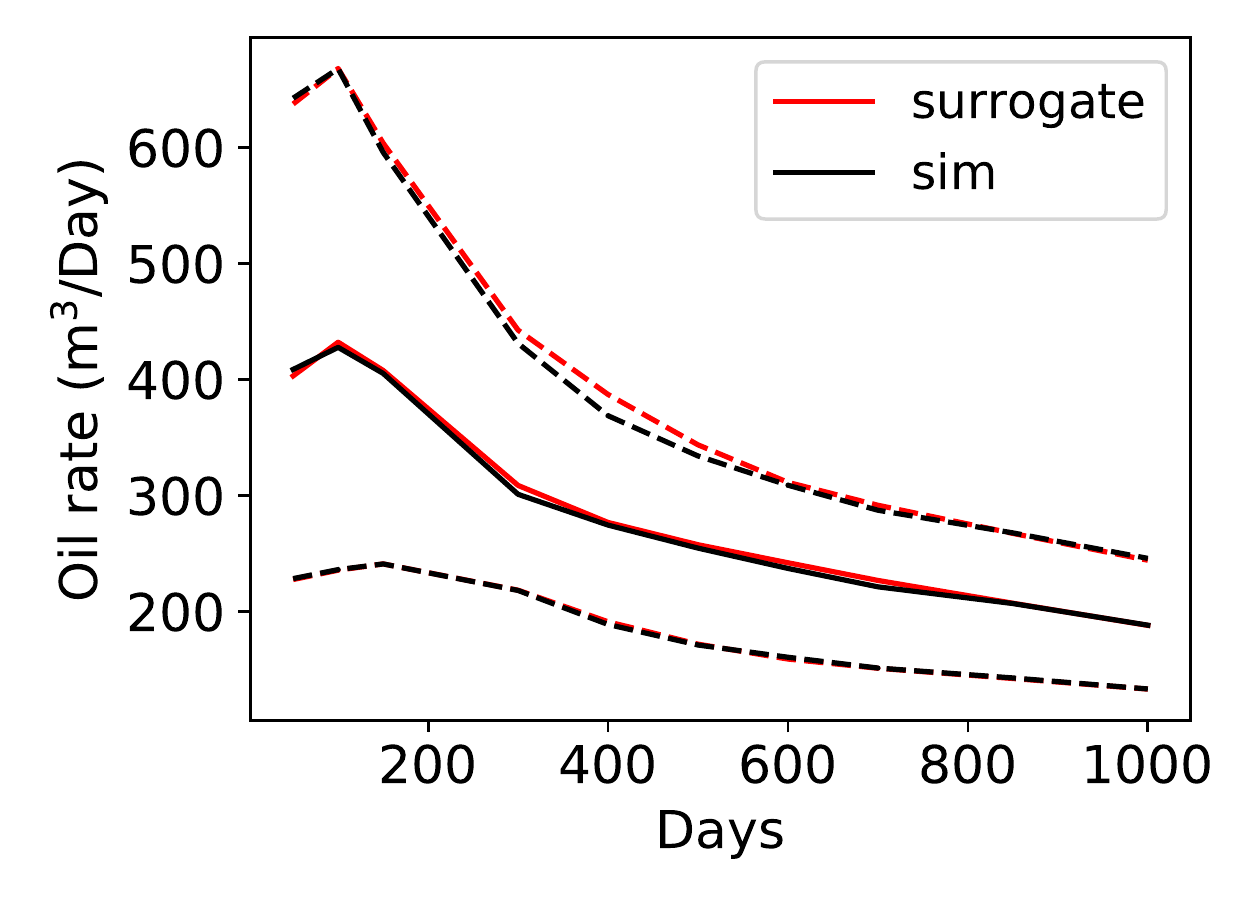}
         \caption{P2 oil rate}
         \label{pfs-orate-w2}
     \end{subfigure}
     \begin{subfigure}[b]{0.36\textwidth}
         \centering
         \includegraphics[width=\textwidth]{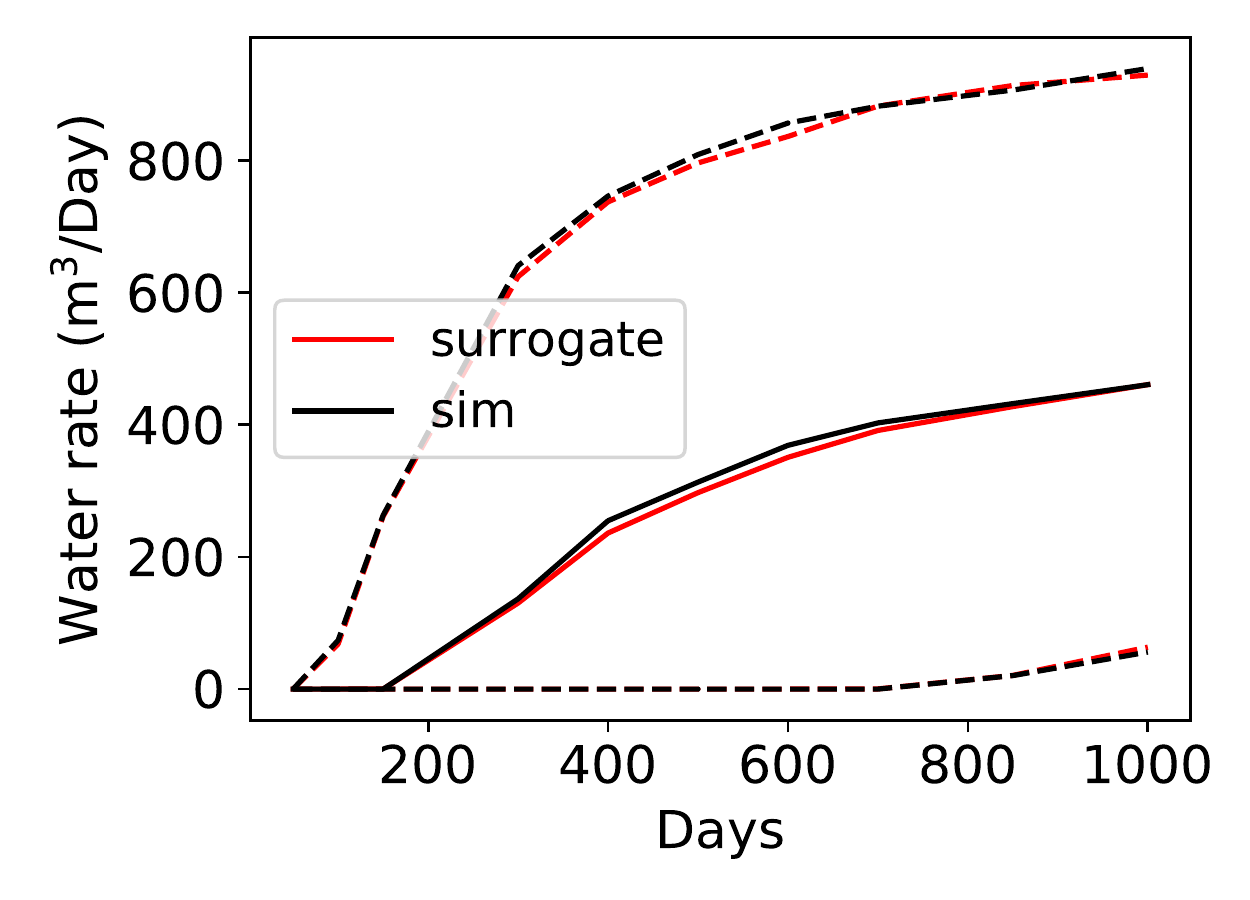}
         \caption{P2 water rate}
         \label{pfs-wrate-w2}
     \end{subfigure}
     
    \begin{subfigure}[b]{0.36\textwidth}
         \centering
         \includegraphics[width=\textwidth]{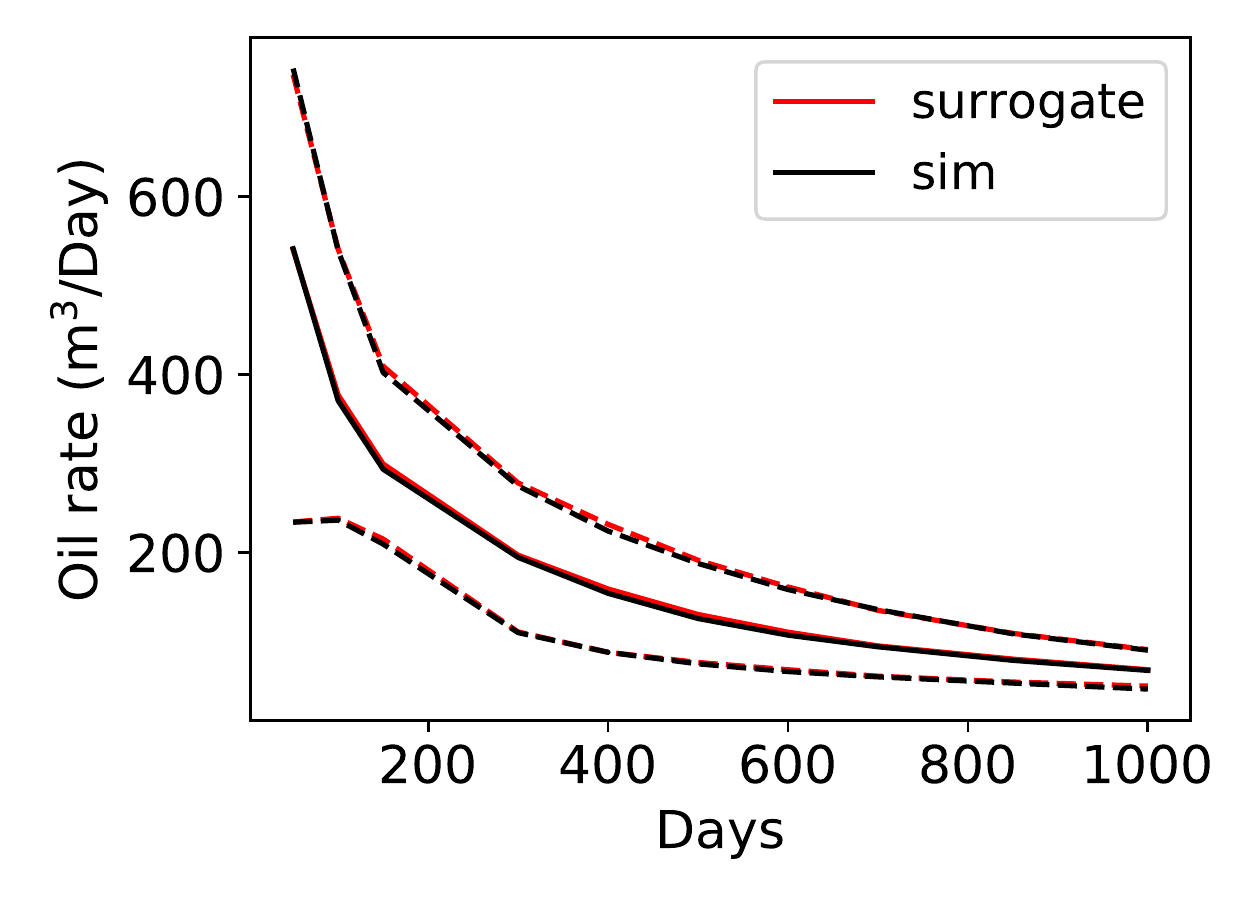}
         \caption{P3 oil rate}
         \label{pfs-orate-w3}
     \end{subfigure}
     \begin{subfigure}[b]{0.36\textwidth}
         \centering
         \includegraphics[width=\textwidth]{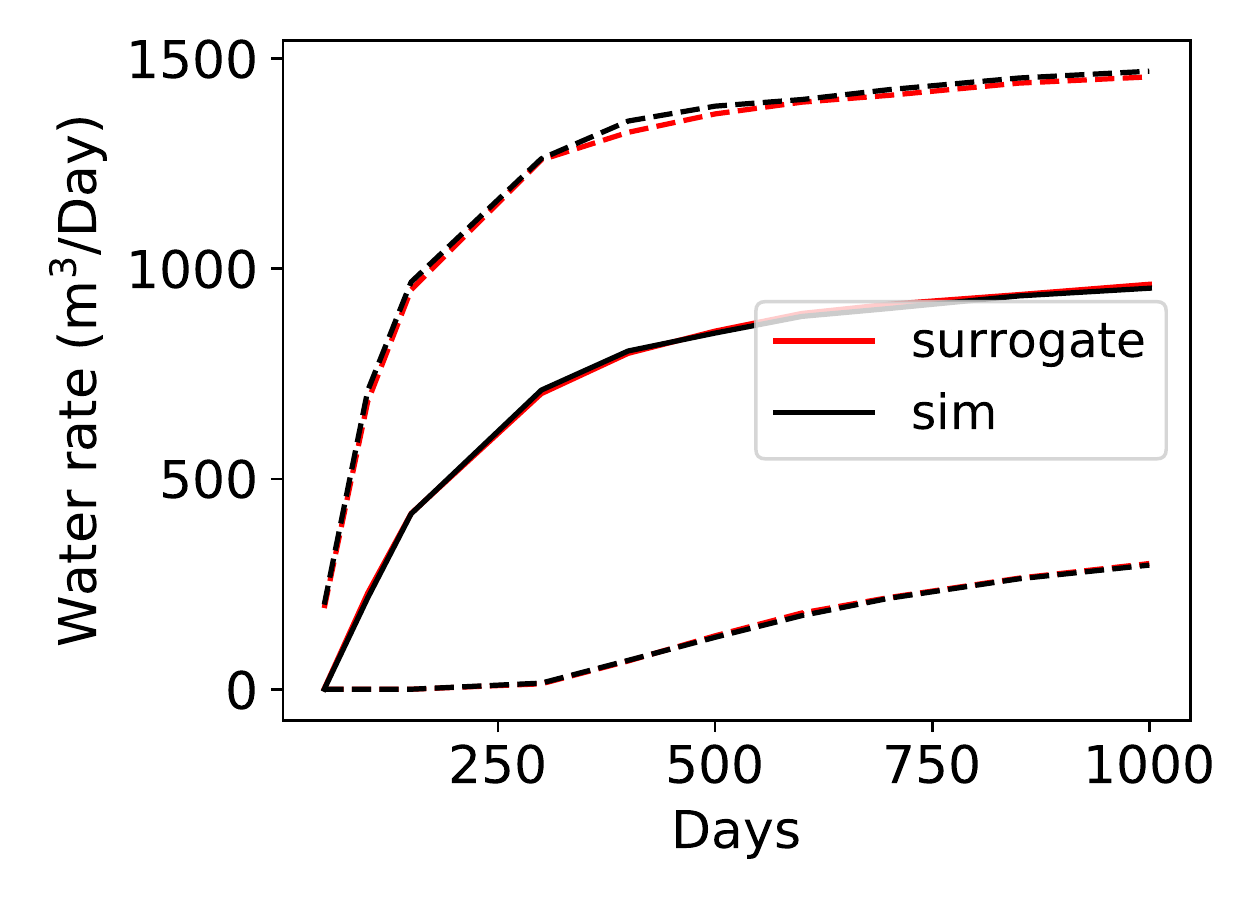}
         \caption{P3 water rate}
         \label{pfs-wrate-w3}
     \end{subfigure}
     
     \begin{subfigure}[b]{0.36\textwidth}
         \centering
         \includegraphics[width=\textwidth]{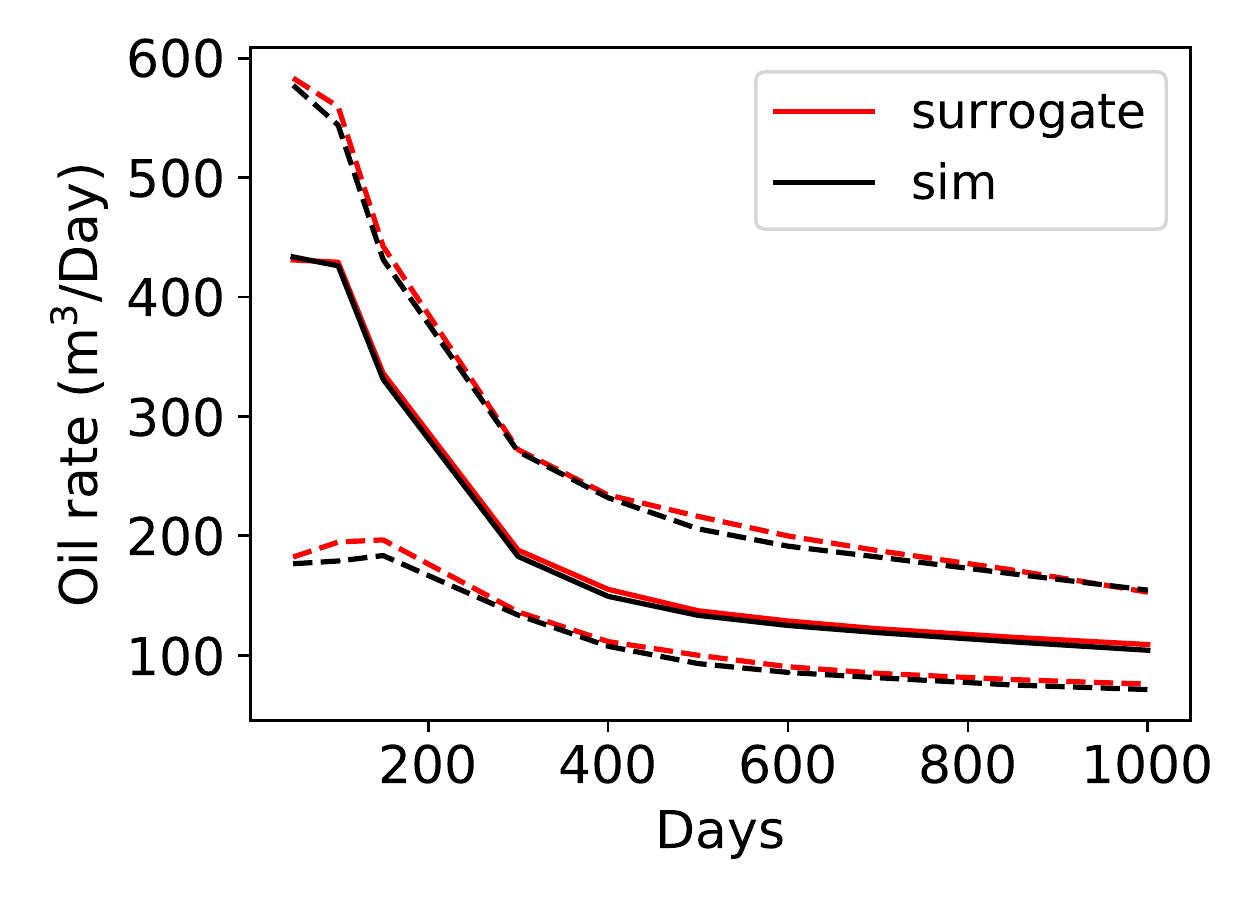}
         \caption{P4 oil rate}
         \label{pfs-orate-w4}
     \end{subfigure}
     \begin{subfigure}[b]{0.36\textwidth}
         \centering
         \includegraphics[width=\textwidth]{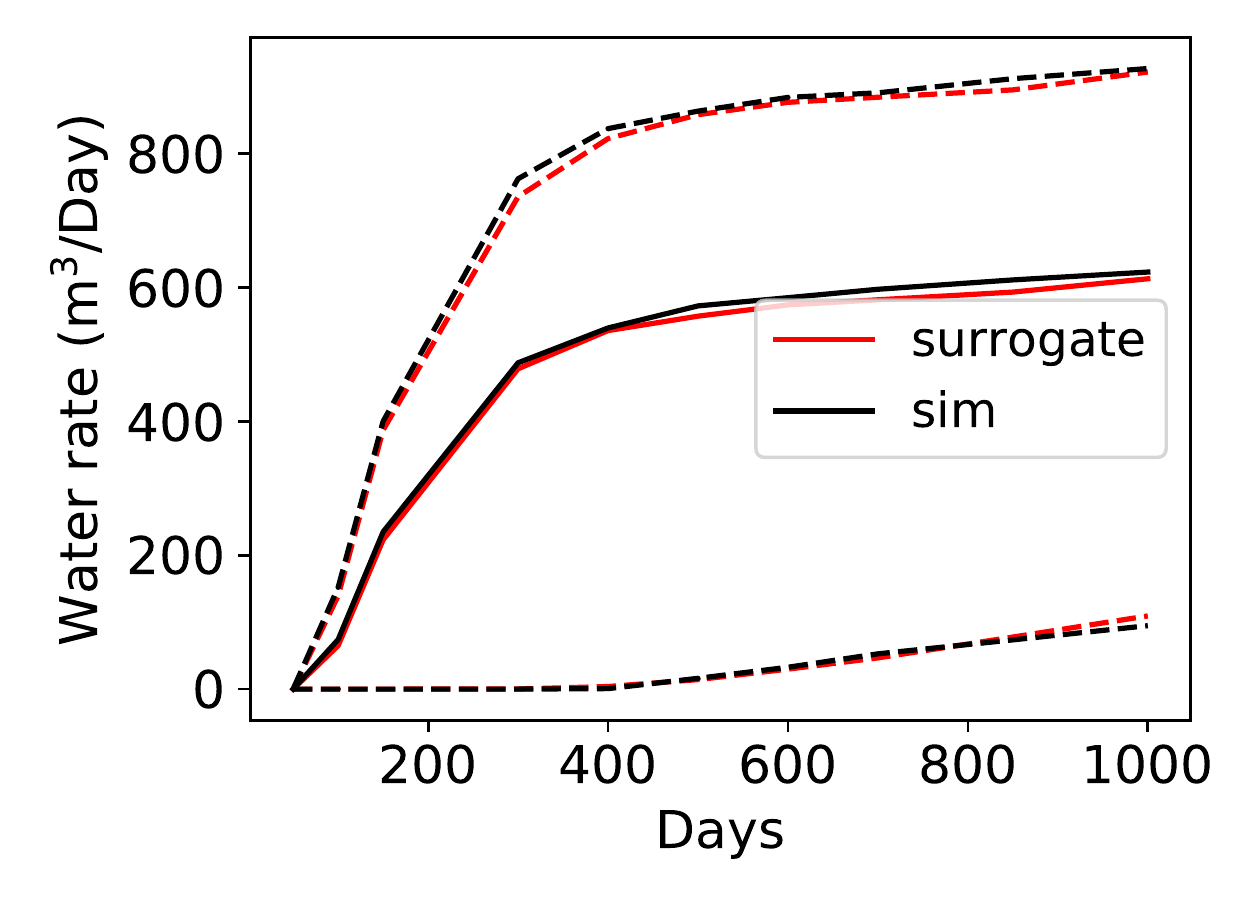}
         \caption{P4 water rate}
         \label{pfs-wrate-w4}
     \end{subfigure}
    
\caption{Comparison of oil (left) and water (right) production rate statistics, for all four production wells, over the full ensemble of 400 test cases. Red and black curves represent results from the recurrent R-U-Net surrogate model and high-fidelity simulation, respectively. Solid curves correspond to $\text{P}_{50}$ results, lower and upper dashed curves to $\text{P}_{10}$ and $\text{P}_{90}$ results.}
    \label{fig:producer-flow-statistics}
\end{figure}

We now present statistical results for well responses over the full set of test cases.
The $\text{P}_{10}$, $\text{P}_{50}$ and $\text{P}_{90}$ oil and water production rates, evaluated over all 400 test cases, are shown in Fig.~\ref{fig:producer-flow-statistics}.
The $\text{P}_{50}$ (solid) curves depict the median response at each time step, while the $\text{P}_{10}$ and $\text{P}_{90}$ (dashed) curves show the responses corresponding to the 10th and 90th percentiles.
Agreement in these statistical quantities is seen to be very close, which again demonstrates the accuracy of the surrogate model. It is noteworthy that accuracy is maintained over large ranges in rates, as is particularly evident in the water rate plots in Fig.~\ref{fig:producer-flow-statistics}.

The average relative errors in oil and water production rates, across all production wells and over all time steps, are given by
\begin{equation}
\delta_{r,j} = \frac{1}{n_{e}n_{p}n_t}\sum_{i=1}^{n_{e}}\sum_{k=1}^{n_p}\sum_{t=1}^{n_{t}} \frac{\norm{{(\hat{r}_{j})}_{i,k}^{t} - {({r}_{j})}_{i,k}^{t}}}{{({r}_{j})}_{i,k}^{t} + \epsilon},
\label{eq:rate-rel-error-time-t}
\end{equation}
where $n_p$ denotes the number of production wells, ${(\hat{r}_{j})}_{i,k}^{t}$ denotes the phase ($j=o,~w$) production rate from the surrogate model for well $k$ at time step $t$ in test sample $i$, and ${({r}_{j})}_{i,k}^{t}$ denotes the corresponding HFS result. A constant $\epsilon = 10$ is introduced in the denominator to avoid division by very small values. Over $n_e = 400$ test samples, the relative errors for oil and water production rates are found to be $\delta_{r,o} = 6.1\%$ and $\delta_{r,w} = 8.8\%$. This water rate error is higher than in the 2D system considered in \cite{tang2020deep} (where $\delta_{r,w} = 5.8\%$), though the error in $\delta_{r,o}$ is about the same. As we will see, this level of accuracy is indeed sufficient for the data assimilation studies considered in the next section.

\section{Use of Deep-Learning Procedures for History Matching}
\label{sec:HM_results}

We now apply the 3D recurrent R-U-Net surrogate model for a challenging data assimilation problem. Two different history matching algorithms are considered -- a rigorous rejection sampling (RS) procedure, and ensemble smoother with multiple data assimilation~(ES-MDA). The 3D CNN-PCA algorithm is used in both cases to generate the geomodels evaluated by the surrogate flow model. 

In RS, 3D CNN-PCA is used only to provide a very large number of prior models; i.e., $\boldsymbol\xi$ is not manipulated to achieve a history match. Overall timing is reduced since the generation of prior models is much faster with 3D CNN-PCA than with geological modeling software. In the ES-MDA framework, by contrast, the low-dimensional variable $\boldsymbol\xi$ is updated by the history matching algorithm. The use of CNN-PCA in this setting ensures that posterior models (for any $\boldsymbol\xi$) are consistent geologically with the original set of object-based Petrel realizations. 


\begin{figure}[htbp]
     \centering
     \begin{subfigure}[b]{0.36\textwidth}
         \centering
         \includegraphics[width=\textwidth]{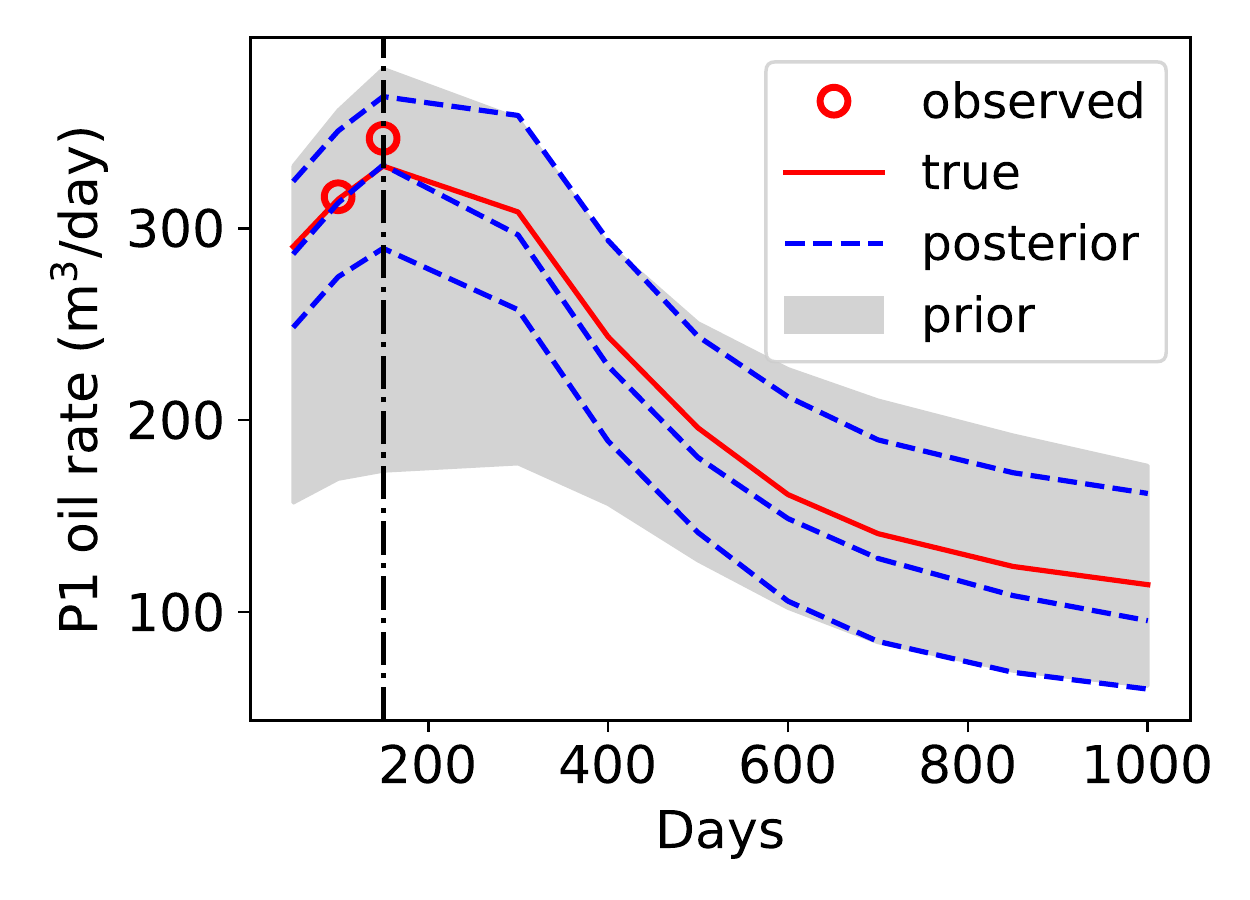}
         \caption{P1 oil rate}
         \label{hm-orate-w1}
     \end{subfigure}
     \begin{subfigure}[b]{0.36\textwidth}
         \centering
         \includegraphics[width=\textwidth]{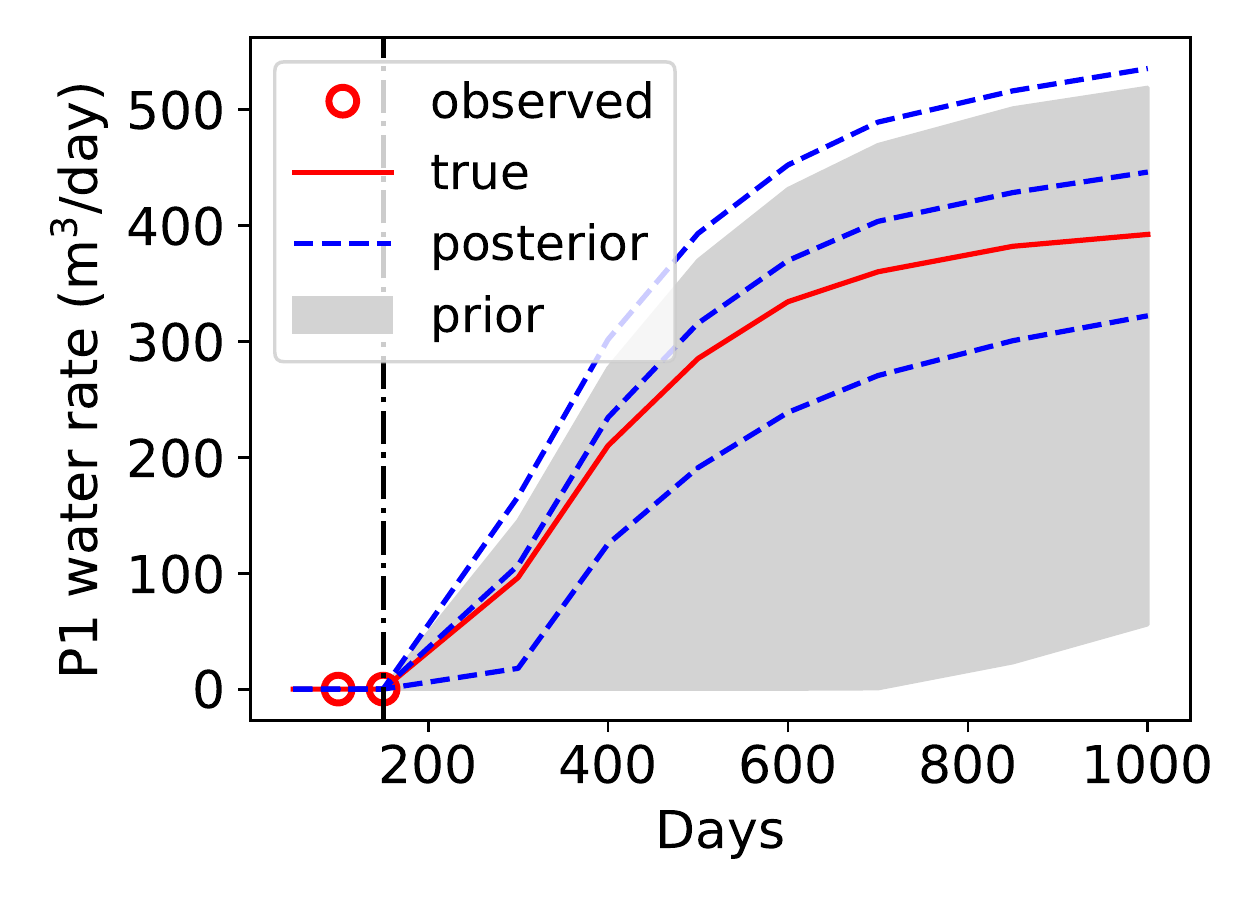}
         \caption{P1 water rate}
         \label{hm-wrate-w1}
     \end{subfigure}
     
     \begin{subfigure}[b]{0.36\textwidth}
         \centering
         \includegraphics[width=\textwidth]{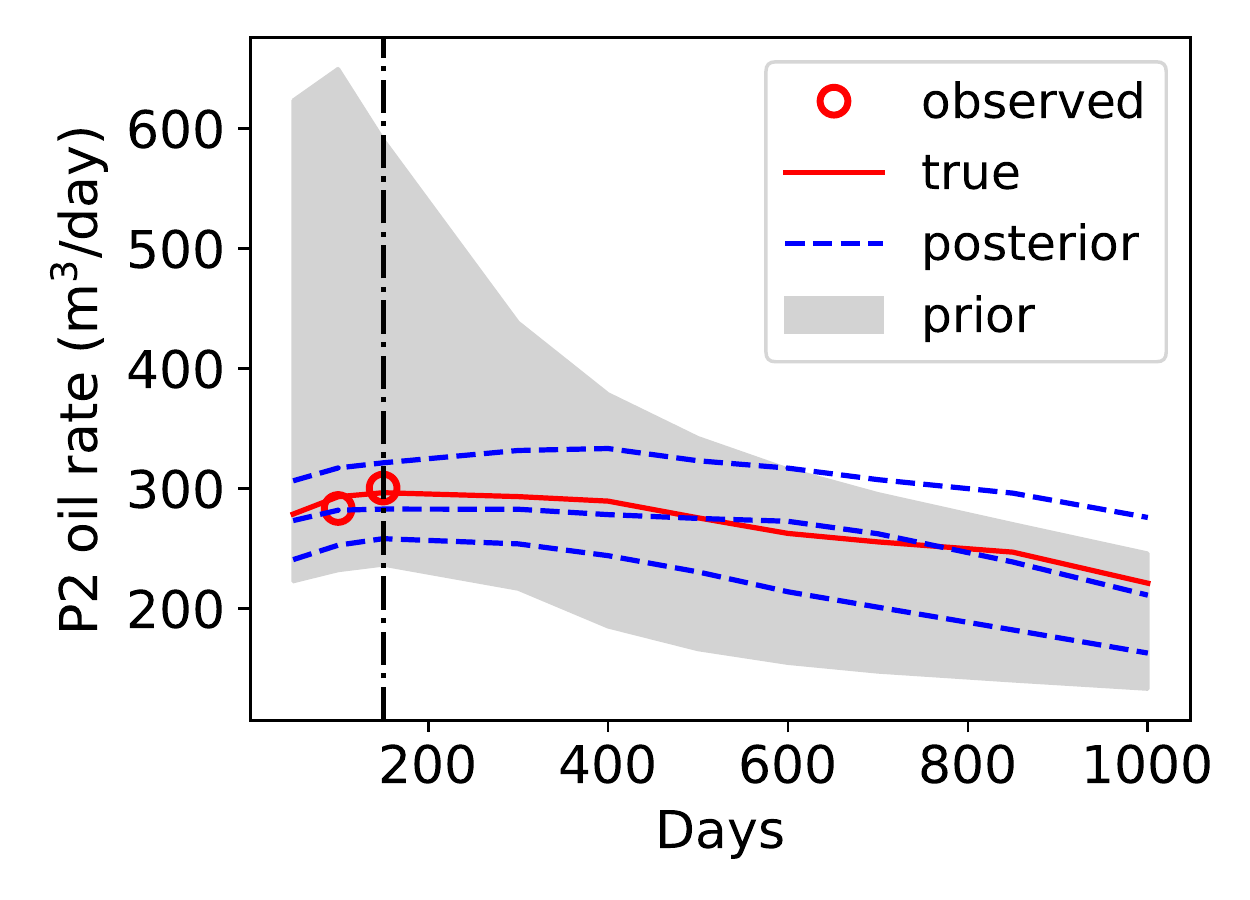}
         \caption{P2 oil rate}
         \label{hm-orate-w14}
     \end{subfigure}
     \begin{subfigure}[b]{0.36\textwidth}
         \centering
         \includegraphics[width=\textwidth]{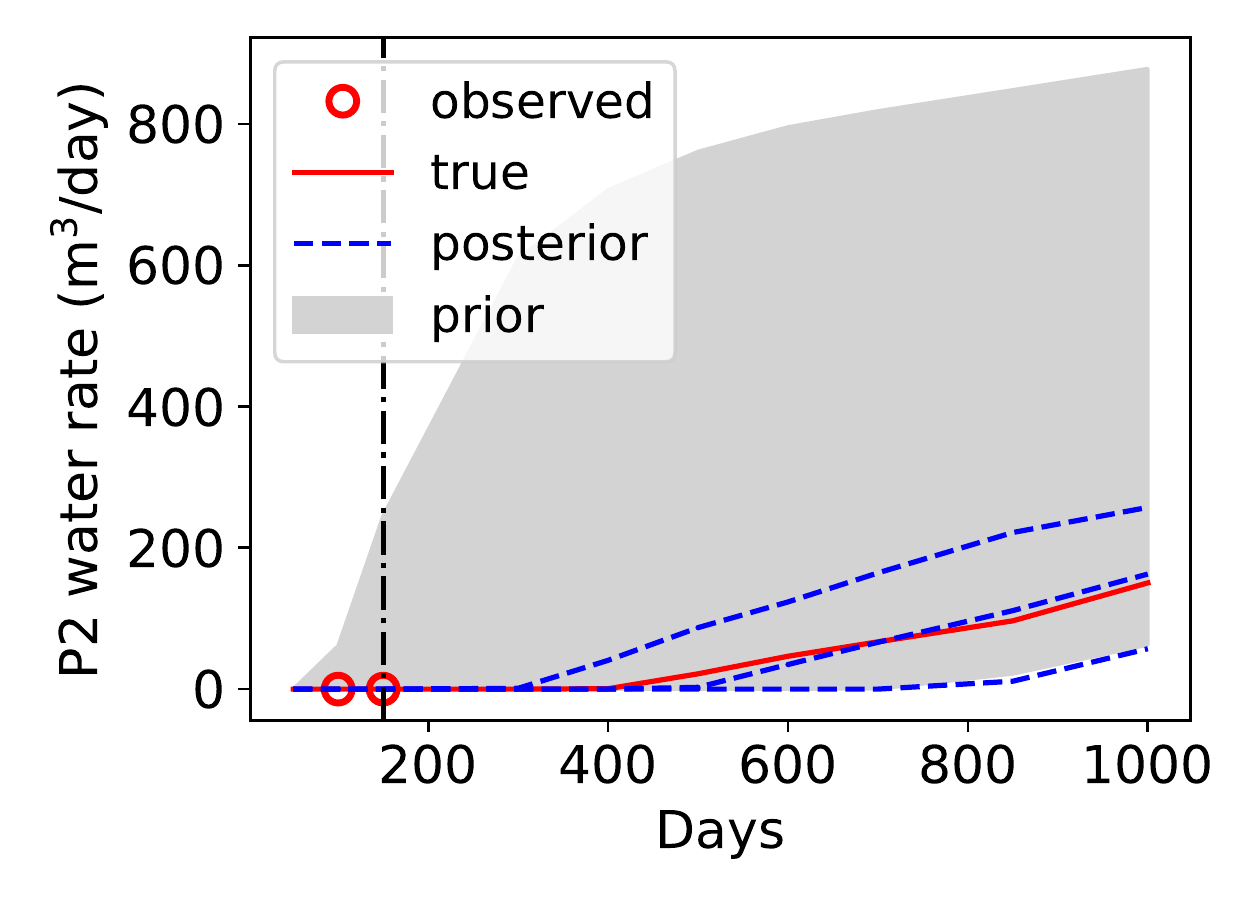}
         \caption{P2 water rate}
         \label{hm-wrate-w14}
     \end{subfigure}
     
    \begin{subfigure}[b]{0.36\textwidth}
         \centering
         \includegraphics[width=\textwidth]{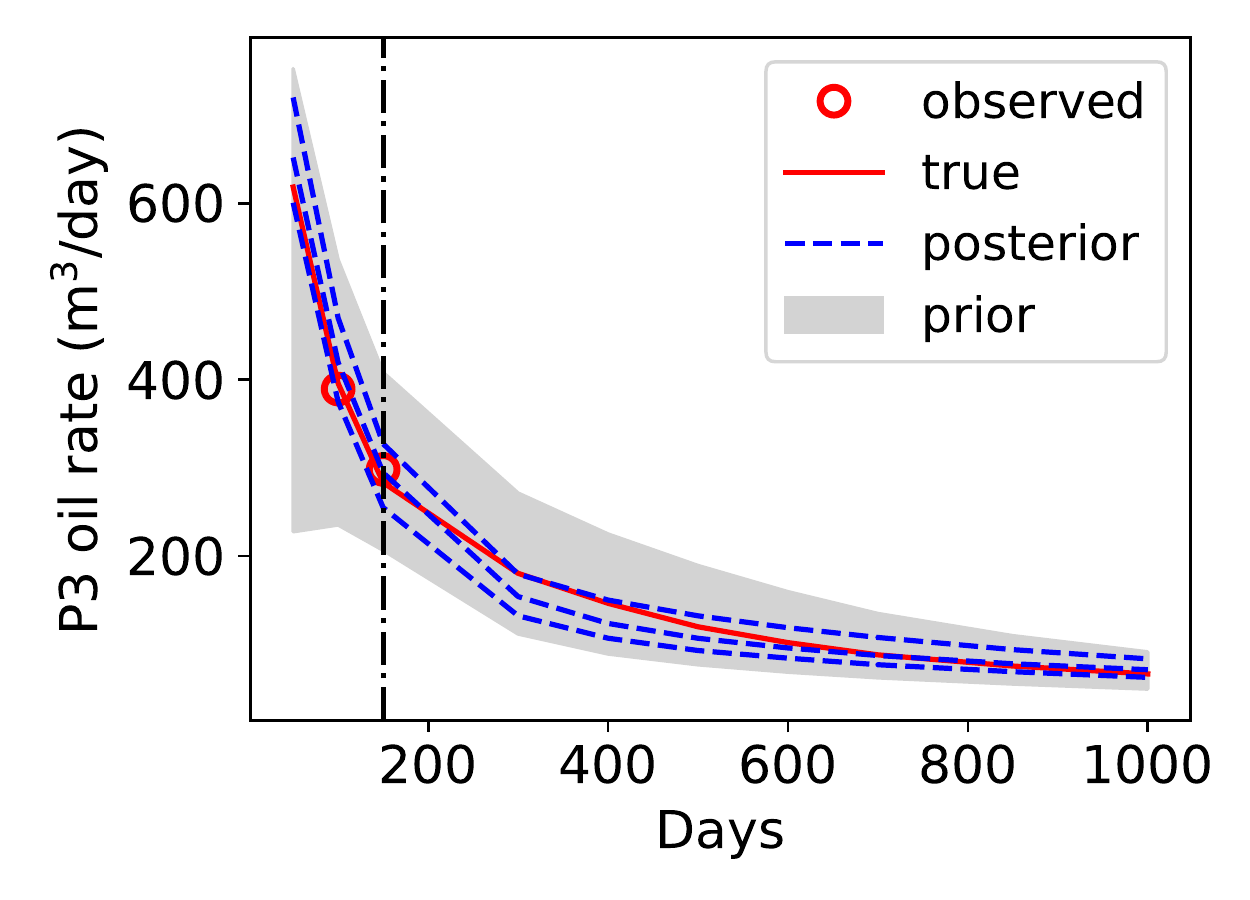}
         \caption{P3 oil rate}
         \label{hm-orate-w17}
     \end{subfigure}
          \begin{subfigure}[b]{0.36\textwidth}
         \centering
         \includegraphics[width=\textwidth]{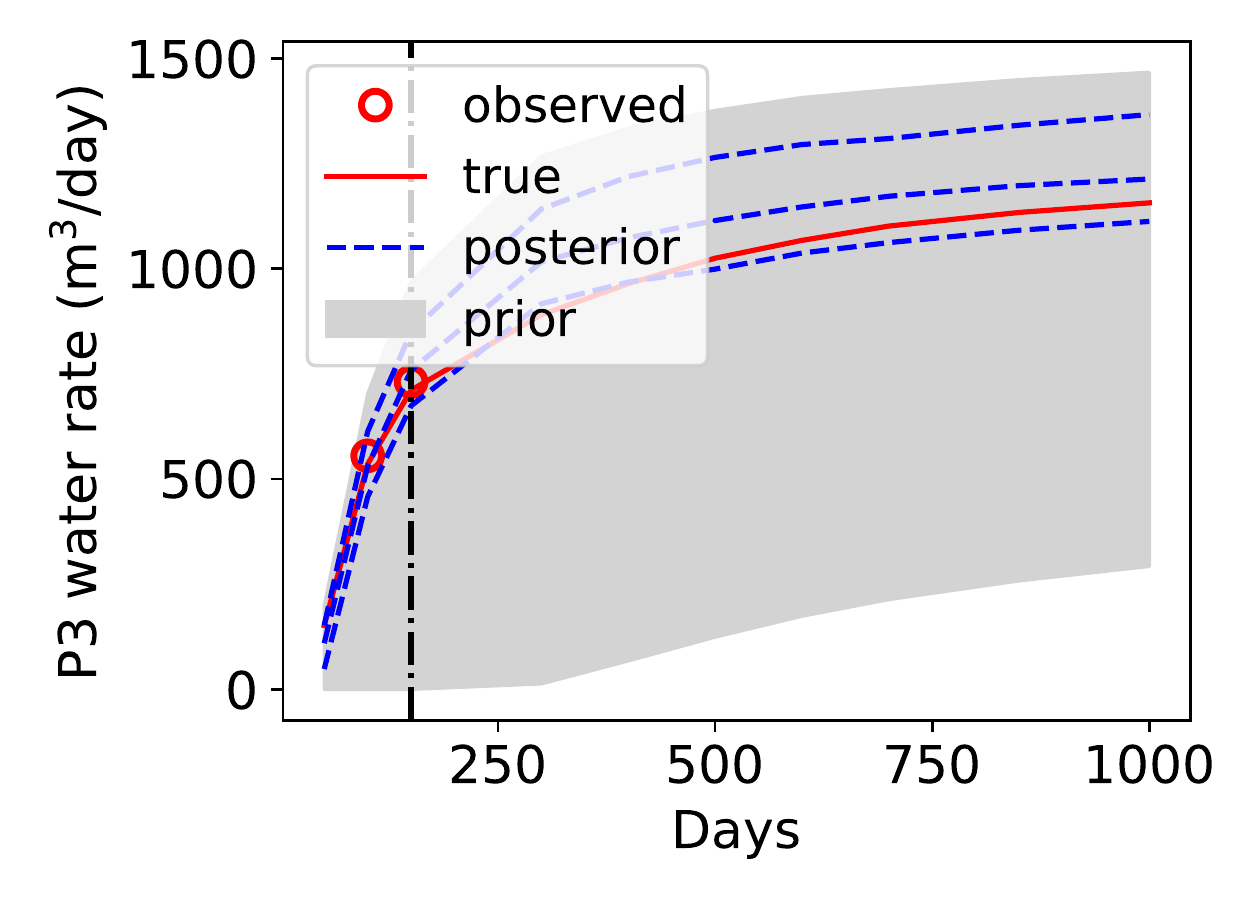}
         \caption{P3 water rate}
         \label{hm-wrate-w17}
     \end{subfigure}

    \begin{subfigure}[b]{0.36\textwidth}
         \centering
         \includegraphics[width=\textwidth]{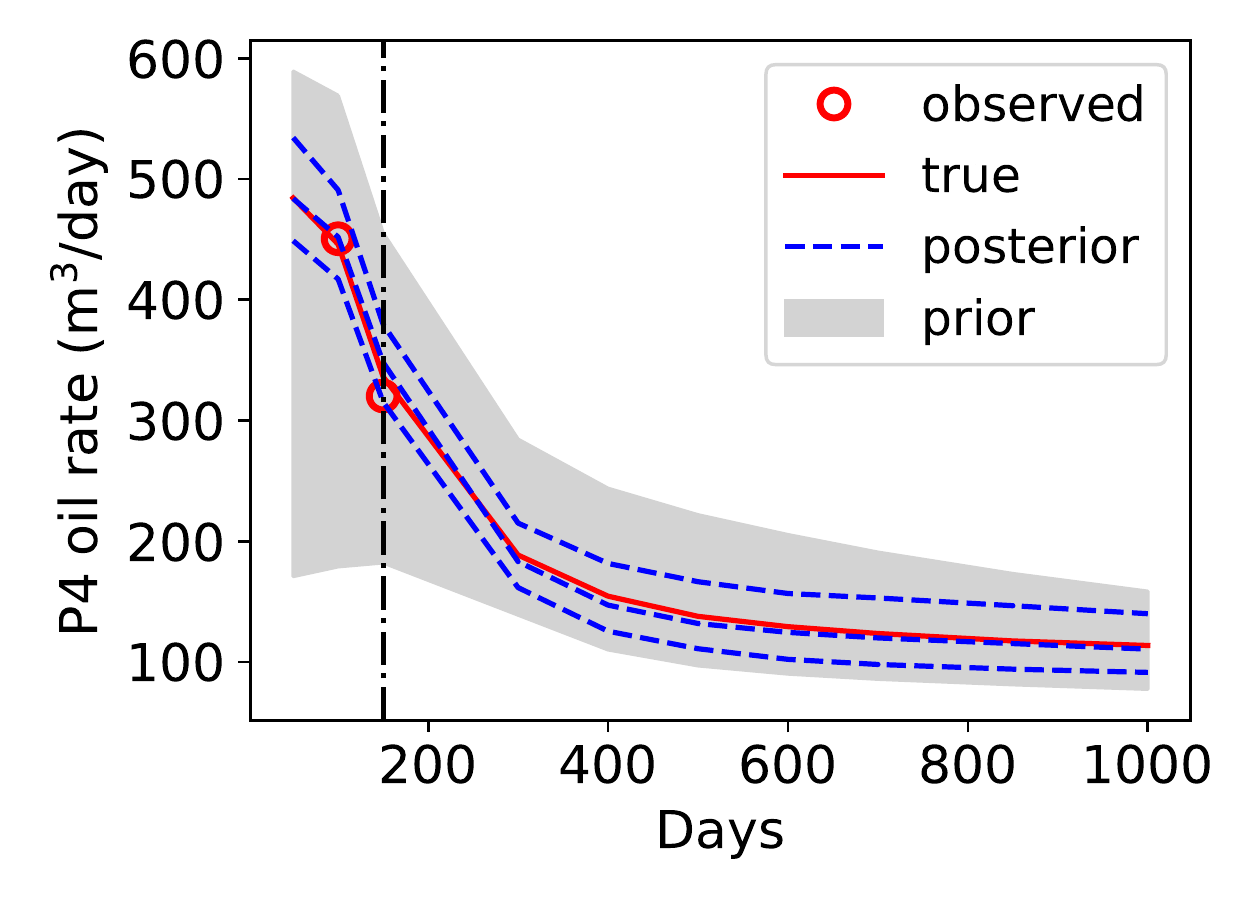}
         \caption{P4 oil rate}
         \label{hm-orate-w18}
     \end{subfigure}
          \begin{subfigure}[b]{0.36\textwidth}
         \centering
         \includegraphics[width=\textwidth]{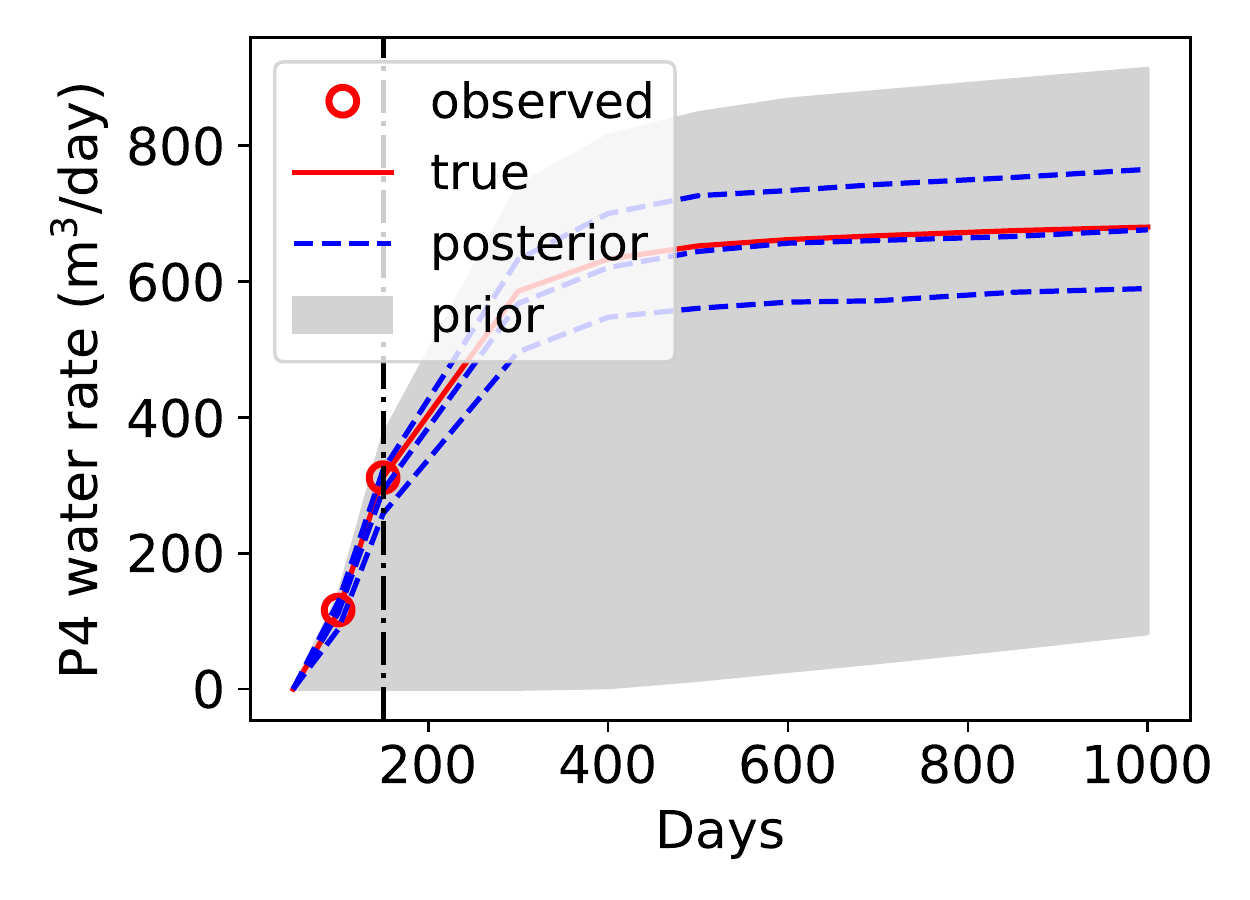}
         \caption{P4 water rate}
         \label{hm-wrate-w18}
     \end{subfigure}
    
    \caption{Oil (left) and water (right) production rates for all four production wells. Gray regions represent the prior $\text{P}_{10}$--$\text{P}_{90}$ range, red circles and curves denote observed and true data, and blue dashed curves denote the RS posterior $\text{P}_{10}$ (lower), $\text{P}_{50}$ (middle) and $\text{P}_{90}$ (upper) predictions. Vertical dashed line indicates the latest time at which data are collected.}
    \label{fig:hm-well-flow}
\end{figure}

\subsection{Rejection Sampling Algorithm}

Rejection sampling can be used to provide a reliable reference for posterior uncertainty. RS is usually very expensive computationally, however, as it requires generating and evaluating an extremely large number of prior samples. Thus the application of RS is clearly facilitated through the use of the 3D recurrent R-U-Net surrogate model and CNN-PCA. RS accepts or rejects prior samples independently, thus assuring that the accepted models are independent. In this work, we combine the RS approach described in \citep{sun2017new} with our deep-learning-based treatments. The overall workflow is as follows:
\begin{itemize}
    \item Sample the low-dimensional variable $\boldsymbol\xi \in \mathbb{R}^l$ from its prior distribution, given by $\mathcal{N}(\mathbf{\mu}_{\boldsymbol\xi}, C_{\boldsymbol\xi})$. Construct $\mathbf{m}_{\text{cnnpca}}(\boldsymbol\xi)$ using CNN-PCA. 
    \item Sample a probability $p$ from a uniform distribution in $[0, 1]$.
    \item Compute the likelihood function $L(\mathbf{m}_{\text{cnnpca}}(\boldsymbol\xi))$, given by 
    \begin{equation}
        L(\mathbf{m}_{\text{cnnpca}}(\boldsymbol\xi)) = c \exp \left(-\frac{1}{2}[\hat{f}(\mathbf{m}_{\text{cnnpca}}(\boldsymbol\xi)) - \mathbf{d}_{\text{obs}}]^{\intercal}C_D^{-1}[\hat{f}(\mathbf{m}_{\text{cnnpca}}(\boldsymbol\xi)) - \mathbf{d}_{\text{obs}}]\right),
    \end{equation}
    where $c$ is a normalization constant, $\hat{f}(\mathbf{m}_{\text{cnnpca}}(\boldsymbol\xi))$ indicates the surrogate model predictions for well rates for geomodel $\mathbf{m}_{\text{cnnpca}}(\boldsymbol\xi)$,
    $\mathbf{d}_{\text{obs}}$ denotes the observed well rate data, and $C_D$ is the covariance matrix of data measurement error. 
    \item Accept $\mathbf{m}_{\text{cnnpca}}(\boldsymbol\xi)$ if $p \leq \frac{L(\mathbf{m}_{\text{cnnpca}}(\boldsymbol\xi))}{S_L}$, where $S_L$ is the maximum likelihood value over all prior models considered. 
\end{itemize}

\subsection{Problem Setup and Rejection Sampling Results}
\label{sec:RS_results}

The `true' model considered here, which is shown in Fig.~\ref{fig:cnn-pca-reals}(b), is a randomly selected 3D CNN-PCA realization. Recall that all realizations (including the true model) are conditioned to hard data at well locations. The true data $\mathbf{d}_{\text{true}}$ are obtained by performing high-fidelity simulation using the true geomodel. The observed data $\mathbf{d}_{\text{obs}}$ comprise the true data with random error added
\begin{equation}
\mathbf{d}_{\text{obs}} = \mathbf{d}_{\text{true}} + \boldsymbol\epsilon,
\end{equation}
where $\boldsymbol\epsilon$ is the measurement error vector, with all components taken to be Gaussian random variables with mean of zero and covariance consistent with the $C_D$ matrix defined above. 

\begin{figure}[htbp]
     \centering
     \begin{subfigure}[b]{0.32\textwidth}
         \centering
         \includegraphics[trim={7cm 5cm 5cm 3cm},clip, scale=0.3]{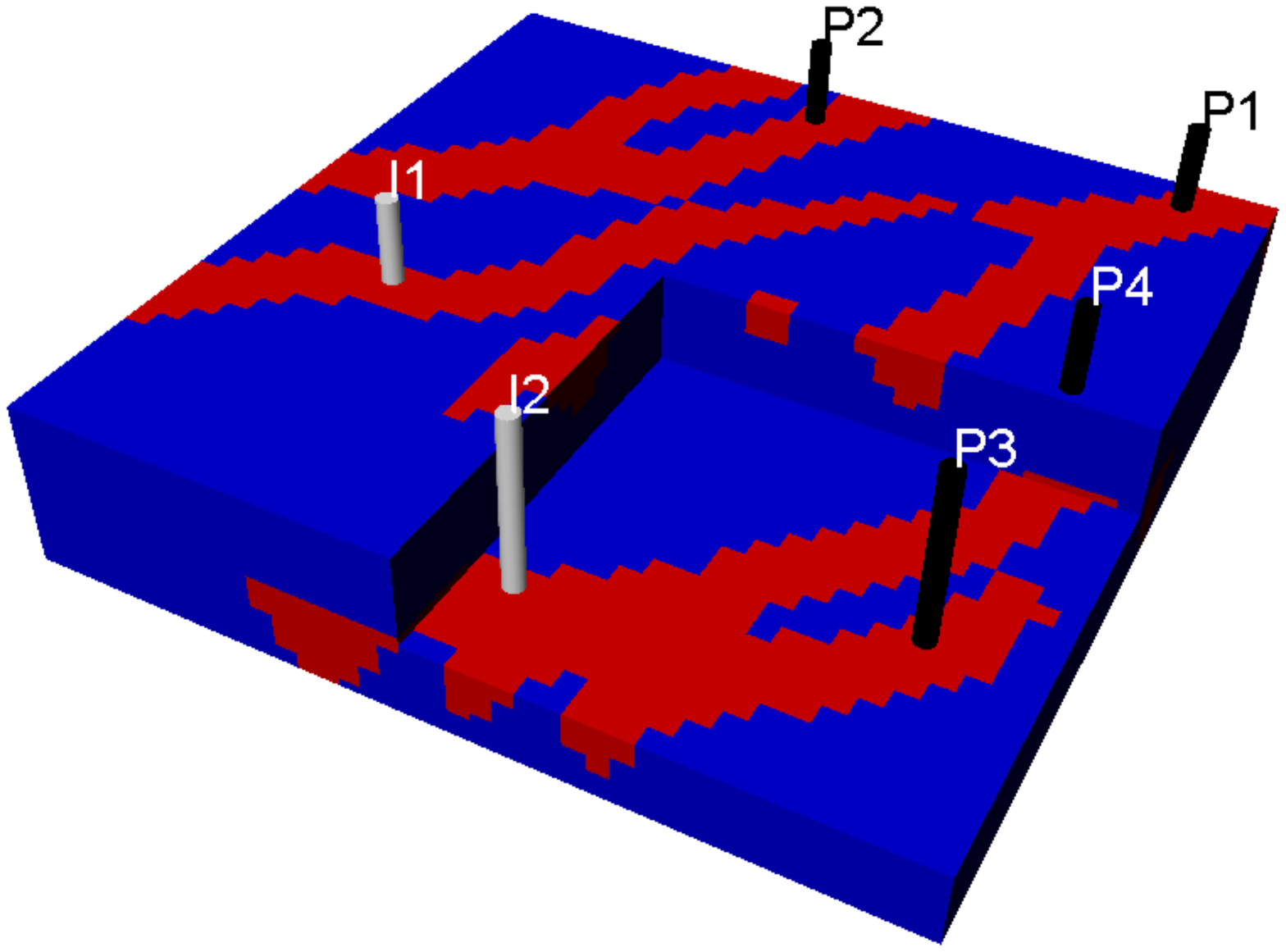}
         \caption{}
         \label{real-34}
     \end{subfigure}
     \begin{subfigure}[b]{0.32\textwidth}
         \centering
         \includegraphics[trim={7cm 5cm 5cm 3cm},clip, scale=0.3]{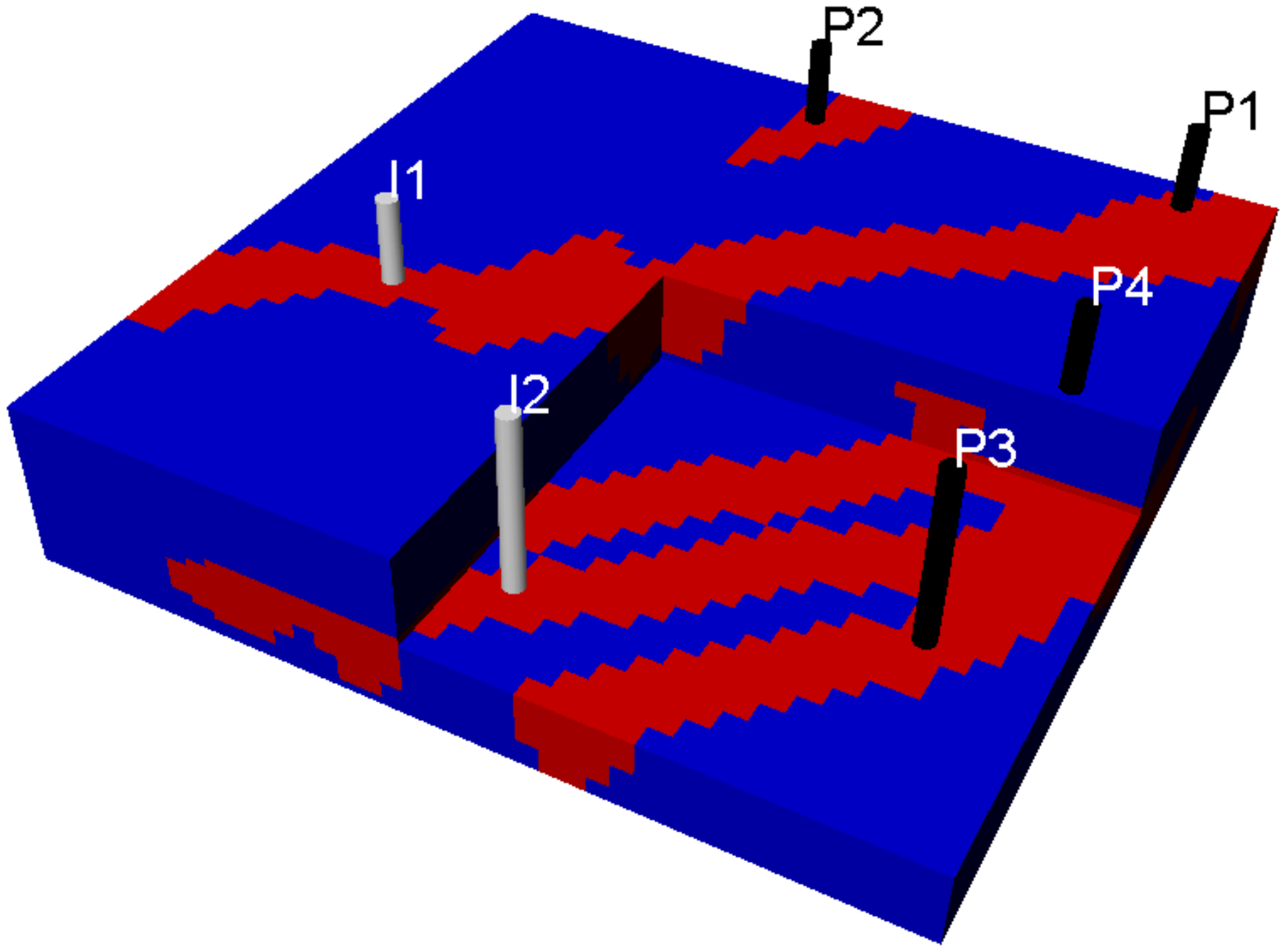}
         \caption{}
         \label{real-44}
     \end{subfigure}
     \begin{subfigure}[b]{0.32\textwidth}
         \centering
         \includegraphics[trim={7cm 5cm 5cm 3cm},clip, scale=0.3]{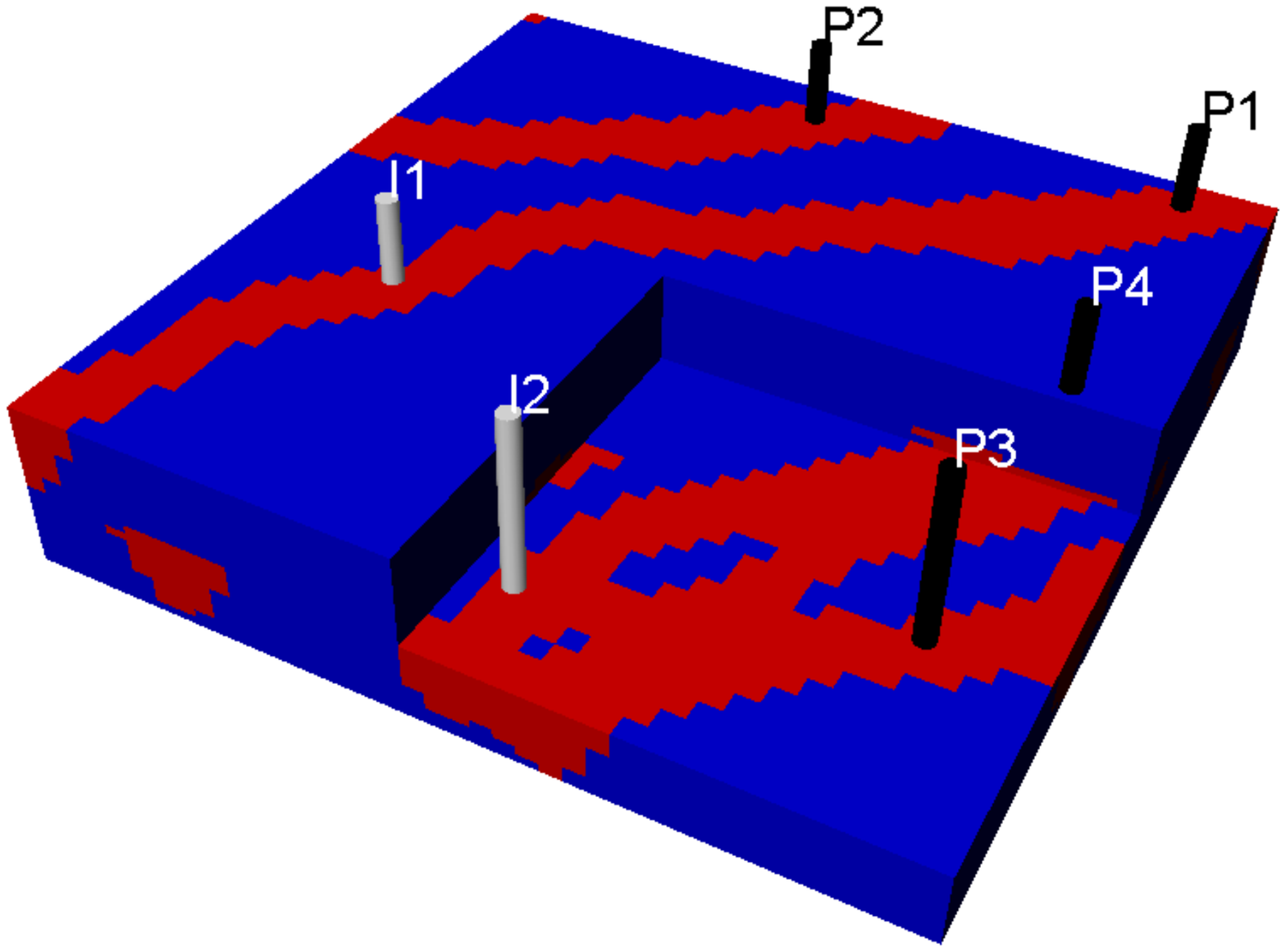}
         \caption{}
         \label{real-54}
     \end{subfigure}

    \caption{Three (randomly selected) 3D CNN-PCA realizations accepted by the rejection sampling procedure.}
    \label{fig:rs-posterior}
\end{figure}

\begin{figure}[htbp]
     \centering
     \begin{subfigure}[b]{0.36\textwidth}
         \centering
         \includegraphics[width=\textwidth]{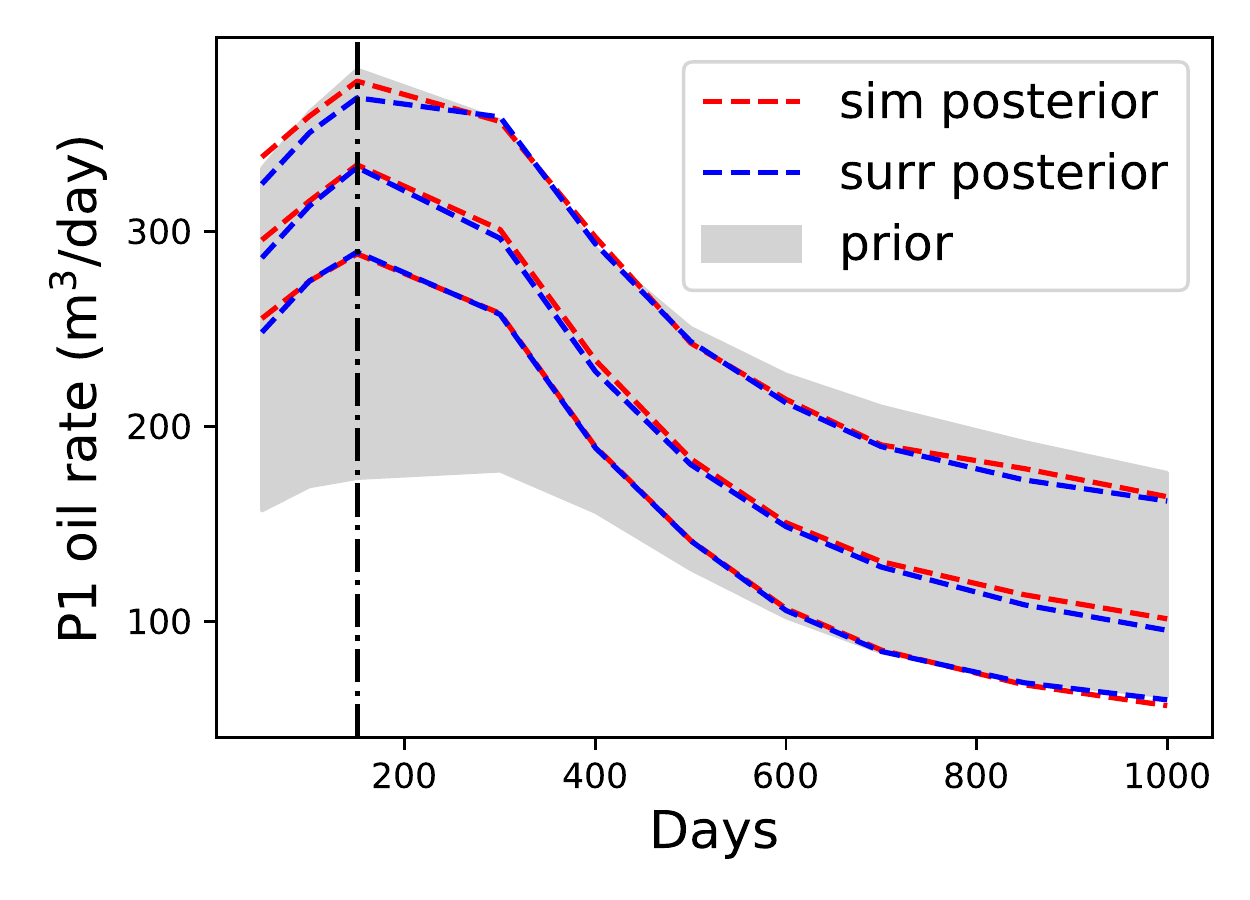}
         \caption{P1 oil rate}
         \label{hm-orate-w11}
     \end{subfigure}
     \begin{subfigure}[b]{0.36\textwidth}
         \centering
         \includegraphics[width=\textwidth]{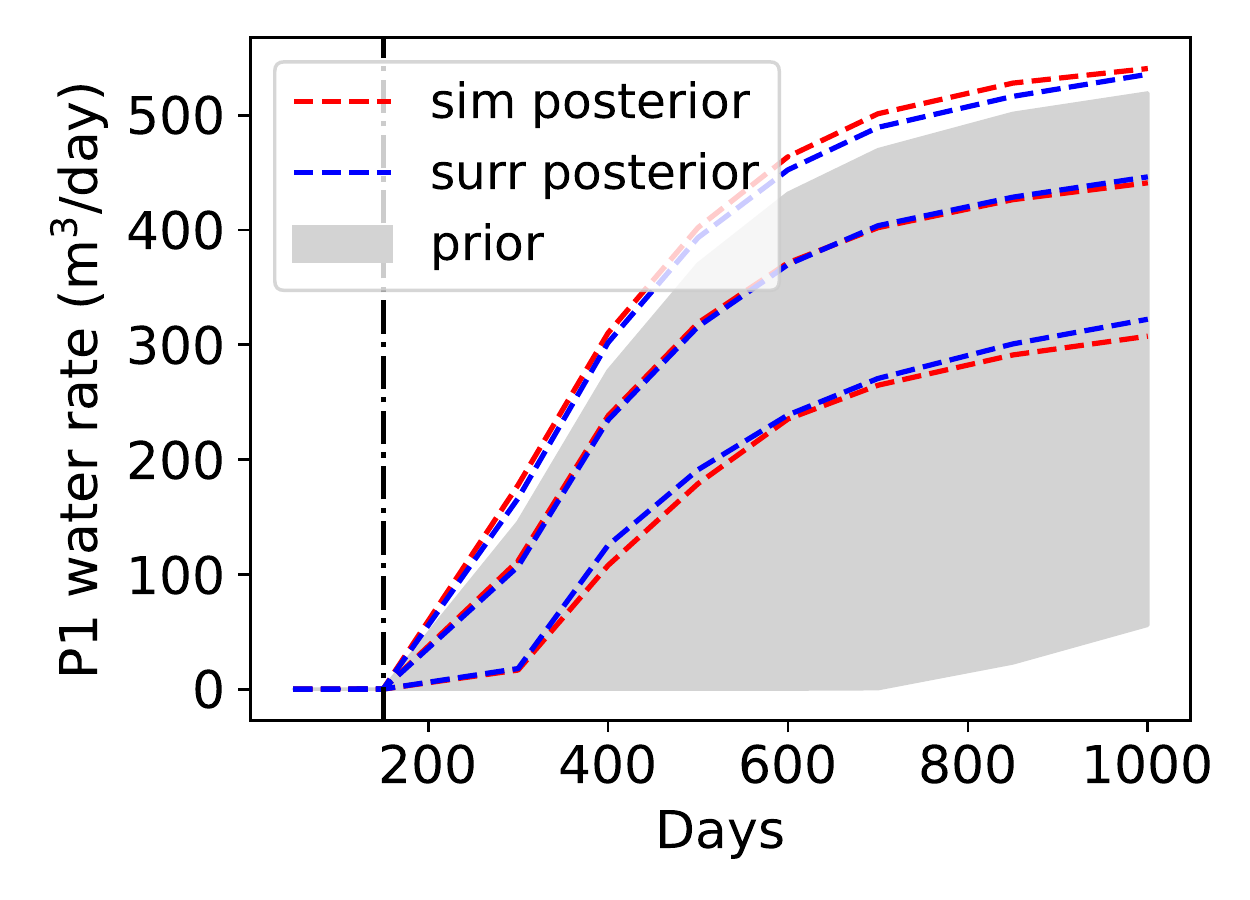}
         \caption{P1 water rate}
         \label{hm-wrate-w11}
     \end{subfigure}
     
     \begin{subfigure}[b]{0.36\textwidth}
         \centering
         \includegraphics[width=\textwidth]{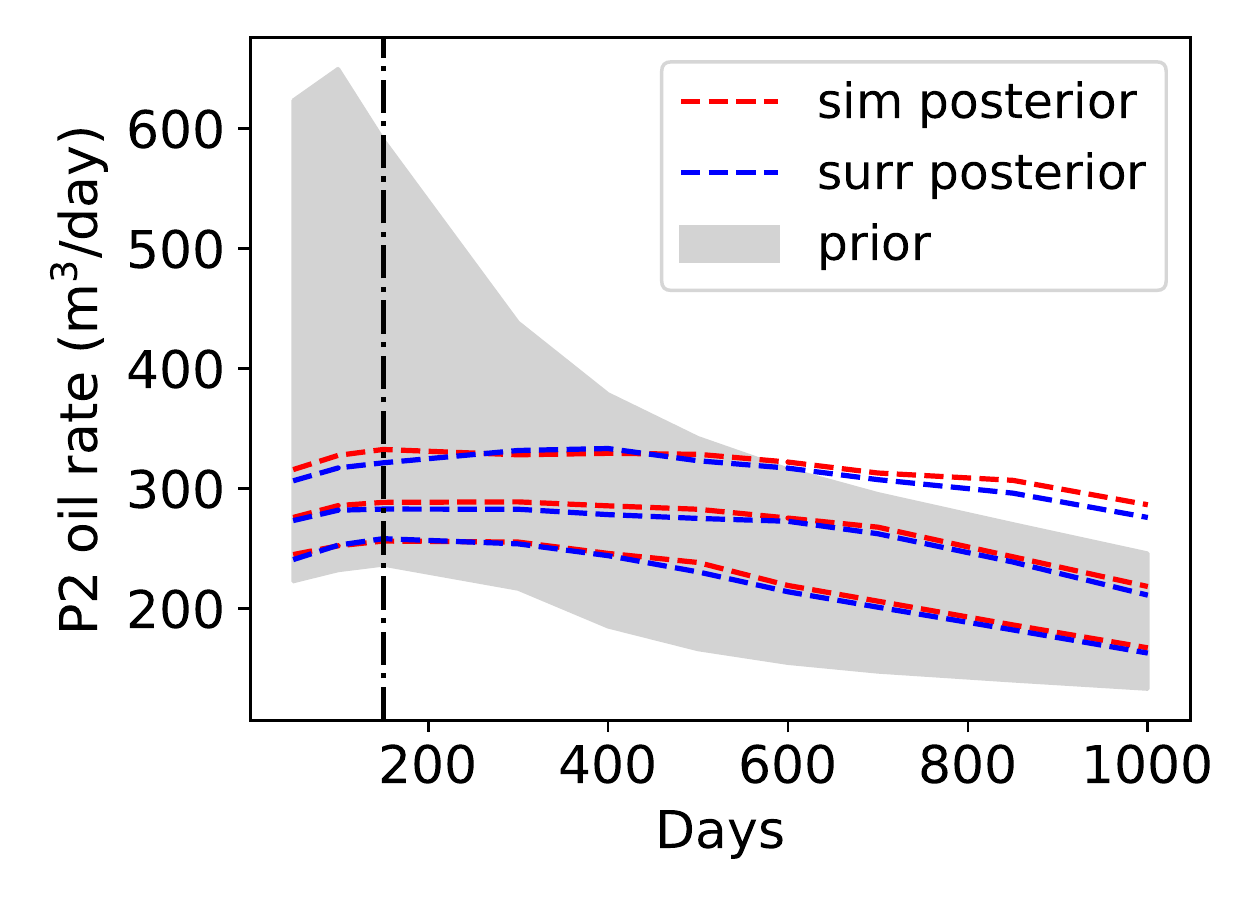}
         \caption{P2 oil rate}
         \label{hm-orate-w141}
     \end{subfigure}
     \begin{subfigure}[b]{0.36\textwidth}
         \centering
         \includegraphics[width=\textwidth]{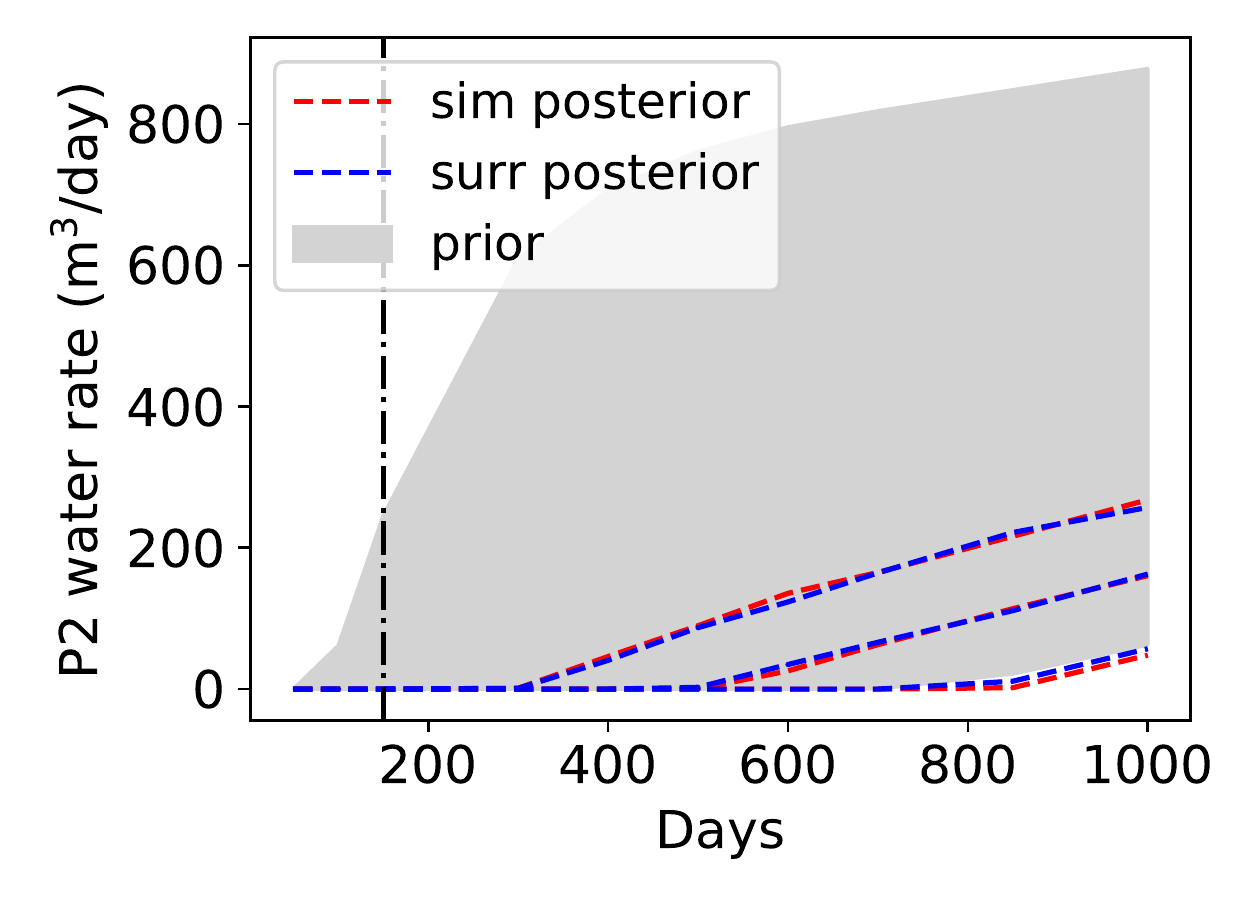}
         \caption{P2 water rate}
         \label{hm-wrate-w141}
     \end{subfigure}
     
    \begin{subfigure}[b]{0.36\textwidth}
         \centering
         \includegraphics[width=\textwidth]{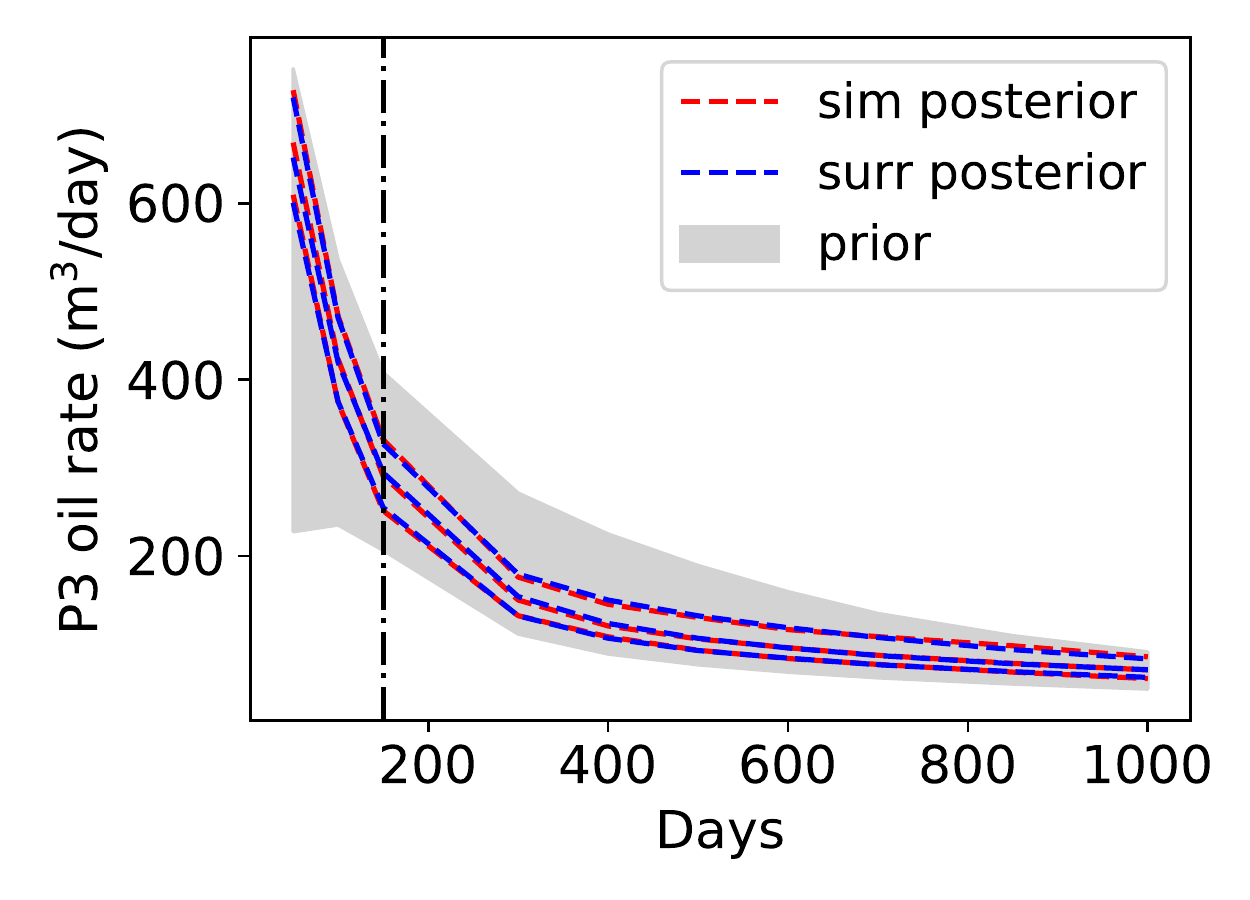}
         \caption{P3 oil rate}
         \label{hm-orate-w171}
     \end{subfigure}
          \begin{subfigure}[b]{0.36\textwidth}
         \centering
         \includegraphics[width=\textwidth]{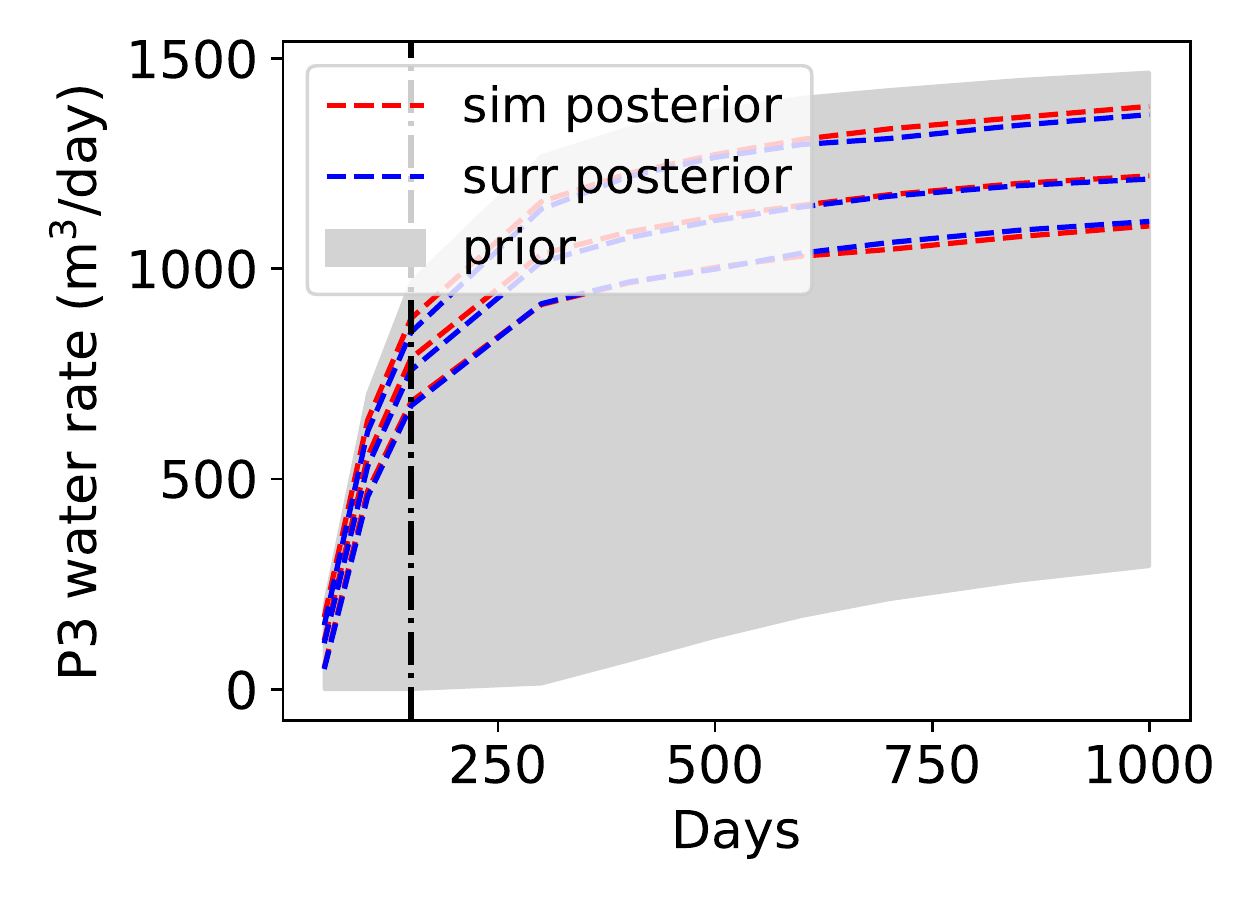}
         \caption{P3 water rate}
         \label{hm-wrate-w171}
     \end{subfigure}

    \begin{subfigure}[b]{0.36\textwidth}
         \centering
         \includegraphics[width=\textwidth]{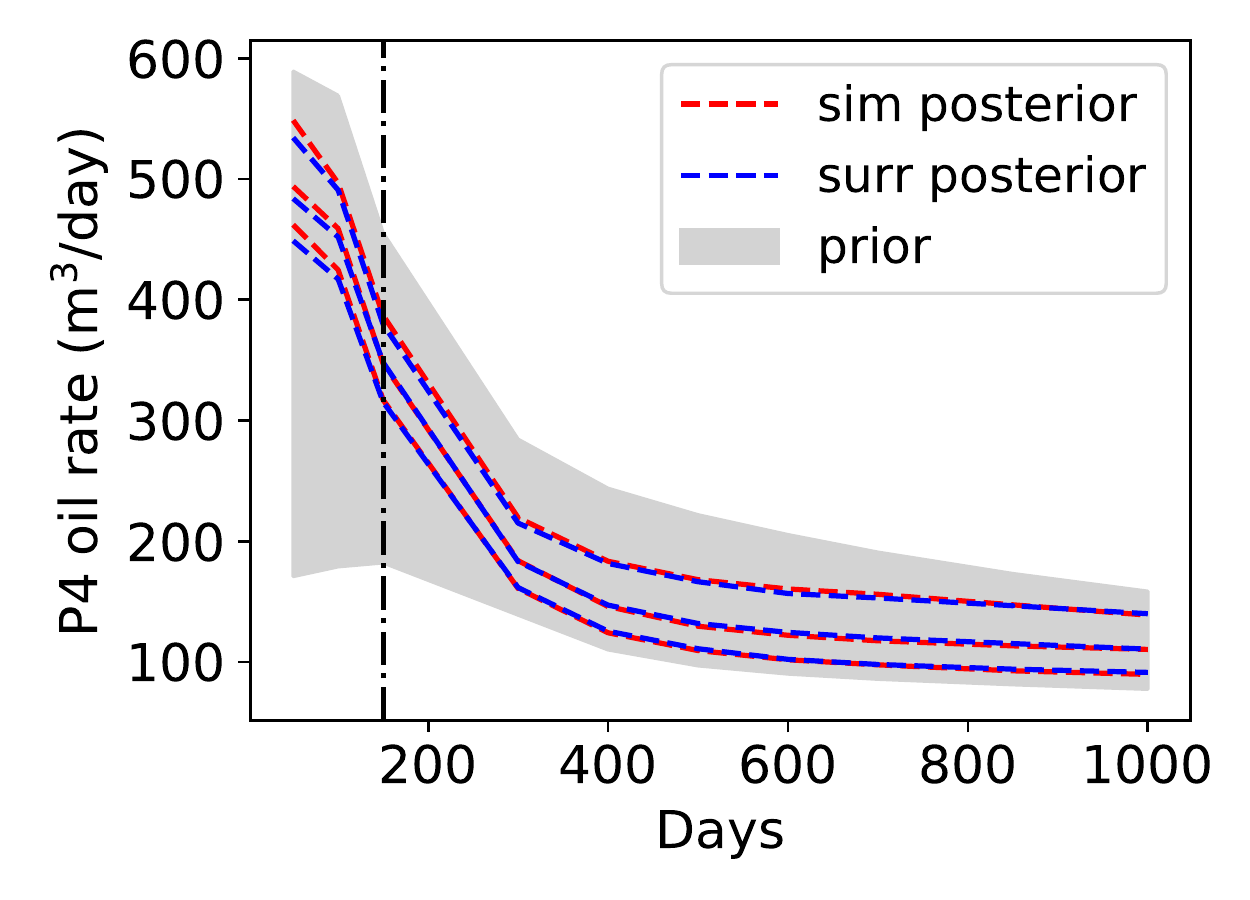}
         \caption{P4 oil rate}
         \label{hm-orate-w181}
     \end{subfigure}
          \begin{subfigure}[b]{0.36\textwidth}
         \centering
         \includegraphics[width=\textwidth]{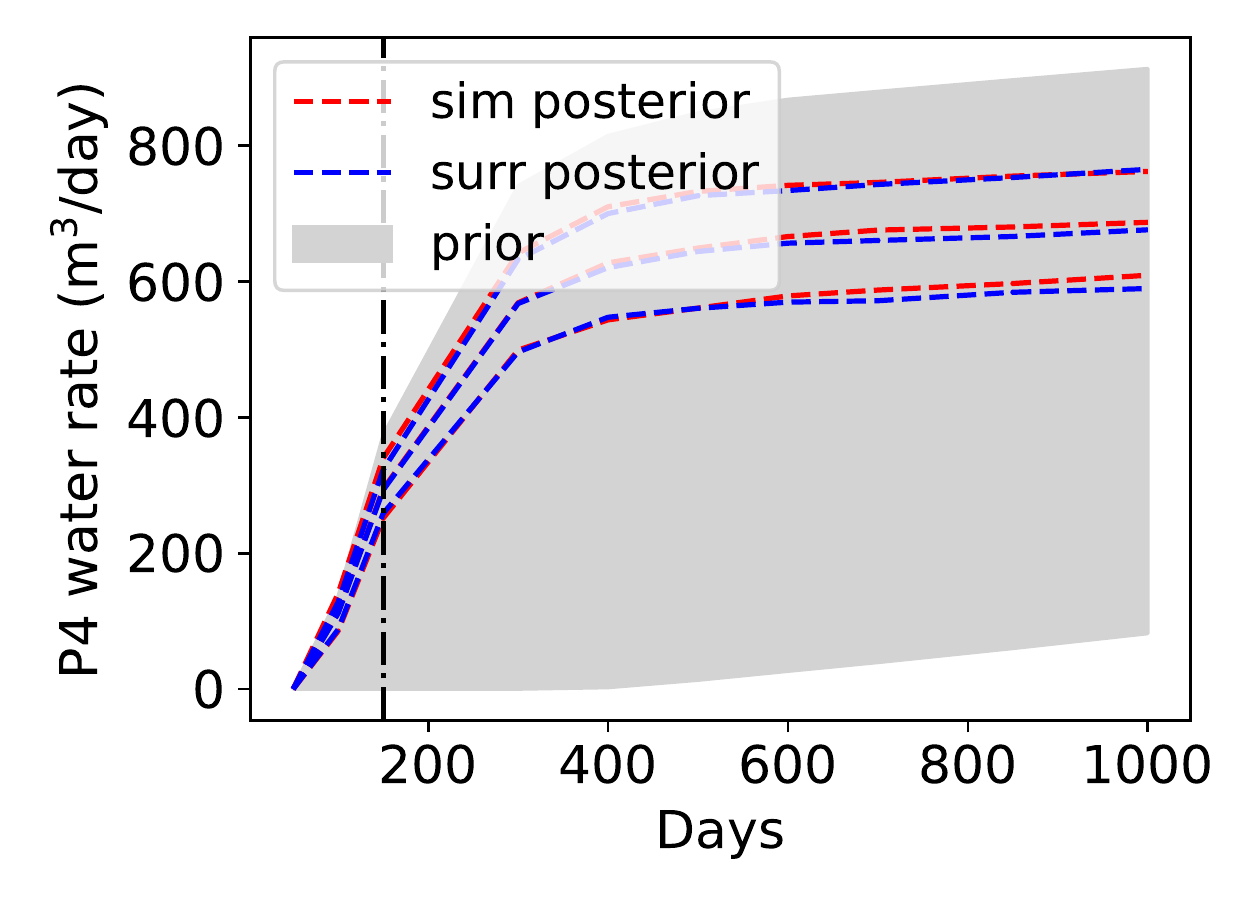}
         \caption{P4 water rate}
         \label{hm-wrate-w181}
     \end{subfigure}
    
    \caption{Oil (left) and water (right) production rates for all four production wells. Gray regions represent the prior $\text{P}_{10}$--$\text{P}_{90}$ range, blue dashed curves denote surrogate model posterior $\text{P}_{10}$ (lower), $\text{P}_{50}$ (middle) and $\text{P}_{90}$ (upper) predictions, red dashed curves denote high-fidelity simulation posterior $\text{P}_{10}$ (lower), $\text{P}_{50}$ (middle) and $\text{P}_{90}$ (upper) predictions. Vertical dashed line indicates the latest time at which data are collected. Surrogate model posterior samples determined using RS.}
    \label{fig:hm-well-flow-comp}
\end{figure}

With RS, very few models will be accepted if we attempt to match a large number of data points, particularly if data error is small. Therefore, we use data at only two time steps, specifically at day~100 and day~150. Oil and water production rates are collected at the four producers, which means we have a total of 16 observations. The standard deviation of the measurement error is set to 15\% of the corresponding true data. One million prior models are generated using CNN-PCA. The 3D recurrent R-U-Net is then applied to provide well response predictions for all of these models. The generation of $10^6$ 3D CNN-PCA models requires about 3~hours, and the $10^6$ surrogate model predictions require about 8~hours. We estimate that high-fidelity simulation for these $10^6$ models would require about 170,000~hours of serial computation (as each HFS requires around 10~minutes). A total of 151 (out of $10^6$) models are accepted by the RS procedure.

The history matching results using this procedure are shown in Fig.~\ref{fig:hm-well-flow}. The gray areas display the $\text{P}_{10}$-$\text{P}_{90}$ interval for the prior models (i.e., 80\% of the prior model results fall within the gray regions). The red points and curves denote the observed data and the true model flow response. The blue dashed curves indicate the $\text{P}_{10}$ (lower), $\text{P}_{50}$ (middle) and $\text{P}_{90}$ (upper) posterior RS results.

The posterior $\text{P}_{10}$--$\text{P}_{90}$ ranges are in most cases significantly smaller than the prior $\text{P}_{10}$--$\text{P}_{90}$ ranges. For this reason, predictions based on the posterior models would be expected to be much more useful than prior forecasts. It is evident that the posterior uncertainty ranges cover the true well responses even when these results are on the edge of the prior $\text{P}_{10}$-$\text{P}_{90}$ interval, e.g., the well~P2 oil rate at late time (Fig.~\ref{fig:hm-well-flow}(c)). We also see that the posterior predictions result in significant uncertainty reduction in well~P1 and P2 water rates (Fig.~\ref{fig:hm-well-flow}(b) and (d)), even though water production at these wells has not yet occurred at day~150.

In Fig.~\ref{fig:rs-posterior} we present three of the realizations accepted by the RS procedure. These realizations all show a lack of connectivity (through sand) between injector I1 and producer P2, at least in the layers that are visible in these images. This is consistent with the true 3D CNN-PCA model (shown in Fig.~\ref{fig:cnn-pca-reals}(b)) and with the low water production rates in Fig.~\ref{fig:hm-well-flow}(d). We also see that injector I2 is connected to producer P3 through sand in all three posterior realizations. This is again consistent with the true model, and with the early breakthrough and high water production rates in P3, evident in Fig.~\ref{fig:hm-well-flow}(f).

In the results above, realizations are accepted or rejected based on the surrogate flow model predictions. It is useful to evaluate the accepted (posterior) models by applying HFS to assure that these models do indeed provide numerical simulation results in agreement with observed data.
We therefore simulate the 151 models accepted by RS using ADGPRS. 
Results for oil and water production rates for the four producers are shown in Fig.~\ref{fig:hm-well-flow-comp}. There we see very close agreement between the recurrent R-U-Net predictions (blue dashed curves) and the high-fidelity simulation predictions (red dashed curves). 
This close correspondence between 3D recurrent R-U-Net and HFS results for posterior models clearly demonstrates the applicability of the 3D recurrent R-U-Net for this challenging history matching problem.

\subsection{ES-MDA Setup and Results}

Ensemble smoother with multiple data assimilation (ES-MDA) \citep{emerick2013ensemble} is an iterative form of the ensemble smoother \citep{evensen2000ensemble} procedure. ES-MDA breaks the single global update applied in ES into multiple data assimilation steps. 
In ES-MDA, observed data are perturbed, consistent with an inflated data covariance matrix, at each data assimilation step. ES-MDA is an efficient and widely used history matching algorithm, though there are no guarantees that it provides correct posterior quantification in complex nonlinear problems. In addition, the method may experience ensemble collapse \citep{morzfeld2017collapse} in some cases. A key usage for the RS results presented in Section~\ref{sec:RS_results} is to assess the performance of much more efficient algorithms such as ES-MDA. We now proceed with this evaluation.


The overall workflow for ES-MDA, with geomodels parameterized using 3D CNN-PCA and flow evaluations performed using the 3D recurrent R-U-Net, is as follows:
\begin{enumerate}
    \item Specify $N_a$, the number of data assimilation steps, and corresponding inflation coefficients $\alpha_i$, $(i = 1, \cdots, N_a)$, where $\sum_{i = 1}^{N_a} \alpha_i^{-1} = 1$. Specify $N_e$, the number of ensemble members.
\end{enumerate}

For each ensemble member:

\begin{enumerate}[resume]

    \item Sample the low-dimensional variable $\boldsymbol\xi \in \mathbb{R}^l$ from its prior distribution $\mathcal{N}(\mathbf{\mu}_{\boldsymbol\xi}, C_{\boldsymbol\xi})$.
    \item For $i=1,\cdots, N_a$:
    \begin{itemize}
    \item Construct $\mathbf{m}_{\text{cnnpca}}(\boldsymbol\xi)$ using 3D CNN-PCA. Then generate surrogate flow predictions $\hat{f}(\mathbf{m}_{\text{cnnpca}}(\boldsymbol\xi))$. 
    \item Perturb the observation data through application of $\mathbf{d}_{\text{uc}} = \mathbf{d}_{\text{obs}} + \sqrt{\alpha_i}C_D^{1/2}\mathbf{z}_d $ and $\mathbf{z}_d \sim \mathcal{N}(0, I)$.
    \item Update low-dimension variable ${\boldsymbol\xi}^i$ to ${\boldsymbol\xi}^{i+1}$ using
    \begin{equation}
       {\boldsymbol\xi}^{i+1} = {\boldsymbol\xi}^{i} + C_{MD}(C_{DD} + \alpha_i C_D)^{-1} [\mathbf{d}_{\text{uc}} - \hat{f}(\mathbf{m}_{\text{cnnpca}}( \boldsymbol\xi^i))],
    \end{equation}
where $C_{MD}$ is the cross-covariance matrix between ${\boldsymbol\xi}^{i}$ and predicted data $\hat{f}(\mathbf{m}_{\text{cnnpca}}({\boldsymbol\xi}^{i}))$, $C_{DD}$ is the auto-covariance matrix of the predicted data, and $C_D$ is the covariance matrix of data measurement error. 
    \end{itemize}
\end{enumerate}

\begin{figure}[htbp]
     \centering
     \begin{subfigure}[b]{0.36\textwidth}
         \centering
         \includegraphics[width=\textwidth]{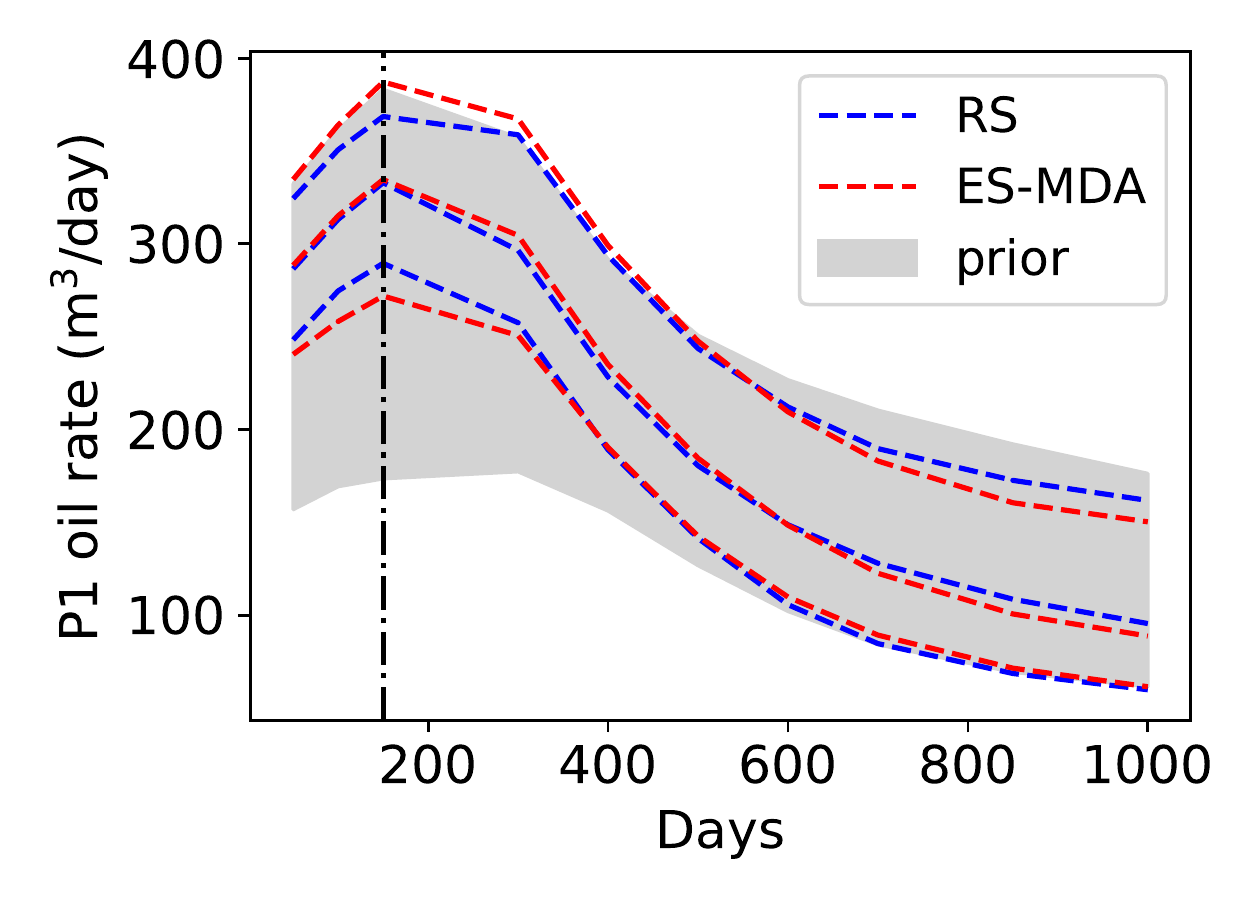}
         \caption{P1 oil rate}
         \label{hm-orate-w12}
     \end{subfigure}
     \begin{subfigure}[b]{0.36\textwidth}
         \centering
         \includegraphics[width=\textwidth]{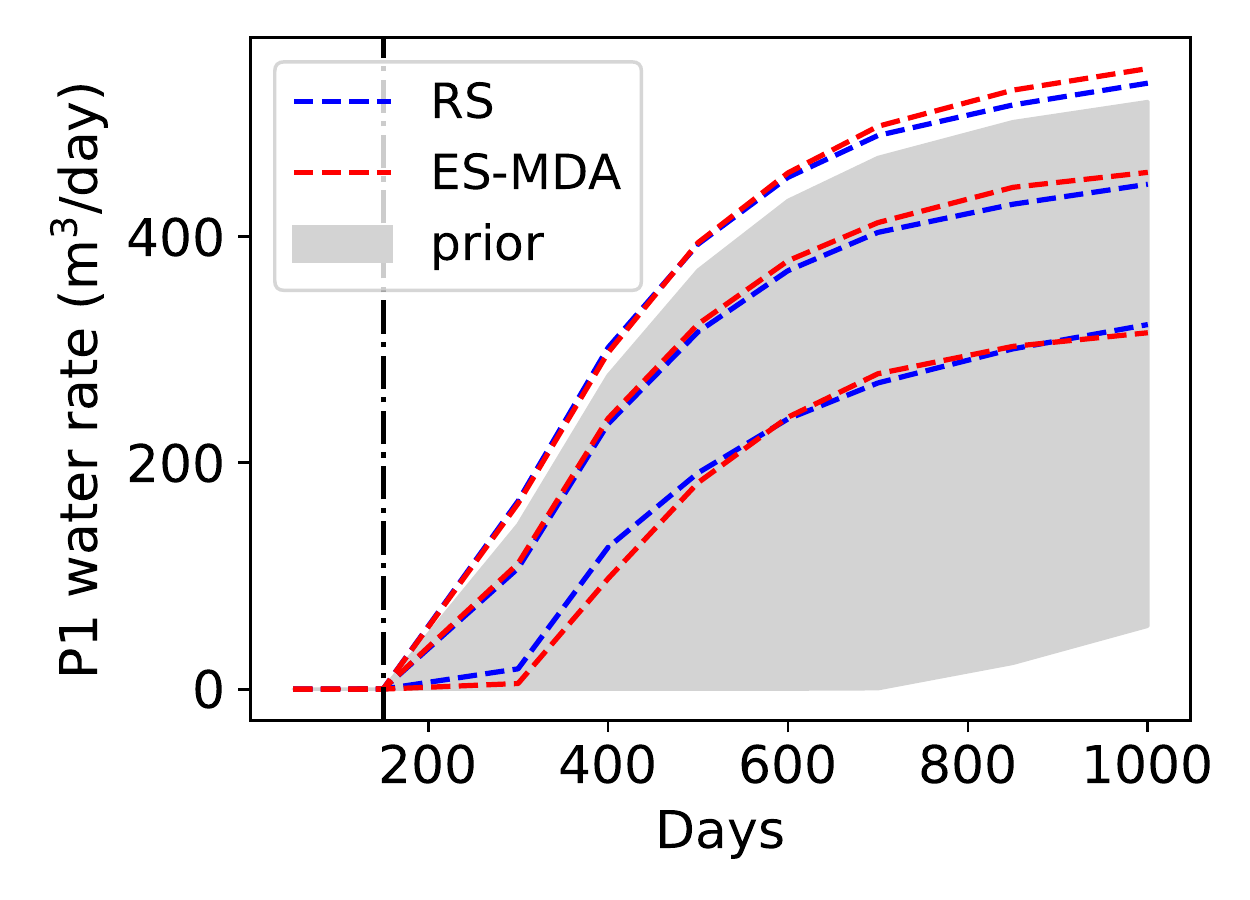}
         \caption{P1 water rate}
         \label{hm-wrate-w12}
     \end{subfigure}
     
     \begin{subfigure}[b]{0.36\textwidth}
         \centering
         \includegraphics[width=\textwidth]{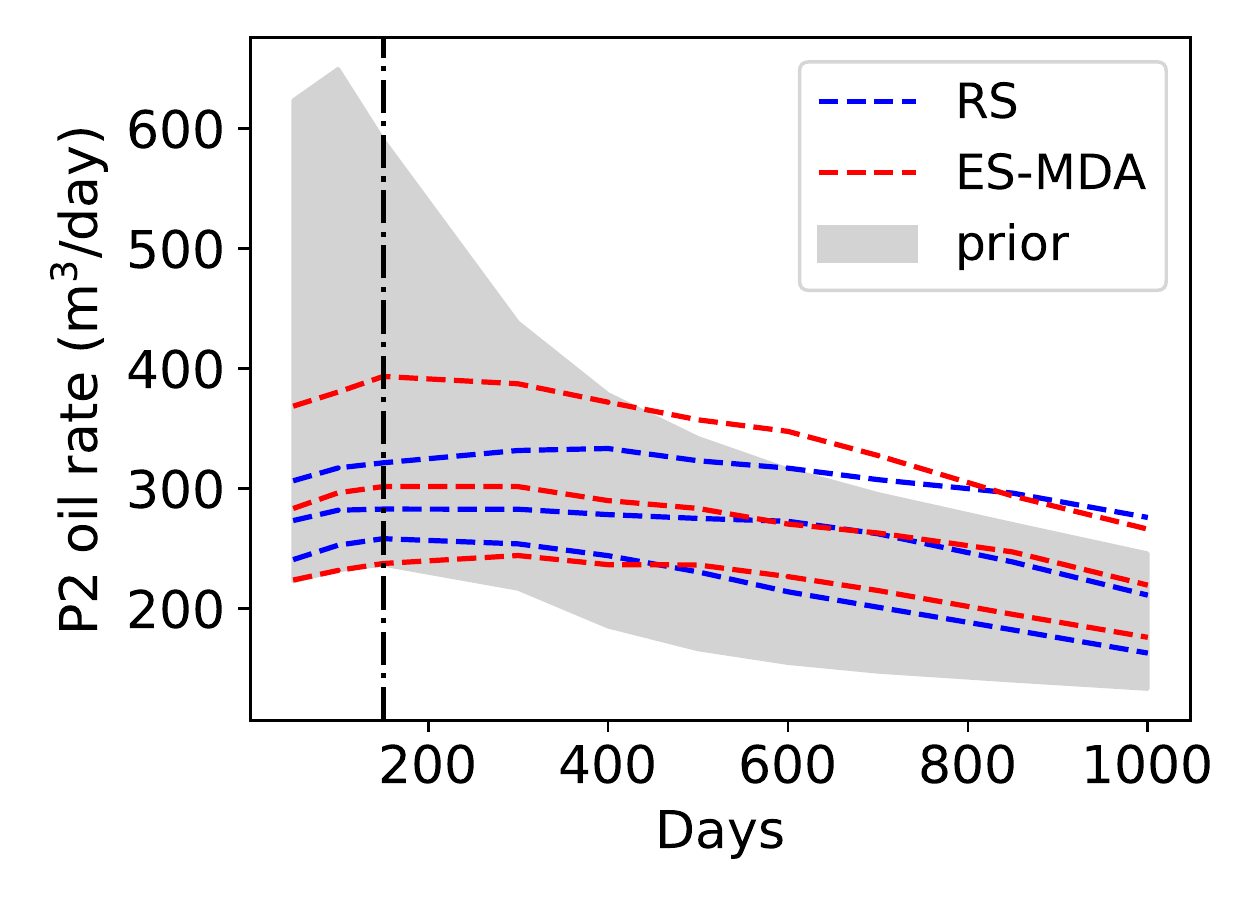}
         \caption{P2 oil rate}
         \label{hm-orate-w142}
     \end{subfigure}
     \begin{subfigure}[b]{0.36\textwidth}
         \centering
         \includegraphics[width=\textwidth]{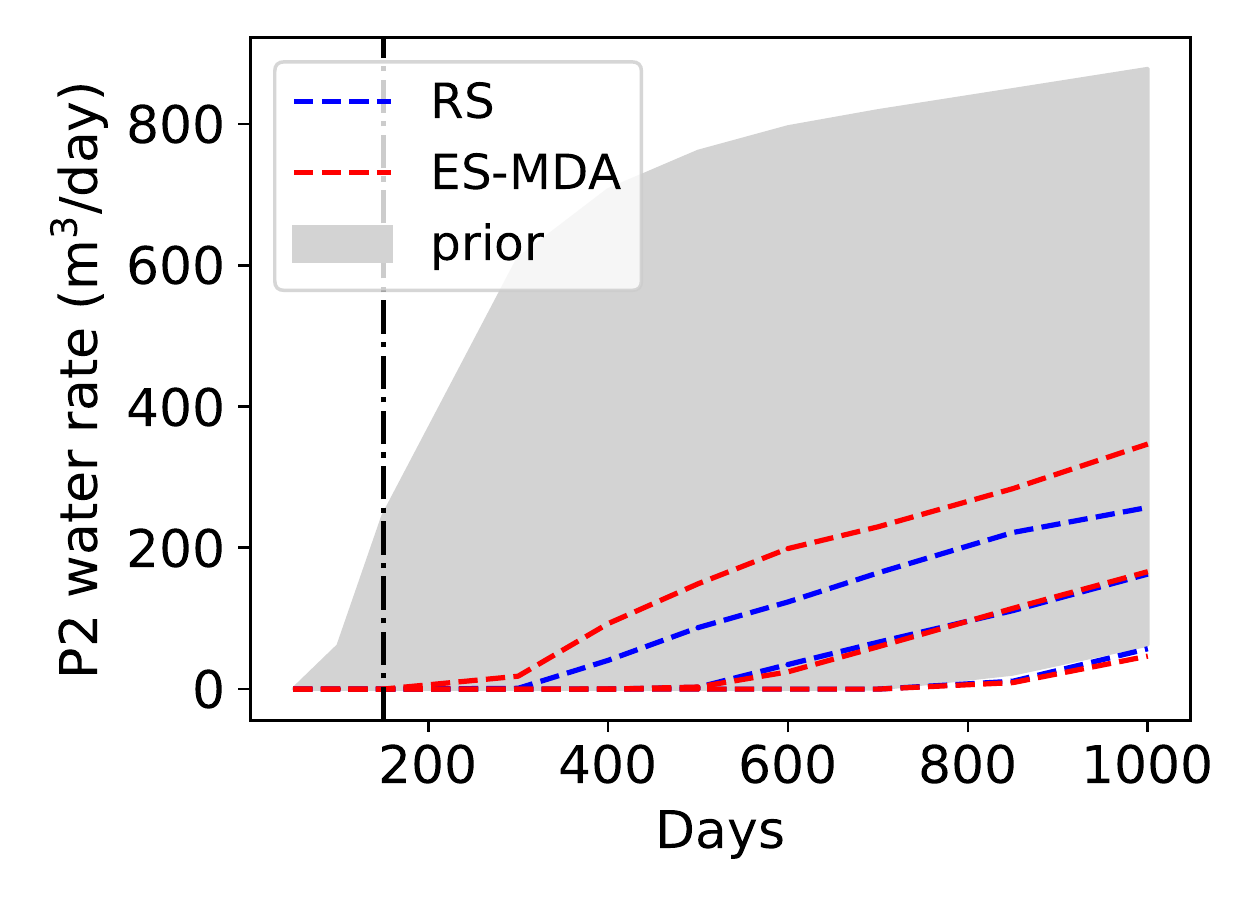}
         \caption{P2 water rate}
         \label{hm-wrate-w142}
     \end{subfigure}
     
    \begin{subfigure}[b]{0.36\textwidth}
         \centering
         \includegraphics[width=\textwidth]{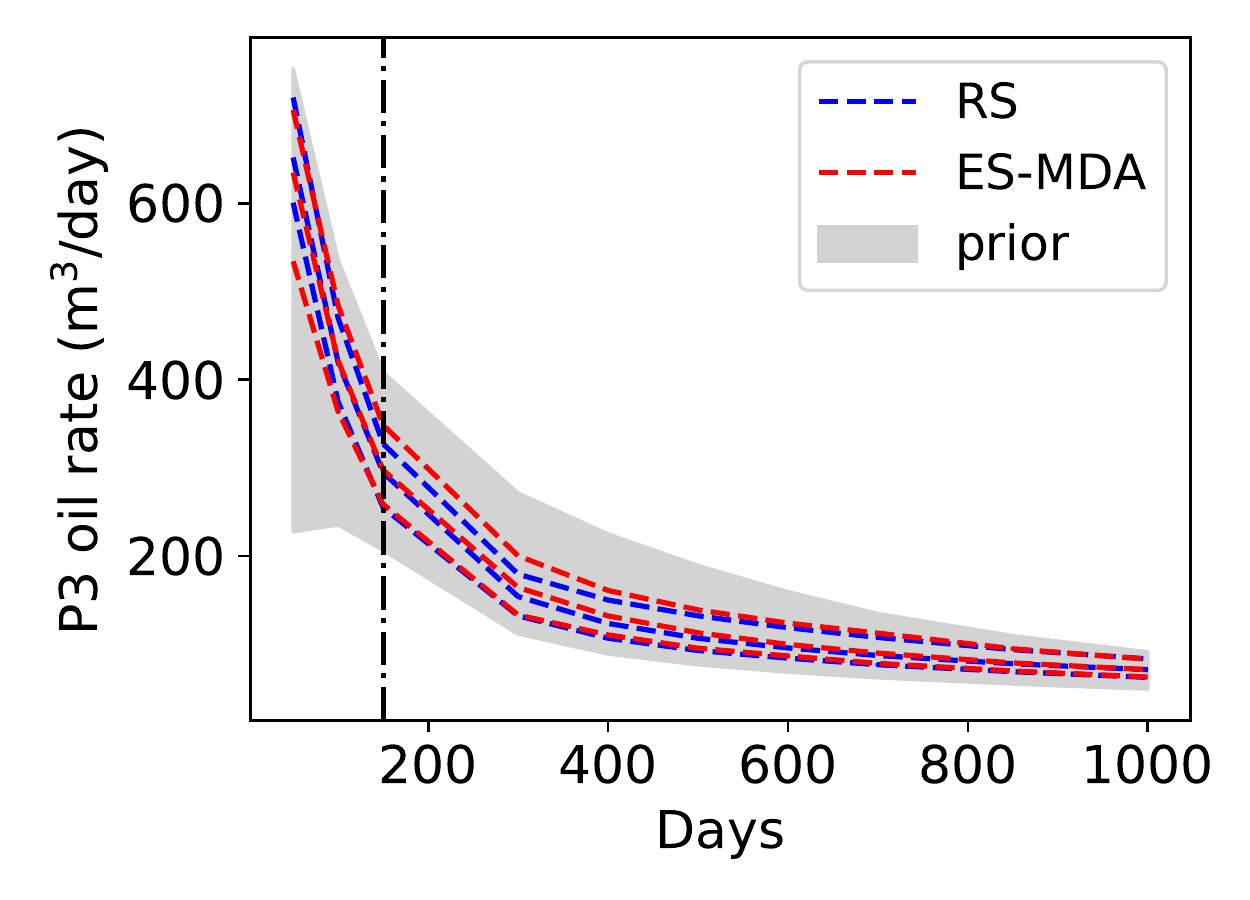}
         \caption{P3 oil rate}
         \label{hm-orate-w172}
     \end{subfigure}
          \begin{subfigure}[b]{0.36\textwidth}
         \centering
         \includegraphics[width=\textwidth]{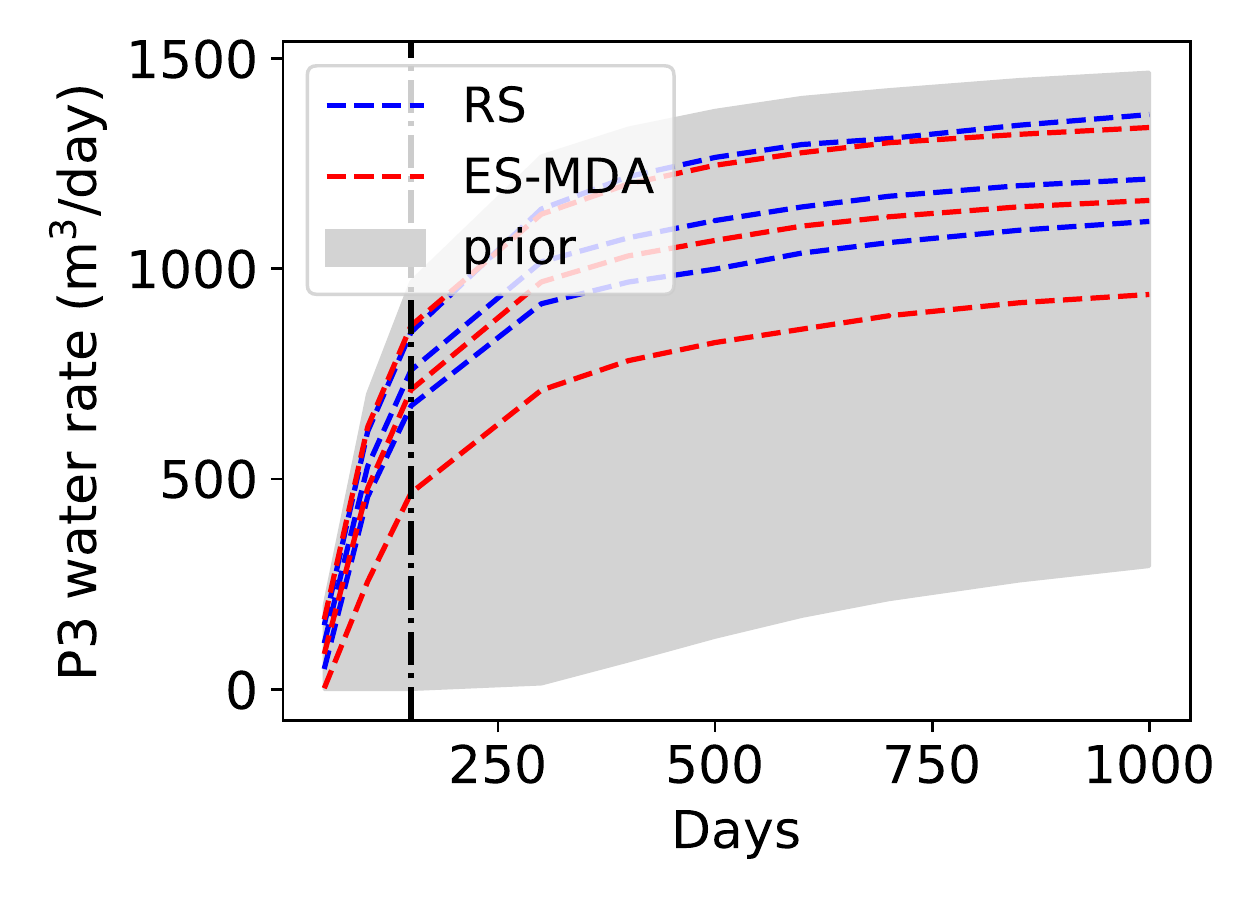}
         \caption{P3 water rate}
         \label{hm-wrate-w172}
     \end{subfigure}

    \begin{subfigure}[b]{0.36\textwidth}
         \centering
         \includegraphics[width=\textwidth]{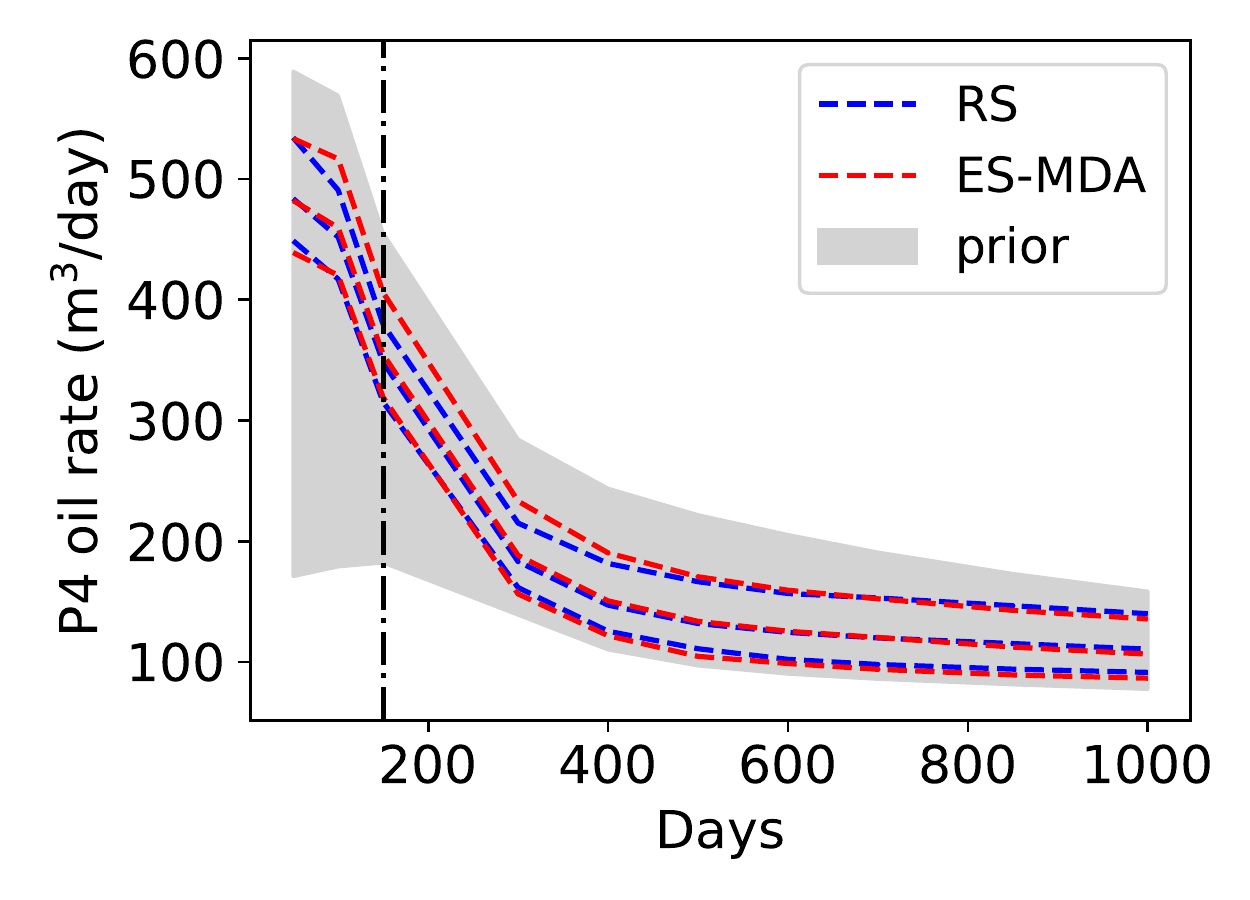}
         \caption{P4 oil rate}
         \label{hm-orate-w182}
     \end{subfigure}
          \begin{subfigure}[b]{0.36\textwidth}
         \centering
         \includegraphics[width=\textwidth]{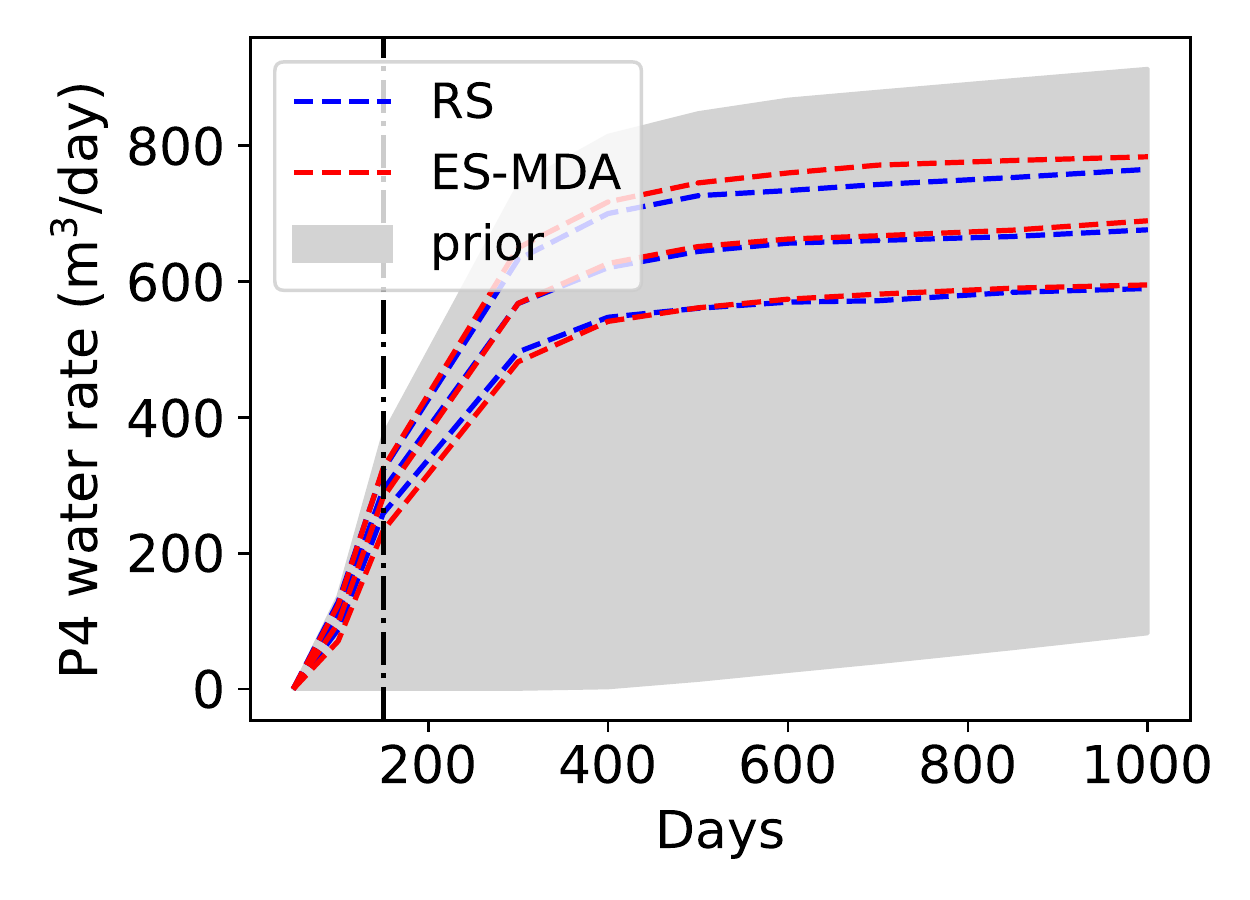}
         \caption{P4 water rate}
         \label{hm-wrate-w182}
     \end{subfigure}
    
    \caption{Oil (left) and water (right) production rates for all four production wells. Gray regions represent the prior $\text{P}_{10}$--$\text{P}_{90}$ range, blue dashed curves denote rejection sampling posterior $\text{P}_{10}$ (lower), $\text{P}_{50}$ (middle) and $\text{P}_{90}$ (upper) predictions, red dashed curves denote ES-MDA posterior $\text{P}_{10}$ (lower), $\text{P}_{50}$ (middle) and $\text{P}_{90}$ (upper) predictions. Vertical dashed line indicates the latest time at which data are collected.}
    \label{fig:hm-well-flow-comp-edmda}
\end{figure}

We specify $N_e=200$ and $N_a=10$. The same true model, the same 16 data observations, and the same standard deviation of measurement error as in Section~\ref{sec:RS_results} are used here.
A comparison of the ES-MDA and (reference) RS history matching results is presented in Fig.~\ref{fig:hm-well-flow-comp-edmda}. We see that ES-MDA is able to provide posterior estimates of many well quantities that are in close agreement with the RS results. ES-MDA does, however, overestimate posterior uncertainty in some well responses, as is evident in Fig.~\ref{fig:hm-well-flow-comp-edmda}(c), (d) and (f). 

We note that ES-MDA performance can be sensitive to the values specified for $N_a$ and the inflation coefficients $\alpha_i$. The determination of `optimal' choices for these parameters can be facilitated through comparisons to reference RS results, which would be very difficult to generate in the absence of a surrogate model for flow. We plan to assess different history matching algorithms, along with parameter tuning procedures, in future work.




\section{Concluding Remarks}
\label{sec:concl}

In this work the recurrent R-U-Net surrogate model, which is capable of predicting dynamic state quantities in subsurface flow systems, was extended to 3D. The method was applied in conjunction with 3D CNN-PCA, a deep-learning-based geological parameterization technique, to enable efficient data assimilation using different history matching algorithms. The 3D CNN-PCA formulation used here is a simplification of the more general approach described in \citep{liu20203d}. Specifically, the current implementation includes only reconstruction and hard data loss terms, which are sufficient for the system considered in this work.


The 3D recurrent R-U-Net was shown to capture the saturation and pressure evolution accurately for 3D channelized models generated using CNN-PCA. This was established both visually and through aggregate error statistics. Time-varying well responses for vertical wells perforated in multiple blocks were also assessed. Recurrent R-U-Net well responses for a particular realization, in addition to well-flow-rate statistics evaluated over the full set of 400 test cases, were shown to agree closely with reference high-fidelity simulation results.

The deep-learning-based treatments were then applied, in combination, for history matching. The rejection sampling algorithm was considered first, with new (proposed) models generated very efficiently using 3D CNN-PCA, and evaluated in terms of flow response with the 3D recurrent R-U-Net. A total of $10^6$ models were generated and evaluated (which would have been extremely time consuming using geological modeling software and HFS), and 151 were accepted, for the problem considered. Significant uncertainty reduction was achieved, and high-fidelity simulation of the accepted geomodels provided results that closely tracked 3D recurrent R-U-Net predictions. ES-MDA was then applied. The low-dimensional 3D CNN-PCA variables were updated iteratively, and flow evaluations were again accomplished using the 3D recurrent R-U-Net. RS provides `target' posterior predictions, and the tuning of ES-MDA parameters can be accomplished with reference to these results.

There are many important directions that should be considered in future work in this general area. A key challenge is the extension of the methods presented here to practical models containing, e.g., $O(10^5-10^6)$ cells. Because training time can scale with problem size, this may require the development and evaluation of different network architectures. It will also be of interest to extend our surrogate model to treat coupled flow-geomechanics systems. These problems can be very expensive to simulate, so surrogate models could be highly useful in this context. Finally, additional history matching algorithms, including those considered too expensive for use in traditional settings, can be evaluated and tuned using our deep-learning-based parameterization and surrogate flow modeling capabilities.

\section{Acknowledgements}
We are grateful to the Stanford Smart Fields Consortium and to Stanford--Chevron CoRE for partial funding of this work. We also thank the Stanford Center for Computational Earth \& Environmental Science (CEES) for providing computational resources.

%
\section*{Appendix A.~Model Architecture Details}

\subsection*{Appendix A.1.~3D CNN-PCA Transform Net Architecture}

The architecture of the 3D CNN-PCA transform net is given in Table~\ref{tab-3d-fw}. Here, $n_x$, $n_y$ and $n_z$ refer to the geomodel dimensions, `conv' represents a 3D convolutional layer followed by batch normalization and ReLU nonlinear activation, while `deconv' denotes a 3D deconvolutional (upsampling) layer followed by batch normalization and ReLU. The last `conv layer' only contains one 3D convolutional layer. A `residual block' contains a stack of two convolutional layers, each with 128 filters of size $3 \times 3 \times 3$ and stride (1, 1, 1). Within each residual block, the first 3D convolutional layer is followed by a batch normalization and a ReLU nonlinear activation. The second convolutional layer is followed only by a 3D batch normalization. The final output of the residual block is the sum of the input to the first 3D convolutional layer and the output from the second 3D convolutional layer.

\begin{table}[!htb]
\caption{Network architecture for the 3D CNN-PCA model transform net}
\centering
\begin{tabular}{ c | c  }
  \textbf{Layer} & \textbf{Output size} \\
  \hline
  Input & ($n_x$, $n_y$, $n_z$, $1$)  \\
  \\[-1em]
  conv, 32 filters of size $9 \times 9 \times 3 \times 1$, stride (1, 1, 1) & ($n_x$, $n_z$, $n_y$, $32$)  \\
  \\[-1em]
  conv, 64 filters of size $3 \times 3 \times 3 \times32$, stride (2, 2, 2) & ($n_x/2$, $n_y/2$, $n_z/2$, $64$)  \\
  \\[-1em]
  conv, 128 filters of size $3 \times 3 \times 3 \times64$, stride (2, 2, 1) & ($n_x/4$, $n_y/4$, $n_z/2$, $128$)  \\
    \\[-1em]
  residual block, 128 filters & ($n_x/4$, $n_y/4$, $n_z/2$, $128$)  \\
      \\[-1em]
  residual block, 128 filters & ($n_x/4$, $n_y/4$, $n_z/2$, $128$)  \\
      \\[-1em]
  residual block, 128 filters & ($n_x/4$, $n_y/4$, $n_z/2$, $128$)  \\
      \\[-1em]
  residual block, 128 filters & ($n_x/4$, $n_y/4$, $n_z/2$, $128$)  \\
      \\[-1em]
  residual block, 128 filters & ($n_x/4$, $n_y/4$, $n_z/2$, $128$)  \\
    \\[-1em]
  deconv, 64 filters of size $3 \times 3 \times 3 \times 128$, stride (2, 2, 1) & ($n_x/2$, $n_y/2$, $n_z/2$, $64$)  \\
      \\[-1em]
  deconv, 32 filters of size $3 \times 3 \times 3 \times 64$, stride (2, 2, 2) & ($n_x$, $n_y$, $n_z$, $32$)  \\
        \\[-1em]
  conv layer, 1 filter of size $9 \times 9 \times 3 \times 64$, stride (1, 1, 1) & ($n_x$, $n_y$, $n_z$, $1$)  \\
  \hline  
\end{tabular}
\label{tab-3d-fw}
\end{table}

\subsection*{Appendix A.2.~3D R-U-Net Schematic}

\begin{figure}[htbp]
  \centering
  \includegraphics[trim={2cm 3.5cm 2cm 4cm},clip, scale=0.8]{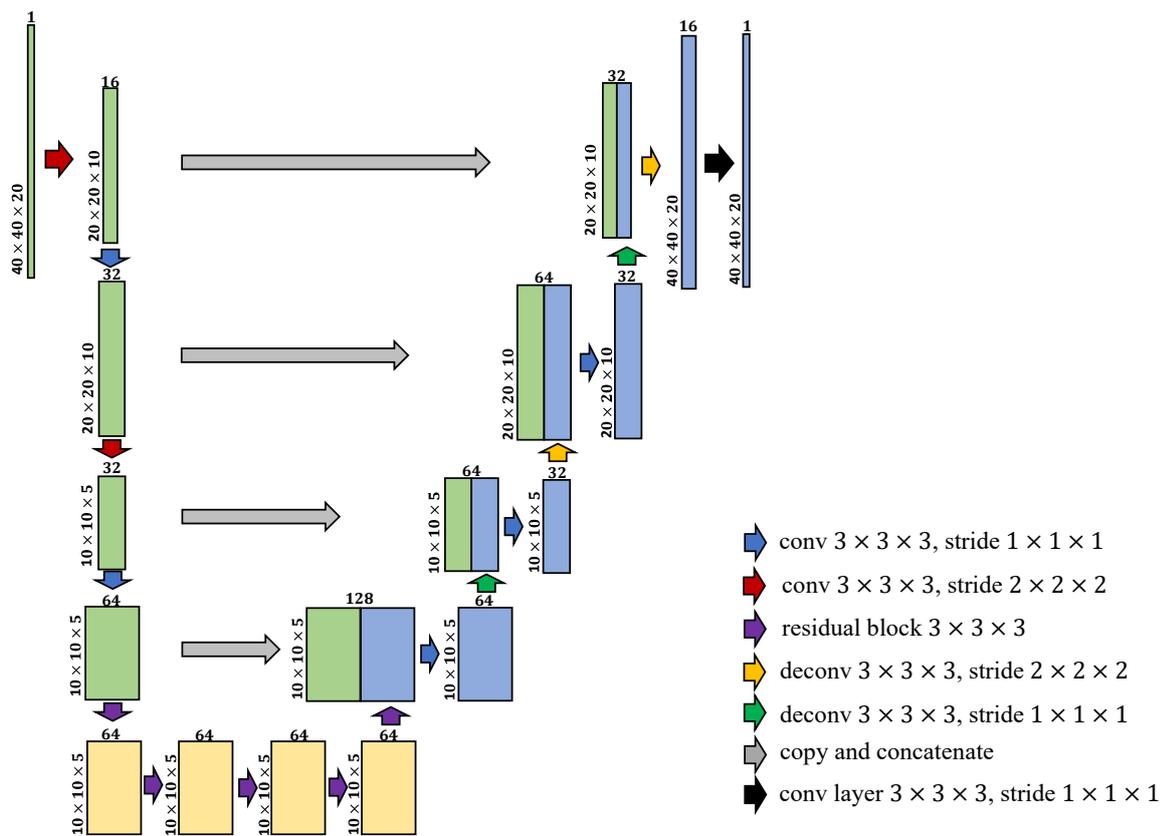}
  \caption{Schematic of the 3D R-U-Net. Boxes represent different multichannel feature maps, with values at the left indicating feature map dimensions and values above providing the number of channels (figure modified from \citep{tang2020deep}). }
  \label{fig:detailed-r-u-net}
\end{figure}

A detailed schematic of the 3D R-U-Net used in this study is provided in Fig.~\ref{fig:detailed-r-u-net}. The boxes represent 3D multichannel feature maps and the arrows denote the various operations. The values to the left of the boxes give the corresponding 3D feature map dimensions, and the values above the boxes indicate the number of channels. The operations `conv,' `residual block,' `deconv' and `conv layer' are as defined in Appendix~A.1, though here they have different numbers of filters and strides. Fig.~\ref{fig:detailed-r-u-net} shows that the extracted features in the encoding net (left) are copied and concatenated to the upsampled features in the decoding net (right).

\subsection*{Appendix A.3.~3D Recurrent R-U-Net Architecture}

The detailed architecture of the recurrent R-U-Net is shown in Table~\ref{table:r-u-net architecture}. 
In the table, we use a single number to represent stride because they are the same in all directions. The convLSTM3D block, which employs 64 filters of size $3\times3\times 3$, performs all of the LSTM gate operations. The convLSTM net generates $(n_x/4, n_y/4, n_z/4, 64)$ activation maps for all $n_t$ time steps. The decoder processes these $n_t$ activation maps separately to produce the state predictions.

\begin{table}[H]
\begin{center}
\caption{3D recurrent R-U-Net architecture details}
  \label{table:r-u-net architecture}
  \begin{tabular}{ c | c | c  }
    \hline
    \textbf{Net} &  \textbf{Layer}  & \textbf{Output size}\\ 
     \hline
    \multirow{8}{4em}{Encoder}& Input  & $(n_x, n_y, n_z, 1)$ \\
    & conv, 16 filters of size $3\times3\times3$, stride 2  & $(n_x/2, n_y/2, n_z/2, 16)$ \\
    & conv, 32 filters of size $3\times3\times3$, stride 1  & $(n_x/2, n_y/2, n_z/2, 32)$ \\
    &conv, 32 filters of size $3\times3\times3$, stride 2  & $(n_x/4, n_y/4, n_z/4, 32)$ \\
    &conv, 64 filters of size $3\times3\times3$, stride 1  & $(n_x/4, n_y/4, n_z/4, 64)$ \\
    &residual block, 64 filters of size $3\times3\times 3$, stride 1& $(n_x/4, n_y/4, n_z/4, 64)$ \\
    &residual block, 64 filters of size $3\times3\times 3$, stride 1& $(n_x/4, n_y/4, n_z/4, 64)$ \\ \hline
    ConvLSTM&convLSTM3D, 64 filters of size $3\times3 \times 3$, stride 1& $(n_x/4, n_y/4, n_z/4, 64, n_t)$ \\ \hline
    \multirow{8}{4em}{Decoder}&residual block, 64 filters of size $3\times3 \times 3$, stride 1& $(n_x/4, n_y/4, n_z/4, 64, n_t)$ \\
    &residual block, 64 filters  of size $3\times3 \times 3$, stride 1& $(n_x/4, n_y/4, n_z/4, 64, n_t)$ \\
    &deconv, 64 filters of size $3\times3 \times 3$, stride 1  & $(n_x/4, n_y/4, n_z/4, 64, n_t)$ \\
    &deconv, 32 filters of size $3\times3 \times 3$, stride 2  & $(n_x/2, n_y/2, n_z/2, 32, n_t)$ \\
    &deconv, 32 filters of size $3\times3 \times 3$, stride 1  & $(n_x/2, n_y/2, n_z/2, 32, n_t)$ \\
    &deconv, 16 filters of size $3\times3 \times 3$, stride 2  & $(n_x, n_y, n_z, 16, n_t)$ \\
    &conv layer, 1 filter of size $3\times3 \times 3$, stride 1  & $(n_x, n_y, n_z, 1, n_t)$ \\
   
    \hline
  \end{tabular}
  \end{center}
\end{table}

\newpage

\bibliographystyle{elsarticle-num-names}
\bibliography{reference, deep_rp}

\end{document}